\documentclass[11pt,a4paper]{article}

\usepackage{cite}
\usepackage{graphicx}
\usepackage{amssymb}
\usepackage{amsmath}
\usepackage{amsfonts}
\usepackage{dsfont}
\usepackage{mathtools}
\usepackage{array}
\usepackage{comment}
\usepackage[dvipsnames]{xcolor}
\usepackage{rotating}
\usepackage{bbold,amsfonts}
\allowdisplaybreaks

\usepackage[utf8]{inputenc}
\usepackage{bm}
\usepackage{xcolor}
\usepackage{float}
\usepackage{braket}
\usepackage{placeins}
\usepackage[height=8.8in,width=6.45in]{geometry}
\usepackage[font=small,labelfont=bf]{caption}
\usepackage[hidelinks]{hyperref}
\usepackage{booktabs,float,slashed}
\usepackage{standalone}
\usepackage{caption}
\usepackage{subcaption}
\usepackage{makecell}
\usepackage[normalem]{ulem}
\numberwithin{equation}{section}
\usepackage{lipsum}
\newcommand\blfootnote[1]{%
  \begingroup
  \renewcommand\thefootnote{}\footnote{#1}%
  \addtocounter{footnote}{-1}%
  \endgroup
}
\newcommand{\vev}[1]{{\left\langle #1 \right\rangle}}

\newcommand{\beq}{\begin{equation}}
\newcommand{\eeq}{\end{equation}}

\newcommand{\overbar}[1]{\mkern 1.5mu\overline{\mkern-1.5mu#1\mkern-1.5mu}\mkern 1.5mu}

\DeclareMathOperator{\Tr}{Tr}
\DeclareMathOperator{\tr}{tr}

\newcommand{\ii}{\mathrm{i}}

\makeatletter
\newcommand*{\letterdef@}{}
\newcommand*{\letterdef}[3]{%
	\def\letterdef@##1{\expandafter\newcommand\csname #1\endcsname{#2{##1}}}%
	\@tfor\@tempa :=#3\do{\expandafter\letterdef@\expandafter{\@tempa}}}
\makeatother
\letterdef{c#1} {\mathcal}{ABCDEFGHIJKLMNOPQRSTUVWXYZ} 
\letterdef{rm#1}{\mathrm} {dDeimM} 
\newcommand{\tmb}[1]{{\mbox{\tiny{#1}}}}
\newcommand{\gym}{g_\tmb{YM}}

\newcommand{\D}{{\scriptscriptstyle{\mathbf{D}}}}
\newcommand{\DS}{{\scriptscriptstyle{\mathbf{D}^*}}}
\newcommand{\NS}{{\scriptscriptstyle{\,\mathcal{N}=2^*}}}
\newcommand{\NN}{{\scriptscriptstyle{\,\mathcal{N}=4}}}

\newdimen\tableauside\tableauside=1.0ex
\newdimen\tableaurule\tableaurule=0.4pt
\newdimen\tableaustep
\def\phantomhrule#1{\hbox{\vbox to0pt{\hrule height\tableaurule
			width#1\vss}}}
\def\phantomvrule#1{\vbox{\hbox to0pt{\vrule width\tableaurule
			height#1\hss}}}
\def\sqr{\vbox{%
		\phantomhrule\tableaustep
		\hbox{\phantomvrule\tableaustep\kern\tableaustep\phantomvrule\tableaustep}%
		\hbox{\vbox{\phantomhrule\tableauside}\kern-\tableaurule}}}
\def\squares#1{\hbox{\count0=#1\noindent\loop\sqr
		\advance\count0 by-1 \ifnum\count0>0\repeat}}
\def\tableau#1{\vcenter{\offinterlineskip
		\tableaustep=\tableauside\advance\tableaustep by-\tableaurule
		\kern\normallineskip\hbox
		{\kern\normallineskip\vbox
			{\gettableau#1 0 }%
			\kern\normallineskip\kern\tableaurule}%
		\kern\normallineskip\kern\tableaurule}}
\def\gettableau#1 {\ifnum#1=0\let\next=\null\else
	\squares{#1}\let\next=\gettableau\fi\next}
\newcommand{\Yfund}{\tableau{1}}
\newcommand{\Ysymm}{\tableau{2}}
\newcommand{\Yasymm}{\tableau{1 1}}

\makeatletter
\newsavebox{\@brx}
\newcommand{\llangle}[1][]{\savebox{\@brx}{\(\m@th{#1\langle}\)}%
  \mathopen{\copy\@brx\kern-0.5\wd\@brx\usebox{\@brx}}}
\newcommand{\rrangle}[1][]{\savebox{\@brx}{\(\m@th{#1\rangle}\)}%
  \mathclose{\copy\@brx\kern-0.5\wd\@brx\usebox{\@brx}}}

\begin{document}

\begin{titlepage}

\begin{flushright}
\small
\texttt{QMUL-PH-25-32}
\end{flushright}

\vspace*{3mm}
\begin{center}
{\LARGE \bf 
$\mathcal{N}=2$ Universality at Strong Coupling
}

\vspace*{10mm}

{\large L. De Lillo${}^{\,a,b}$, Z. Duan${}^{\,c,d}$, M. Frau${}^{\,a,b}$, F. Galvagno${}^{\,a,b,d}$,\\[3mm] A. Lerda${}^{\,e,b}$, P. Vallarino${}^{\,f}$, C. Wen${}^{\,d}$}

\vspace*{8mm}

${}^a$ Dipartimento di Fisica, Universit\`a di Torino,
			Via P. Giuria 1, I-10125 Torino, Italy
			\vskip 0.3cm
			
${}^b$   I.N.F.N. - sezione di Torino,
			Via P. Giuria 1, I-10125 Torino, Italy 
			\vskip 0.3cm

${}^c$  Section de Math\'ematiques, Universit\'e de Gen\`eve, 1211 Gen\`eve 4, Switzerland
            \vskip 0.3cm
			
${}^d$ Centre for Theoretical Physics, Department of Physics and Astronomy, \\
Queen Mary University of London, London E1 4NS, UK
			\vskip 0.3cm
            
${}^e$  Dipartimento di Scienze e Innovazione Tecnologica, Universit\`a del Piemonte Orientale,\\
			Viale T. Michel 11, I-15121 Alessandria, Italy
			\vskip 0.3cm
       
${}^f$ Theoretische Natuurkunde, Vrije Universiteit Brussel and The International Solvay Institutes,\\ 
            Pleinlaan 2, B-1050 Brussels, Belgium

\vspace*{0.8cm}
\end{center}

\begin{abstract}
We present a detailed analysis of integrated correlators for an $\mathcal{N}=2$ superconformal field theory on a squashed sphere with SU$(N)$ gauge group and fundamental/anti-symmetric matter. Employing the matrix model arising from supersymmetric localisation, we compute derivatives of the partition function $\mathcal{Z}$ with respect to the fundamental mass ($\mu$), the anti-symmetric mass ($m$) and the squashing parameter ($b$), corresponding to integrated insertions of the $\mathcal{N}=2$ flavour-current and stress-tensor multiplets, which are holographically dual to
gluon and graviton scatterings in the presence of D7-branes. For correlators dual to only graviton scatterings, we confirm the planar-limit equivalence with $\mathcal{N}=4$ SYM. Our main result is a remarkable universality for the mixed gluon-graviton scattering amplitudes off D7-branes, obtained from $\partial_\mu^2 \partial_m^2 \log \mathcal{Z}$ and $\partial_\mu^2 \partial_b^2 \log \mathcal{Z}$. We show that the leading and sub-leading large-$N$ contributions in the strong-coupling regime are governed by universal asymptotic series, identical to those found for integrated giant-graviton correlators in $\mathcal{N}=4$ SYM. We also propose an $\text{SL}(2, \mathbb{Z})$-invariant completion of these results in terms of non-holomorphic Eisenstein series. This completion provides exact constraints on higher-derivative terms in the dual $\text{AdS}_5$ brane-string amplitudes and highlights an unexpected universality across distinct superconformal theories at strong coupling.
\end{abstract}
    \blfootnote{E-mail:\,\,\texttt{lorenzo.delillo, marialuisa.frau, francesco.galvagno@unito.it; Zhihao.Duan@unige.ch; \\ lerda@to.infn.it; Paolo.Vallarino@vub.be; c.wen@qmul.ac.uk}}
\end{titlepage}
\setcounter{tocdepth}{2}
\tableofcontents

\section{Introduction and summary of results}
\label{sec:intro}
In the study of non-perturbative phenomena in Quantum Field Theories (QFTs), localisation has emerged as a very powerful framework for obtaining exact results in supersymmetric theories \cite{Pestun:2007rz, Pestun:2016zxk}, thanks to the reduction of the QFT path-integral to a finite dimensional matrix model. Classic examples of such exact computations using localisation include the vacuum expectation value of $\frac{1}{2}$-BPS Wilson loops \cite{Pestun:2007rz} and extremal correlators of $\frac{1}{2}$-BPS local operators \cite{Gerchkovitz:2016gxx} in four-dimensional conformal theories with extended supersymmetry ($\mathcal{N}\geq 2$). In these cases, the spacetime dependence 
of the observables is completely fixed by symmetries, and localisation provides the exact coupling-dependent coefficients multiplying these fixed structures.

This idea has recently been generalized to observables with  nontrivial spacetime dependence, such as higher-point correlators of local operators \cite{Binder:2018yvd, Binder:2019jwn, Binder:2019mpb, Chester:2020dja} and correlators in the presence of defects \cite{Pufu:2023vwo, Billo:2023ncz, Dempsey:2024vkf, Billo:2024kri}. This generalization is achieved by considering {\emph{integrated correlators}}, in which the spacetime dependence is integrated against specific supersymmetry-preserving measures. From the localisation perspective, this corresponds to deforming the Super-Conformal Field Theory (SCFT) by suitable supersymmetric operators and taking derivatives of the partition function with respect to the deformation parameters.

When applied to $\cN=4$ SYM, this procedure yields integrated constraints for the four-point correlators of $\frac{1}{2}$-BPS scalar operators belonging to the $\cN=4$ stress-tensor multiplet
or higher-dimensional chiral primaries. These results have provided deep insights into the $\cN=4$ non-perturbative dynamics, modular structure and S-duality properties which have been extensively studied both analytically and numerically \cite{Binder:2019jwn, Chester:2020dja,Chester:2019jas, Chester:2020vyz, Green:2020eyj, Dorigoni:2021guq, Dorigoni:2021bvj, Alday:2021vfb, Dorigoni:2022zcr, Collier:2022emf, Paul:2022piq, Dorigoni:2023ezg, Dorigoni:2022zcr, Alday:2023pet,Chester:2021aun, Caron-Huot:2024tzr, Chester:2024bij}. 
Moreover, the $\cN=4$ integrated correlators at finite charges have been used in the AdS$_5$/CFT$_4$ correspondence to obtain constraints on four-graviton scattering processes (and higher Kaluza-Klein modes) in AdS space. Recently, also the integrated correlators involving heavy operators have been considered. For operators of conformal dimensions larger than $N^2$, localisation combined with semiclassical techniques has provided exact results for the heavy-heavy-light-light correlators \cite{Paul:2023rka, Brown:2023why, Caetano:2023zwe, Brown:2024yvt,Grassi:2024bwl,Brown:2025cbz}. Particularly relevant to this paper are the cases involving operators with conformal dimensions scaling as $N$, which are called giant gravitons and are dual to D3-branes (wrapping an $S^3$ inside either the $S^5$ or the AdS$_5$ factor of the AdS$_5 \times S^5$). The corresponding integrated correlators provide exact constraints on the scattering of two gravitons off D3-branes~\cite{Brown:2024tru, Brown:2025huy}.

Analogous procedures have been applied to theories with reduced supersymmetry, including several classes of $\mathcal{N}=2$ SCFTs, which allow for a broader range of constructions and physical setups
\cite{Chester:2022sqb,Fiol:2023cml,Behan:2023fqq, Billo:2023kak,Pini:2024uia,Billo:2024ftq,Pini:2024zwi, DeLillo:2025hal,Chester:2025ssu,DeSmet:2025mbc,Damia:2025eed,DeLillo:2025eqg}. The matrix model integrals arising from localisation in $\mathcal{N}=2$ SCFTs are typically more involved than in $\mathcal{N}=4$ SYM, and the resulting expressions are correspondingly more complex. Nevertheless, explicit examples \cite{Beccaria:2020azj,Beccaria:2021hvt,Billo:2021rdb,Billo:2022xas,Billo:2022fnb,Billo:2022lrv,Billo:2023kak}
have already revealed an intriguing phenomenon: certain observables in special $\mathcal{N}=2$ SCFTs coincide with those of $\mathcal{N}=4$ SYM in the 't Hooft large-$N$ limit, a property often referred to as planar equivalence. 

In this work, we further explore these connections, both between $\mathcal{N}=4$ SYM and $\mathcal{N}=2$ SCFTs, and among different $\mathcal{N}=2$ theories.
We focus primarily on a special $\cN=2$ theory, known as the \textbf{D}-theory, originally considered in \cite{Billo:2019fbi,Beccaria:2020hgy,Beccaria:2021ism}.
This is an $\cN=2$ SCFT with gauge group SU($N$) and a matter content consisting of two anti-symmetric and four fundamental hypermultiplets. It can be engineered in Type IIB string theory with $N$ D3-branes in the presence of a $\mathbb{Z}_2$-orbifold probing an O7-orientifold background with $(4+4)$ D7-branes which give rise to a U(4)$\,\subset\,$SO(8) flavour symmetry \cite{Park:1998zh,Ennes:2000fu}. 
A closely related $\mathcal{N}=2$ SCFT with Sp($N$) gauge group, one anti-symmetric and four fundamental matter hypermultiplets and an SO(8) flavour symmetry, has been recently considered in \cite{Beccaria:2021ism,Beccaria:2022kxy,Behan:2023fqq,Alday:2024yax,Chester:2025ssu}. 
In both setups, localisation provides exact field-theoretic results that can be interpreted holographically as constraints on gluon and graviton scattering amplitudes in AdS in the presence of D7-branes. These observables can be compared with the scattering of gravitons off a D3 brane in $\cN=4$ SYM, arising from heavy insertions in the large-$N$ limit as discussed above. 
Remarkably, despite their distinct microscopic origins and very different weak-coupling expressions, we find that their strong-coupling expansions are governed by the same asymptotic series, revealing a surprising universality among these observables. 

\subsection{Summary of results and outline}
We now briefly summarize our findings. 
One of the main advances of this paper is at the level of the matrix-model computations. 
Previous studies of integrated correlators in $\mathcal{N}=2$ SCFTs have focused on mass deformations, corresponding to integrated insertions of the so-called moment-map operator. Here, we introduce also a \textit{squashing deformation} of the four-sphere, corresponding to integrated insertions of the $\mathcal{N}=2$ stress-tensor multiplet.
In particular, we consider 
a deformation of the \textbf{D}-theory, which we call the $\mathbf{D}^*$-theory.
This deformed theory is defined on an ellipsoid with squashing parameter $b$, where the two anti-symmetric and the four fundamental hypermultiplets acquire masses $m$ and $\mu$, respectively. 
Although they arise from distinct supermultiplets on the field-theory side, in the holographic dual both the $b$- and $m$-deformations correspond to closed-string excitations  in AdS that probe graviton scattering processes. In contrast, the $\mu$-deformation corresponds to open-string excitations in AdS probing the scattering of gluons on the D7-brane world-volume.

We derive in detail the matrix model for the $\mathbf{D}^*$-theory, paying particular attention to its dependence on $b$ and clarifying some subtleties in its construction.
Denoting by $\cZ^{\mathbf{D}^*}$ the partition function of the $\mathbf{D}^*$-theory, we then study the following quantities: 
\begin{equation}
   \partial^4_m \log \cZ^{\mathbf{D}^*}\big|_{\D}~, \qquad    \big(\partial^4_b -15 \partial^2_b\big) \log \cZ^{\mathbf{D}^*}\big|_{\D}~, \qquad   \partial^2_m\partial^2_b \log \cZ^{\mathbf{D}^*}\big|_{\D}\, ,
\end{equation}
where the notation $|_\D$ means that the derivatives are evaluated in the undeformed \textbf{D}-theory. 
In the planar limit, we find that these quantities are identical to those in $\mathcal{N}=4$ SYM~\cite{Chester:2020vyz}. 
This fact points to a universal structure underlying the integrated correlators of both the \textbf{D}-theory and $\mathcal{N}=4$ SYM.
 
Next, we consider other types of integrated correlators, corresponding to
\begin{equation}
   \partial^2_m \partial^2_\mu \log \cZ^{\mathbf{D}^*}\big|_{\D}~, \qquad  \quad   \partial^2_\mu \partial^2_b \log \cZ^{\mathbf{D}^*}\big|_{\D}~,
   \label{mixed_0}
\end{equation} 
which are holographically dual to mixed scattering amplitudes involving two gluons and two gravitons. 
At large $N$, both these observables admit a topological expansion of the form
\begin{equation}
   \mathcal{F}(\lambda',N) = \sum_{g=0}^\infty N^{1-g} \, \cF_g (\lambda') ~,
\end{equation}
where $\lambda'$ denotes the shifted 't Hooft coupling, $1/\lambda' = 1/\lambda+ \log(2)/(2\pi^2 N)$.
While the weak-coupling series exhibit intricate structures, the strong-coupling expansions simplify dramatically. 
Indeed, in both cases the leading contribution $\mathcal{F}_0$
takes the form
\begin{align} \label{eq:Ft0}
    \mathcal{F}_0 (\lambda') \underset{\lambda' \rightarrow \infty}{\sim} \,&\,   a_0 + \frac{\pi^2 a_1}{\lambda'}-\sum_{n=1}^\infty\frac{64\,n\,\Gamma(n-\frac{1}{2})^2\,\Gamma(n+\frac{1}{2})\,\zeta(2n+1)}{\pi^{3/2}\,\Gamma(n)\,(\lambda')^{n+1/2}}~,
\end{align}
while the sub-leading term $\mathcal{F}_1$ is given by
\begin{align} \label{eq:Ft1}
    \mathcal{F}_1 (\lambda') ~\underset{\lambda' \rightarrow \infty}{\sim}   b_0 + \frac{3  a_1}{8} \log(\lambda')  +\sum_{n=1}^\infty \frac{16\,n\,\Gamma(n-\frac{1}{2})\,\Gamma(n+\frac{1}{2})^2\,\zeta(2n+1)}{\pi^{3/2}\,\Gamma(n)\,(\lambda')^{n+1/2}}~,
\end{align}
where $a_0, a_1$ and $b_0$ are constants. These constants are the only non-universal terms that depend on the specific deformation.
Remarkably, the same asymptotic expansions \eqref{eq:Ft0} and \eqref{eq:Ft1} also describe the integrated giant-graviton correlators in $\mathcal{N}=4$ SYM \cite{Brown:2024tru, Brown:2025huy}\,\footnote{In the case of giant-graviton correlators, there is no redefinition of the 't Hooft coupling, and the ratio between the coefficients of the $1/\lambda$ and $\log(\lambda)$ terms is $4\pi^2/3$ (rather than $8\pi^2/3$ as in \eqref{eq:Ft0} and \eqref{eq:Ft1}). This factor of $2$ becomes relevant in the SL$(2,\mathbb{Z})$ completion discussed below and in details in Section\,\ref{secn:verystrong}, and is compensated by the fact that the complexified YM coupling in $\mathcal{N}=2$ theories, as given in \eqref{eq:n2YM}, is twice that of $\mathcal{N}=4$ SYM.},
as well as analogous integrated correlators in the  $\mathcal{N}=2$ SCFT with Sp($N$) gauge group \cite{Chester:2025ssu}. 
The presence of common asymptotic series at strong coupling across distinct observables in different theories is highly nontrivial and surprising.

Finally, considering the large-$N$ expansion with fixed YM coupling $\tau_2 = 8\pi N/\lambda'$, we show that the combination $N\mathcal{F}_1+\mathcal{F}_0$ admits an ${\rm SL}(2, \mathbb{Z})$-invariant completion in terms of the non-holomorphic Eisenstein series 
given by
  \begin{align}
& c_0+ \frac{3 a_1}{8}   E(1;\tau,\bar{\tau})-\frac{E\big(\frac{3}{2};\tau,\bar{\tau}\big)}{\sqrt{2}\,N^{1/2}} -\frac{3\,E\big(\frac{5}{2};\tau,\bar{\tau}\big)-4\,E\big(\frac{3}{2};\tau,\bar{\tau}\big)}{32\sqrt{2}\,N^{3/2}}\\[2mm]
&-\frac{405\,E\big(\frac{7}{2};\tau,\bar{\tau}\big)-288\,E\big(\frac{5}{2};\tau,\bar{\tau}\big)+\ldots}{8192\sqrt{2}\,N^{5/2}}-\frac{7875\,E\big(\frac{9}{2};\tau,\bar{\tau}\big)-4050\,E\big(\frac{7}{2};\tau,\bar{\tau}\big)+\ldots}{131072\sqrt{2}\,N^{7/2}}+O(N^{-9/2})~, \notag
\end{align}
where $c_0$ is a coupling-independent constant 
and the ellipses denote additional non-holomorphic Eisenstein series of lower half-integer index. 
We observe that the appearance of the Eisenstein series
$E(1; \tau, \bar{\tau})$ and  $E(\frac{3}{2}; \tau, \bar{\tau})$ in the first two orders is fully consistent with the leading higher-derivative corrections arising from the $\alpha'$-expansion of the dual superstring amplitudes \cite{Bachas:1999um, Green:2000ke, Kiritsis:2000zi, Lin:2015ixa}. 

This paper is organized as follows. Section\,\ref{secn:General} reviews how integrated correlators in $\mathcal{N}=2$ SCFTs can be obtained as derivatives of the partition functions with respect to supersymmetric deformations, such as masses and squashing.  
Section\,\ref{sec:squashing} presents the matrix model of the $\mathbf{D}^*$-theory and the derivation of the main inputs for our subsequent computations. In Section\,\ref{secn:integrated} we study integrated correlators dual to four-graviton amplitudes and show that at the leading order in the large-$N$ limit they are identical to those of $\mathcal{N}=4$ SYM. Section\,\ref{secn:integratedmixed} extends the analysis to mixed amplitudes involving two gluons and two gravitons, using both the Lie-algebra and the topological recursion approaches to obtain their large-$N$ expansion. In Section\,\ref{secn:strongcoupling} we analyze the strong-coupling regime and show that all observables exhibit a universal asymptotic behavior that matches the one of integrated giant-graviton correlators in $\mathcal{N}=4$ SYM. In Section\,\ref{secn:verystrong} we consider the large-$N$ expansion at fixed YM coupling, where instanton effects become relevant, and show that our results can be completed by non-holomorphic Eisenstein series, yielding an SL(2, $\mathbb{Z}$)-invariant form. Finally, Section\,\ref{sec:concl} summarizes our conclusions and outlines future research directions. Several appendices collect technical details and complementary results.

\section{SUSY deformations and integrated correlators}
\label{secn:General}
In this section, we review how integrated correlators in SCFTs with extended supersymmetry can be computed using localisation.

The procedure begins by placing the theory on a compact curved space (typically a four-sphere $S^4$ or an ellipsoid) and introducing suitable deformations that preserve $\mathcal{N}=2$ supersymmetry.
The integrated correlators are then obtained by differentiating the partition function with respect to the deformation parameters. This differentiation effectively inserts into the path-integral local operators integrated over their spacetime positions. 
On the other hand, the deformed partition function on such curved spaces, when at least $\mathcal{N}=2$ supersymmetry is preserved, can be computed exactly by means of supersymmetric localisation \cite{Pestun:2007rz,Pestun:2016zxk}. Therefore, the integrated correlators take the following schematic form:
\begin{align} \label{eq:int_corr_schem}
   \int \!d \mu(x_i) \, \big\langle \cO_1(x_1) \cO_2(x_2) \cO_3(x_3) \cO_4(x_4) \big\rangle = \partial_{h_1}  \partial_{h_2}  \partial_{h_3}  \partial_{h_4} \, \log \mathcal{Z}(h_i) \big|_{h_i = 0} ~, 
\end{align}
where $\mu(x_i)$ denotes the supersymmetry-preserving integration measure and $\mathcal{Z}(h_i)$ is the deformed partition function depending on the deformation parameters $h_i$.
Since the right-hand side of \eqref{eq:int_corr_schem} is exactly computable via localisation, this relation provides a set of integral constraints on correlation functions, valid for arbitrary values of the theory's parameters.

\subsection{SUSY-preserving deformations in SCFTs}
We now illustrate this general procedure by first reviewing the most extensively studied supersymmetric deformations in $\cN=2$ SCFTs, namely the chiral/anti-chiral Coulomb-branch deformations and the mass deformations. Later we extend the discussion to integrated correlators corresponding to squashing deformations.

The deformations associated with the Coulomb-branch chiral and anti-chiral operators $\mathcal{A}_p$ and $\bar{\mathcal{A}}_p$ arise from the corresponding couplings $\kappa_p$ and $\bar{\kappa}_p$. In particular, $\cA_2$ corresponds to the exactly marginal coupling
\begin{equation} \label{eq:n2YM}
  \kappa_2\equiv\tau= \frac{\theta}{\pi} + \ii \frac{8\pi}{\gym^2} \, ,
\end{equation}
where $\gym$ is the YM coupling and $\theta$ the vacuum angle.
Following the prescription of \cite{Gerchkovitz:2016gxx}, the integrated insertions of these chiral and anti-chiral operators can be implemented by placing $\mathcal{A}_p$ at the North pole and $\bar{\mathcal{A}}_p$ at the South pole of $S^4$, thus adding to the action the terms
\begin{equation}
    S_{\kappa_p}=\kappa_p \,\cA_p(N)~,\qquad S_{\bar\kappa_p}=\bar\kappa_p\, \bar\cA_p(S)~.
\end{equation}

For the mass deformations, instead, we can follow the recent approach of \cite{Dempsey:2024vkf}. 
Consider an $\mathcal{N}=2$ SCFT with a $\mathrm{U}(1)$ flavour symmetry (usually embedded in a larger non-Abelian group). The associated flavour current multiplet contains an $\mathfrak{su}(2)_R$ triplet of scalars $\Phi^{ij}=\Phi^{ji}$ (with $i,j=1,2$) satisfying the reality condition $(\Phi^{ij})^*=\epsilon_{ik}\,\epsilon_{j\ell}\,\Phi^{k\ell}$, an $\mathfrak{su}(2)_R$ doublet of chiral and anti-chiral fermions $X_{i\alpha}$ and $\overbar{X}_i^{\,\dot\alpha}$, two real scalars $P$ and $\overbar{P}$ of opposite $\mathfrak{u}(1)_R$ charges, and the conserved flavour current $j_\mu$.
A supersymmetry-preserving deformation can be constructed by coupling this flavour current multiplet to an off-shell background vector multiplet, whose components we denote as $(A_\mu, \lambda_{i\alpha}, \overbar{\lambda}_i^{\,\dot\alpha},\varphi,\bar{\varphi},Y_{ij})$.
Assigning nontrivial vacuum expectation values to the scalar and auxiliary fields in this background yields a supersymmetric mass deformation. A configuration preserving the desired supersymmetry on $S^4$ takes the form \cite{Dempsey:2024vkf}
\begin{equation}\label{eq:mass_back}
\varphi = \frac{m\, \rme^{\ii\vartheta}}{2}~,\qquad \overbar\varphi = \frac{m \,\rme^{-\ii\vartheta}}{2}~,\qquad Y_{ij} = \pm \ii \,\frac{m}{2r}\,\delta_{ij}~,
\end{equation}
with all other components vanishing. Here, $m$ is the mass parameter, $\vartheta$ is one of the angular coordinates of $S^4$ required to maintain supersymmetry and $r$ is the radius of the sphere. The corresponding mass deformation of the action is:
\begin{equation}\label{eq:mass_deform}
    S_{m} = \frac{m}{2} \int \!d^4 x\,
\sqrt{g} \,\Big[ \rme^{\ii\vartheta} P + \rme^{-\ii\vartheta} \overbar P\pm \frac{\ii}{r} \big(\Phi^{11}+\Phi^{22}\big) \Big]~,
\end{equation}
where $g$ is the determinant of the metric of $S^4$. Differentiating with respect to $m$ yields integrated correlators of $P$, $\overbar P$ and $\Phi^{ij}$, which are related among themselves by supersymmetric Ward identities. This procedure determines the explicit form of the integration measure in \eqref{eq:int_corr_schem}\,\footnote{See \cite{Binder:2019jwn} and \cite{Chester:2020dja} for the cases involving two and four mass-derivatives, respectively, and \cite{Billo:2023ncz,Dempsey:2024vkf,Billo:2024kri} for the analogous derivation for two mass derivatives in presence of a line defect.}.

The best studied cases of integrated correlators arising from mass deformations are in $\cN=4$ SYM, which can be regarded as an $\cN=2$ SCFT with a single adjoint hypermultiplet. In this theory, the mass deformation breaks the R-symmetry according to $\mathrm{SU}(4)_R \to \mathrm{SU}(2)_F\times \mathrm{SU}(2)_R\times \mathrm{U}(1)_R$, with $\mathrm{SU}(2)_F$ being the flavour symmetry. The associated flavour current multiplet belongs to the $\mathcal{N}=4$ stress-tensor multiplet, whose top-component is the well-known $\mathbf{20^\prime}$ operator $\cO_2$. In this setting, two classes of integrated correlators can be obtained from the mass-deformed theory (also known as the $\mathcal{N}=2^*$ theory) via
\begin{equation}
\partial_{\kappa_p}\partial_{\bar\kappa_p}\partial^2_m\log \cZ^{\NS}\big|_{0}~, \qquad \partial^4_m \log \cZ^{\NS}\big|_{0}~,
\end{equation}
where the notation $|_0$ denotes evaluation in the undeformed $\mathcal{N}=4$ SYM, namely at $\kappa_p=\bar\kappa_p=m=0$. These observables provide
integrated constraints on the $\cN=4$ correlators $\vev{\cO_p \cO_p \cO_2 \cO_2}$ and $\vev{\cO_2 \cO_2 \cO_2 \cO_2}$, with $\cO_p$ a gauge-invariant scalar operator of dimensions $p$. 
The explicit integration measures for these correlators are collected in Appendix~\ref{app:A}, and many detailed computations can be found in
\cite{Binder:2019jwn, Chester:2020dja,Chester:2019jas, Chester:2020vyz, Green:2020eyj, Dorigoni:2021guq, Dorigoni:2021bvj, Alday:2021vfb, Dorigoni:2022zcr, Collier:2022emf,Paul:2022piq, Dorigoni:2023ezg, Dorigoni:2022zcr, Alday:2023pet,Chester:2021aun, Caron-Huot:2024tzr, Chester:2024bij}.

Another supersymmetry-preserving deformation of $\cN=2$ SCFTs on $S^4$
is the squashing deformation, characterized by a parameter $b$ chosen in such a way that the round-sphere limit corresponds to $b\to1$.
Defining the theory on the squashed sphere and taking derivatives with respect to $b$, we generate a new class of integrated correlators involving the $\cN=2$ stress-energy tensor. For $\cN=4$ SYM,
where the operator associated to the $b$-deformation is again part of the larger $\cN=4$ stress-tensor multiplet, two notable classes of such correlators are\,\footnote{The specific combination
$\big(\partial_b^4-15\partial_b^2\big)$ yields a scheme-independent result as pointed out in \cite{Chester:2020vyz}.}
\begin{equation}
    \partial^2_m\partial^2_b\log \cZ^{\NS}\big|_{0}~, \qquad \big(\partial^4_b -15 \partial^2_b\big) \log \cZ^{\NS}\big|_{0}~,
\end{equation}
where now $\cZ^\NS_b$ is the partition function of the $\cN=2^*$ theory in presence of the $b$-deformation and the notation $|_0$ means evaluation at $m=0$ and $b=1$.
Integrated correlators arising from the squashing deformation remain largely unexplored. While localisation-based computations of the $b$-derivatives in the maximally supersymmetric case have been considered in \cite{Chester:2020vyz}, the derivation of the corresponding integration measure in the left-hand side of \eqref{eq:int_corr_schem} is still missing. 

We now outline the conceptual framework needed to study integrated correlators arising from $b$-deformations in general $\mathcal{N}=2$ settings.
To this end, we first consider $\cN=2$ SCFTs on four-dimensional ellipsoids preserving rigid supersymmetry, following the general approach of \cite{Festuccia:2011ws,Klare:2013dka} and the specific construction of \cite{Hama:2012bg}.
The four-dimensional ellipsoid is defined as the surface in $\mathbb{R}^5$ described 
by the equation
\begin{equation}
	\label{eq:def_ellipsoid}
		\frac{x_1^2+x_2^2}{\ell^2}+\frac{x_3^2+x_4^2}{\widetilde\ell^{\,2}}+\frac{x_5^2}{r^2}=1~.
\end{equation}
The two radii, $\ell$ and $\widetilde{\ell}$, can be conveniently parametrized as \cite{Hama:2012bg}
\begin{align}
\ell = b\, r~, \qquad \widetilde{\ell} = \frac{r}{b}~,
\label{b}
\end{align}
where $b = (\ell/ \widetilde{\ell})^{1/2}$ is the dimensionless squashing parameter. Since \eqref{eq:def_ellipsoid} is invariant under $b\leftrightarrow 1/b$, corresponding to the exchange $(x_1,x_2)\leftrightarrow(x_3,x_4)$, one may restrict to $b\in(0,1]$. In the limit $b\to1$, the ellipsoid reduces to the round sphere of radius $r$, which can be set to $1$ without losing generality.

The construction proceeds analogously to the mass deformation. Defining a SCFT on the ellipsoid requires coupling it to an off-shell conformal supergravity multiplet 
\begin{equation}
	\label{eq:sugra_mult}
    M~,~~
    \eta^{i}~,~~
    \mathsf K_{\mu\nu}~,~~ 
	\overbar {\mathsf K}_{\mu\nu}~,~~
	 V_{\mu}^0~,~~
	(V_{\mu})^{i}_{j}~,~~
    \psi^i_\mu~,~~
	G_{\mu\nu}~,
\end{equation}
where $M$ is a scalar field, $\eta^{i}$ is the dilatino (with $i=1,2$ as before), $\mathsf K_{\mu\nu}$ and $\overbar {\mathsf K}_{\mu\nu}$ are real self-dual and anti self-dual tensors, $V_{\mu}^0$ and $(V_{\mu})^{i}_{j}$ 
are the gauge fields of the $\mathrm{SU}(2)_R\times \mathrm{U}(1)_R$ R-symmetry, $\psi^{i}_\mu$ is the gravitino and $G_{\mu\nu}$ is the metric on the ellipsoid. The corresponding dual operator is the $\cN=2$ stress-tensor multiplet:
\begin{equation}
	\label{eq:stressT_mult}
    O_2~,~~
    \chi^{i}~,~~
    H_{\mu\nu}~,~~ 
	\overbar{H}_{\mu\nu}~,~~
	 t_{\mu}^0~,~~
	(t_{\mu})^{i}_{j}~,~~
    J^i_\mu~,~~
	T_{\mu\nu}~,
\end{equation}
where $O_2$ is the scalar top-component with conformal dimension 2, neutral under $\mathrm{U}(1)_R$ and singlet under $\mathrm{SU}(2)_R$, $\chi^{i}$ is a chiral fermion with dimension $\frac{5}{2}$ transforming as a doublet of $\mathrm{SU}(2)_R$, $H_{\mu\nu}$ and $\overbar H_{\mu\nu}$ are dimension-3 (anti) self-dual operators, $t_{\mu}^0$ and $(t_{\mu})^{i}_{j}$ are the conserved R-symmetry currents, $J^i_\mu$ is the supercurrent and finally $T_{\mu\nu}$ is the stress tensor.

To preserve supersymmetry on the ellipsoid we must require the vanishing of the gravitino and dilatino variations. This leads to Killing spinor equations that constrain the bosonic background fields in \eqref{eq:sugra_mult} in terms of the geometry.
Their explicit (and rather involved) form can be found in Section 3 of \cite{Hama:2012bg} (see also \cite{Bianchi:2019dlw}), but it is not needed for 
our purposes. Here, it suffices to note that this background configuration depends on the geometric parameters of the ellipsoid, including the squashing parameter $b$, and that it plays a role analogous to the background vector-field configuration in \eqref{eq:mass_back}.
We thus expect the $b$-deformation $S_b$ of any $\cN=2$ SCFT to be qualitatively similar to the mass deformation of \eqref{eq:mass_deform}, though with greater complexity. In particular, for small deformations, all bosonic operators of the stress-tensor multiplet, $O_2$, $H_{\mu\nu}$, $\overbar{H}_{\mu\nu}$, $t_{\mu}^0$, $(t_{\mu})^{i}_{j}$ and $T_{\mu\nu}$, appear linearly in $S_b$ with coefficients proportional to $(b-1)$, so that differentiating with respect to $b$ yields insertions of these operators, integrated over spacetime together with their corresponding geometrical factors. All these insertions are again related to each other by supersymmetry, and the appropriate system of superconformal Ward identities can be used to derive the integration measure for different classes of integrated correlators involving the stress tensor. Since the explicit derivation of this measure lies beyond the scope of this paper, we defer it to future work.

\subsection{Classes of integrated correlators in \texorpdfstring{$\mathcal{N}=2$}{} SCFTs with fundamental matter}

We now introduce the main $\cN=2$ superconformal gauge theory that will be the main focus of this work, together with the integrated correlators studied in subsequent sections.
We begin by considering the class of four-dimensional $\mathcal{N}=2$ theories with gauge group SU($N$) and matter content consisting of hypermultiplets transforming in the representation
\begin{align}
\cR=N_{\mathrm{adj}}\, \big(\mathrm{adjoint}\big)\, \oplus\, N_{\mathrm{F}} ~\big(\Yfund\big)\, \oplus\,N_{\mathrm{A}}~\Big(\Yasymm\Big)\, \oplus\,N_{\mathrm{S}}~\big(\Ysymm\big)~,
\label{Rrep}
\end{align}
whose $\beta$-function is proportional to
\begin{align}
\beta_0=2N\big(1-N_{\mathrm{adj}}\big)-N_{\mathrm{F}}-(N-2)N_{\mathrm{A}}-(N+2)N_{\mathrm{S}}~.
\label{beta0}
\end{align}
In Table\,\ref{tab:1}, we list the integer values of $N_{\mathrm{adj}}$, $N_{\mathrm{F}}$, $N_{\mathrm{A}}$ and $N_{\mathrm{S}}$ that yield superconformal theories with $\beta_0=0$.
\begin{table}[ht]
\begin{center}
\begin{tabular}{ c|c|c|c|c| }
& $\phantom{\Big|}N_{\mathrm{adj}}$ & $\phantom{\Big|}N_{\mathrm{F}}$& $\phantom{\Big|}N_{\mathrm{A}}$ & $\phantom{\Big|}N_{\mathrm{S}}$ \\\hline\hline
$\mathcal{N}=4$ SYM & $\phantom{\Big|}1$& $\phantom{\Big|}0$& $\phantom{\Big|}0$ & $\phantom{\Big|}0$
\\\hline
\textbf{A}-theory (a.k.a. SQCD) & $\phantom{\Big|}0$& $\phantom{\Big|}2N$& $\phantom{\Big|}0$ & $\phantom{\Big|}0$
\\\hline
\textbf{B}-theory & $\phantom{\Big|}0$& $\phantom{\Big|}N-2$& $\phantom{\Big|}0$ & $\phantom{\Big|}1$
\\\hline
\textbf{C}-theory  & $\phantom{\Big|}0$& $\phantom{\Big|}N+2$& $\phantom{\Big|}1$ & $\phantom{\Big|}0$
\\\hline
\textbf{D}-theory  & $\phantom{\Big|}0$& $\phantom{\Big|}4$& $\phantom{\Big|}2$ & $\phantom{\Big|}0$
\\\hline
\textbf{E}-theory  & $\phantom{\Big|}0$& $\phantom{\Big|}0$& $\phantom{\Big|}1$ & $\phantom{\Big|}1$
\\\hline
\end{tabular}
\end{center}
\caption{The superconformal $\cN=2$ theories with group SU($N$), following the nomenclature of \cite{Billo:2019fbi,Beccaria:2020hgy}. In the first case the supersymmetry is enhanced to $\mathcal{N}=4$.}
\label{tab:1}
\end{table}

The most prominent example in this list is $\mathcal{N}=4$ SYM, for which we have previously discussed the integrated correlators. Among the other superconformal models, 
the \textbf{D}- and  \textbf{E}-theories are special since the number of hypermultiplets is independent of the rank of the gauge group. This property greatly simplifies the large-$N$ expansion of various observables, making these theories especially suitable for the study of integrated correlators when supersymmetry is not maximal\,\footnote{See \cite{Billo:2023kak} and \cite{Billo:2024ftq} for the study of some integrated correlators in the \textbf{E}- and \textbf{D}-theories, respectively.}.
 
In this paper we focus on the \textbf{D}-theory whose flavour symmetry group is $\mathrm{SU}(2)_L \times \mathrm{U}(1)_L \times \mathrm{U}(4)$. This model can be engineered  with $N$ D3-branes in Type IIB string theory in the presence of a $\mathbb{Z}_2$-orbifold probing an O7-orientifold background with $(4+4)$ D7-branes \cite{Park:1998zh,Ennes:2000fu}. From the D7-branes viewpoint, the $\mathrm{U}(4)$ symmetry acts as a gauge group, so that the dynamics on the D7-branes world-volume is dual to $\mathrm{U}(4)$ gluon scattering in AdS. From the bulk perspective, instead, the D7-branes introduce nontrivial defects which can be probed by gravitons.

The \textbf{D}-theory allows for several deformations associated with its matter content. In particular, we can assign masses $\mu_{\mathrm{F}}$ (${\mathrm{F}}=1,\ldots,4$) to each of the four fundamental hypermultiplets and masses $m_{\mathrm{A}}$ (${\mathrm{A}}=1,2$) to the two anti-symmetric multiplets, and also deform the four-sphere into an ellipsoid with squashing parameter $b$.
To streamline the discussion, we take all fundamental masses to be equal, $\mu_{\mathrm{F}}\,=\,\mu$ for any $\mathrm{F}$, and similarly all anti-symmetric masses to be equal, $m_{\mathrm{A}}\,=\,m$ for any $\mathrm{A}$. We refer to the theory deformed by these mass parameters and the squashing as the $\mathbf{D}^*$-theory. 
Several classes of integrated correlators can then be studied in the presence of these deformations, which we classify according to their dual interpretation in AdS space. 

\paragraph{Gluons in AdS:} As discussed above, the brane realization of \textbf{D}-theory gives rise to an eight-dimensional SYM theory with $\mathrm{U}(4)\subset \mathrm{SO}(8)$ gauge group on the D7-branes world-volume. The corresponding fields in AdS$_5$ are obtained from a Kaluza-Klein (KK) reduction on $S^3$, yielding a tower of modes labeled by an integer $k=2,3,\dots$. For each $k$ these KK modes transform in the adjoint representation of U(4) and in the spin $j_L=\frac{k}{2}-1$ representation of $\mathrm{SU}(2)_L$.
We refer to these as ``gluon supermultiplets'', as they are associated to open string excitations in AdS. In particular, the $k=2$ mode corresponds to a massless U(4) vector-multiplet in AdS$_5$, which
is dual to the U(4) flavour current multiplet in the CFT. Hence, the integrated correlator obtained by taking derivatives only with respect to the fundamental mass
\begin{equation}\label{eq:IC_d4mu}
    \partial^4_\mu \log \cZ^{\mathbf{D}^*}\big|_{\D}
\end{equation}
provides integrated constraints for the four-gluon scattering amplitude in AdS. The 
large-$N$ expansion of this quantity, including the case of non equal masses $\mu_{\mathrm{F}}$, has been studied in \cite{Billo:2024ftq}.

\paragraph{Gravitons in AdS:}  Analogously, starting from the ten-dimensional graviton, one can perform a KK reduction to AdS$_5$, producing a tower of fields labeled by an integer $p=2,3,\dots$ with Lorentz spins ranging from zero to two. A subset of these organize into $\frac{1}{2}$-BPS supermultiplets similar to the gluon case, but transforming differently under the global symmetries. Such supermultiplets carry $\mathrm{SU}(2)_L$ spin $j_L = p/2$, are singlets of U(4) and are related to closed string excitations. 
We focus on the $p = 2$ mode, corresponding to a massless vector multiplet in five dimensions, dual to the $\mathrm{SU}(2)_L$ flavour current multiplet. 
Following the terminology of \cite{Chester:2025ssu}, we refer to these as ``graviton supermultiplets'', although there are no spin-two fields, to emphasize their origin from the ten-dimensional graviton. 
Conversely, the direct reduction to the massless five-dimensional graviton generates a gravity supermultiplet dual to the stress-tensor multiplet, which is a singlet under $\mathrm{U}(4)\times \mathrm{SU}(2)_L$. Therefore, although the $m$- and $b$- deformations activate different CFT supermultiplets, they are both associated to closed-string modes in the dual description.

\paragraph{Integrated correlators dual to scattering processes:}  We can classify the integrated correlators studied in this paper from their dual interpretation described above. In particular, we refer to the integrated correlators
\begin{equation}\label{eq:IC_dmdb}
   \partial^4_m \log \cZ^{\mathbf{D}^*}\big|_{\D}~, \qquad    \big(\partial^4_b-15\partial_b^2\big) \log \cZ^{\mathbf{D}^*}\big|_{\D}~, \qquad   \partial^2_m\partial^2_b \log \cZ^{\mathbf{D}^*}\Big|_{\D}~,
\end{equation}
which are dual to scattering processes of gravitons, as the integrated correlators with closed string modes. These observables have been studied in $\cN=4$ SYM using matrix-model techniques in \cite{Chester:2020vyz}, where it was shown that only one of them provides nontrivial constraints to the four-graviton scattering amplitude as a consequence of nontrival relations. In Section\,\ref{secn:integrated} we revisit these properties and examine how they are modified for the $\mathbf{D}$-theory. 

The other class of integrated correlators considered in this paper correspond to mixed gluon-graviton scattering process, which at the matrix-model level can be extracted from the mixed derivatives
\begin{equation}\label{eq:IC_dmudm_dmudb}
\partial_\mu^2\,\partial_m^2 \log \mathcal{Z}^\DS\big|_{\D}~, \qquad \partial_\mu^2\,\partial_b^2 \log \mathcal{Z}^\DS\big|_{\D} ~.
\end{equation}
In the following, we compute these quantities using supersymmetric localisation. We emphasize that, while all observables associated to four mass-derivatives have a well-defined CFT derivation (\textit{i.e.} the left-hand side of \eqref{eq:int_corr_schem}), for observables involving
derivatives with respect to the squashing parameter $b$ such a CFT derivation has not yet been provided and in particular the integration measure has not yet been computed. We leave this derivation for future work.

\section{The matrix model for massive \texorpdfstring{$\mathcal{N}=2$}{} theories on a squashed sphere}
\label{sec:squashing}
 
When the $\mathcal{N} = 2$ SYM theory is defined on a compact space $\mathcal{S}$, its partition function $\mathcal{Z}$ can be expressed as an integral over the eigenvalues $a_u$ of a Hermitian matrix $a\in \mathfrak{su}(N)$ according to \cite{Pestun:2007rz}
\begin{align}
\label{Zloc}
    \mathcal{Z}=\int\!\prod_{u=1}^N da_u ~ \rme^{-\frac{8\pi^2}{g_{_{\rm YM}}^2}\,\tr a^2} \, \big|Z_{\mathrm{1-loop}}\,Z_{\mathrm{inst}}\big|^2 \,\delta\bigg( \sum_{u=1}^N a_u \bigg)~.
\end{align}
Here, $g_{_{\rm YM}}^2$ denotes the YM coupling, $Z_{\mathrm{1-loop}}$ encodes the fluctuations around the localisation points, and $Z_{\mathrm{inst}}$ represents the instanton contribution \cite{Nekrasov:2002qd}. Since we will primarily work in the ’t Hooft planar limit, instanton effects can be neglected, allowing us to set $Z_{\mathrm{inst}} = 1$. The precise form of $Z_{\mathrm{1-loop}}$ depends on both the matter content of the theory and the geometry of the compact manifold $\mathcal{S}$. In the present work, we take $\mathcal{S}$ to be an ellipsoid with squashing parameter $b$ (see (\ref{eq:def_ellipsoid})).
A further ingredient that plays a crucial role in the expression for $Z_{\mathrm{1\text{-}loop}}$ is the possibility of assigning a generic mass to the matter hypermultiplets. While such a deformation explicitly breaks conformal invariance, it preserves $\mathcal{N}=2$ supersymmetry. To illustrate this point, let us first consider the massive deformation of $\mathcal{N}=4$ SYM, {\it{i.e.}} the $\mathcal{N}=2^*$ SYM theory. Although this case has already been analyzed in \cite{Chester:2020vyz}, we briefly review it here using a slightly different approach to introduce notation that will be employed in subsequent sections.

\subsection{\texorpdfstring{$\mathcal{N}=2^*$}{} SYM}
\label{subsecn:N2*}
This is an $\mathcal{N}=2$ gauge theory with one massive hypermultiplet in the adjoint representation of SU($N$). 
When the theory is put on a squashed sphere with parameter $b$, the 1-loop part appearing in the matrix-model partition function is\,\footnote{The factor $\Upsilon^\prime_b(0)=\partial_x\Upsilon_b(x)|_{x=0}$ 
was missing in \cite{Hama:2012bg} and has been added in \cite{Chester:2020vyz}.}
\begin{align}
\label{Z1loopIN2*}
\big|Z^{\NS}_{\mathrm{1-loop}}\big|^2= \frac{ \displaystyle{\Upsilon_b^\prime(0)^{N-1} \prod_{u<v=1}^N \!\Upsilon_b\big(\ii\,a_{uv}\big)\,\Upsilon_b\big(\!-\ii\,a_{uv}\big)}}{\displaystyle{\Upsilon_b\Big(\frac{Q}{2}+\ii\,m\Big)^{N-1}\prod_{u\neq v=1}^N \!\Upsilon_b\bigg(\frac{Q}{2}+\ii\,a_{uv}+\ii m\bigg)}}~,
\end{align}
where $m$ is the mass, $a_{uv}\equiv a_u-a_v$, $Q\equiv b+\frac{1}{b}$ and the function $\Upsilon_b(x)$, related to the Barnes double Gamma functions, is defined in \cite{Hama:2012bg}. This function, which was first introduced in \cite{Zamolodchikov:1995aa} to study the structure constants of the conformal Liouville field theory with coupling $b$, satisfies 
\begin{align}
    \Upsilon_b(x)=\Upsilon_{1/b}(x)\qquad\mbox{and}\qquad \Upsilon_b\bigg(\frac{Q}{2}+x\bigg)=\Upsilon_b\bigg(\frac{Q}{2}-x\bigg) ~.
\end{align}
Consequently, the right-hand side of (\ref{Z1loopIN2*}) is invariant under $b\leftrightarrow1/b$ and $m\leftrightarrow-m$. We recall that, in principle, the regularized partition function of the $\mathcal{N}=2^*$ theory is ambiguous due to the presence of divergences \cite{Bobev:2013cja}. However, the physical observables obtained from the partition function (in particular the integrated correlators that we will consider) are free of such ambiguities \cite{Chester:2020vyz}. The specific regularized representation given in \eqref{Z1loopIN2*} corresponds to the ellipsoid version of the 1-loop factor of the matrix model originally found in \cite{Pestun:2007rz} for the $\mathcal{N}=2^*$ theory on a round sphere.

In the undeformed theory ($b=1,m=0$), one has
\begin{align}
\label{Z1loop0}
\big|Z^{\NS}_{\mathrm{1-loop}}\big|^2 \,\Big|_{0}\,= \, \Delta(a) ~,
\end{align}
where $\Delta(a)$ is the Vandermonde determinant, so that the partition function reduces to the one of the free Gaussian matrix model describing $\mathcal{N}=4$ SYM on $S^4$. Since for the computation of integrated correlators one considers the fluctuations around $m=0$ and $b=1$, it is convenient to define the functions\,\footnote{These functions are related (but not identical) to the functions $H_v$ and $H_h$
introduced in \cite{Mitev:2015oty}.}
\begin{align}
    H_{\mathrm{v}}(x;b)&:=\frac{\Upsilon_b(\ii\,x)\,\Upsilon_b(-\ii\,x)}{\Upsilon_1(\ii\,x)\,\Upsilon_1(-\ii\,x)}~,
\qquad    
    H_{\mathrm{h}}(x;b,m)
    :=\frac{\Upsilon_b\Big(\displaystyle{\frac{Q}{2}}+\ii\,m+\ii \,x\Big)}{\Upsilon_1\big(1+\ii \,x\big)\phantom{\Big|}}~,\label{Hfunctions}
\end{align}
which, by construction, satisfy $H_{\mathrm{v}}(x;1)=1$ and $H_{\mathrm{h}}(x;1,0)=1$. 
Then, it is straightforward to see that (\ref{Z1loopIN2*}) can be recast in the form
\begin{align}
\label{Z1loopIb}
\big|Z^{\NS}_{\mathrm{1-loop}}\big|^2 = \Delta(a)\,\frac{ \displaystyle{\Upsilon_b^\prime(0)^{N-1}\,\prod_{u<v=1}^N \!H_{\mathrm{v}}(a_{uv};b)}}{\displaystyle  \Upsilon_b\Big(\frac{Q}{2}+\ii\,m\Big)^{N-1}\prod_{u\neq v=1}^N \!H^{\frac{1}{2}}_{\mathrm{h}}(a_{uv};b,+m)\,H^{\frac{1}{2}}_{\mathrm{h}}(a_{uv};b,-m)}~.
\end{align}
Introducing the rescaling
\begin{align}
a\rightarrow \sqrt{\frac{\lambda}{8\pi^2N}}\,\,a \, , 
\label{rescaling}
\end{align}
with $\lambda = N g_{_{\rm YM}}^2$ the ’t Hooft coupling, adopting the full Lie-algebra approach of \cite{Billo:2017glv}, and exploiting the properties of the functions $H_{\mathrm{v}}$ and $H_{\mathrm{h}}$ listed in Appendix\,\ref{App:Hfunctions}, the partition function of the $\mathcal{N}=2^*$ theory can be written, up to an overall normalization constant, 
as\,\footnote{Decomposing $a = a^b T_b$ with $b=1,\dots,N^2-1$, where $T_b$ are the generators of $\mathrm{SU}(N)$ in the fundamental representation normalized by $\tr(T_b T_c) = \tfrac{1}{2}\delta_{bc}$, the integration measure is
\begin{align*}
da = \prod_{b=1}^{N^2-1} \frac{da^b}{\sqrt{2\pi}}
\quad \text{so that} \quad \int \! da~ \rme^{-\tr a^2} = 1~.
\end{align*}
}
\begin{align}
\mathcal{Z}^{\NS} =\int\! da~ \rme^{-\,\mathrm{tr} a^2-S^{\NS} } \, , 
\label{ZN2*}
\end{align}
where\,\footnote{Even if (\ref{SN2*}) is expressed in terms of the eigenvalues $a_u$, it is easy to rewrite it using powers of $\tr a$, as we will explicitly see in Section\,\ref{secn:integrated}.}
\begin{align}
S^{\NS} &=(N-1)\bigg[\log\Upsilon_b\bigg(\frac{Q}{2}+\ii\,m\bigg)-\log\Upsilon_b^\prime(0)\bigg] -\sum_{u<v=1}^{N} \log H_{\mathrm{v}}\bigg(\sqrt{\frac{\lambda}{8\pi^2N}}\,\,a_{uv};b\bigg) \notag \\
& \quad+ \sum_{u\neq v=1}^{N} \log H_{\mathrm{h}}\bigg(\sqrt{\frac{\lambda}{8\pi^2N}}\,\,a_{uv};b,m\bigg)~.
\label{SN2*}
\end{align}
We are interested in the fluctuations around the undeformed point $(b=1,m=0)$, at which $S^{\NS}$ vanishes. Exploiting the invariance under $b \leftrightarrow 1/b$ and $m \leftrightarrow -m$, the expansion of $S^{\NS}$ takes the form
\begin{align}
S^{\NS}&=m^2\,\mathcal{M}^{\NN}_2+\left[(b-1)^2-(b-1)^3\right]\,\mathcal{B}^{\NN}_2+m^4\,\mathcal{M}^{\NN}_4\notag\\[1mm]
&\quad+m^2(b-1)^2\,\mathcal{C}^{\NN}+(b-1)^4\,\mathcal{B}_4^{\NN}+\ldots \,  ,
\label{SN2*exp}
\end{align}
where the coefficients $\mathcal{M}^{\NN}_2$, $\mathcal{B}^{\NN}_2$, etc. are nontrivial functions of the coupling $\lambda$ which are related to integrated correlators in $\mathcal{N}=4$ SYM.

\subsection{\texorpdfstring{$\mathbf{D}^*$}{}-theory}
\label{subsecn:D*}
The above analysis easily extends to the $\mathbf{D}$-theory (see Tab.\,\ref{tab:1}) on a squashed sphere with massive hypermultiplets, which we have called the $\mathbf{D}^*$-theory.
In this case we have
\begin{align}
\label{Z1loopIbD*}
\big|Z^{\DS}_{\mathrm{1-loop}}\big|^2 &= 
\rme^{-S^\D}\,\Delta(a)\,\,\Upsilon_b^\prime(0)^{N-1}\!\!\prod_{u<v=1}^N \!H_{\mathrm{v}}(a_{uv};b)
\,\Bigg[\prod_{\mathrm{F}=1}^4\prod_{u=1}^N H^{-\frac{1}{2}}_{\mathrm{h}}(a_u;b,+\mu_{\mathrm{F}})\,H^{-\frac{1}{2}}_{\mathrm{h}}(a_u;b,-\mu_{\mathrm{F}})\Bigg]\,\times\notag \\
&\quad\times\,\Bigg[\prod_{{\mathrm{A}}=1}^{2}\prod_{u<v=1}^{N}H^{-\frac{1}{2}}_{\mathrm{h}}(a_u+a_v;b,+m_{\mathrm{A}})\,
H^{-\frac{1}{2}}_{\mathrm{h}}(a_u+a_v;b,-m_{\mathrm{A}})\Bigg]
\end{align}
where $S^\D$ is the interaction action of the matrix model for the \textbf{D}-theory, which will be given shortly.
The terms in the square brackets of the first line of (\ref{Z1loopIbD*}) are associated to the four fundamental multiplets with mass $\mu_{\mathrm{F}}$ (${\mathrm{F}}=1,\ldots,4$), while the terms in the second line correspond to the two anti-symmetric multiplets with mass $m_{\mathrm{A}}$ (${\mathrm{A}}=1,2$).
As written above, we take all masses to be equal, namely
$\mu_{\mathrm{F}}\,=\,\mu$ for any $\mathrm{F}$, and $m_{\mathrm{A}}\,=\,m$ for any $\mathrm{A}$.

Proceeding as described in the previous subsection, after the rescaling \eqref{rescaling} we can write the partition function as
\begin{align}
\mathcal{Z}^{\DS}=\int\! da~ \rme^{-\,\mathrm{tr} a^2-S^{\DS}}\, , 
\label{ZD*}
\end{align}
where
\begin{align}
S^{\DS}&= S^\D - (N-1)\log \Upsilon_b^\prime(0)-\sum_{u<v=1}^{N} \log H_{\mathrm{v}}\bigg(\sqrt{\frac{\lambda}{8\pi^2N}}\,\,a_{uv};b\bigg) \notag \\
& \quad+4\sum_{u=1}^{N} \log H_{\mathrm{h}}\bigg(\sqrt{\frac{\lambda}{8\pi^2N}}\,\,a_u;b,\mu\bigg)+ 2\sum_{u< v=1}^{N} \log H_{\mathrm{h}}\bigg(\sqrt{\frac{\lambda}{8\pi^2N}}\,\,(a_u+a_v);b,m\bigg) ~,
\label{SD*}
\end{align}
with
\cite{Beccaria:2020hgy,Billo:2024ftq}
\begin{align}
    S^\D&=2\,
   \sum_{n=1}^{\infty}\sum_{k=1}^{n-1}(-1)^{n}\,
   \frac{(2n+2)!\,\zeta(2n+1)}{(n+1)(2k+1)!\,(2n-2k+1)!}\,\Big(\frac{\lambda}{8\pi^2N}\Big)^{n+1}\,\tr a^{2k+1}\, \tr a^{2n-2k+1}\notag\\
   &\quad-4\sum_{n=1}^{\infty} (-1)^{n} \frac{(4^{n}-1)\,\zeta(2n+1)}{n+1}\,\Big(\frac{\lambda}{8\pi^2N}\Big)^{n+1}\,\tr a^{2n+2}~.
   \label{Sd}
\end{align}
The action $S^\DS$ admits an expansion similar to (\ref{SN2*exp}), namely
\begin{align}
S^{\DS} &= S^{\D}+m^2\,\mathcal{M}^\D_{2,\mathrm{A}}+\mu^2\,\mathcal{M}^\D_{2,\mathrm{F}}+\left[(b-1)^2-(b-1)^3\right]\mathcal{B}^\D_2+m^4\,\mathcal{M}^\D_{4,\mathrm{A}}+\mu^4\,\mathcal{M}^\D_{4,\mathrm{F}} \notag \\[1mm]
& \quad+m^2(b-1)^2\,\mathcal{C}_{\mathrm{A}}^\D +\mu^2(b-1)^2\,\mathcal{C}_{\mathrm{F}}^\D +(b-1)^4\,\mathcal{B}^\D_4+\ldots~.
\label{SDSexp}
\end{align}
Also in this case, the coefficients $\mathcal{M}^\D_{2,\mathrm{A}}$, $\mathcal{M}^\D_{2,\mathrm{F}}$, etc., which are related to integrated correlators in the \textbf{D}-theory, are nontrivial functions of $\lambda$ which can be explicitly written in terms of derivatives of $\log H_{\mathrm{v}}$ and $\log H_{\mathrm{h}}$, as we will see in the next section.

\section{Integrated correlators with only closed string modes}
\label{secn:integrated}
We now examine how the gauge theory responds to deformations corresponding to the squashing of the four-sphere and to the introduction of a mass parameter $m$, either for the adjoint hypermultiplet in the case of $\mathcal{N}=4$ SYM, or for the anti-symmetric hypermultiplets in the case of the $\mathbf{D}$-theory. As argued in Section\,\ref{secn:General}, both deformations admit an interpretation as closed string excitations from the holographic dual perspective. From the standpoint of the matrix model, instead, the response is simply encoded in the coefficients $\mathcal{M}_2$, $\mathcal{B}_2$, etc., appearing in the expansion of the effective action around the point $(m=0,b=1)$, see (\ref{SN2*exp}) and (\ref{SDSexp}).
To illustrate the structure of this expansion, we first consider the deformation of $\mathcal{N}=4$ SYM into the $\mathcal{N}=2^*$ theory on the ellipsoid.

\subsection{\texorpdfstring{$\mathcal{N}=4$ SYM}{}}
The simplest quantity is $\mathcal{M}_2^\NN$ defined as  
\begin{equation}
    \begin{aligned}
  \mathcal{M}_2^\NN&:=\frac{1}{2}\partial_m^2 S^\NS
        \big|_{0}\label{M2a}\\&=\Bigg\{\frac{N-1}{2}\,\partial_m^2 \log\Upsilon_b\bigg(\frac{Q}{2}+\ii\,m\bigg)+\frac{1}{2}\partial_m^2 \bigg[\sum_{u\neq v=1}^{N} \log H_\mathrm{h}\bigg(\sqrt{\frac{\lambda}{8\pi^2N}}\,\,a_{uv};b,m\bigg)\bigg]\Bigg\}_{\substack{m=0 \\ b=1}}~.
\end{aligned}
\end{equation}
Using the results summarized in Appendix\,\ref{App:Hfunctions}, this expression reduces to
\begin{align}
  \mathcal{M}_2^\NN&=(N^2-1)(1+\gamma) 
 +\sum_{n=1}^\infty\sum_{k=0}^{2n} (-1)^{n+k}\, \frac{(2n+1)!\,\zeta(2n+1)}{(2n-k)!\,k!} \Big(\frac{\lambda}{8\pi^2N}\Big)^{n}\tr a^{2n-k}\, \tr a^{k} \, ,
 \label{M2explicit}
\end{align}
where $\gamma$ is the Euler-Mascheroni constant. A parallel computation for $\mathcal{B}_2^\NN$ yields
\begin{align}
  \mathcal{B}_2^\NN&:=\frac{1}{2}\partial_b^2 S^\NS
        \big|_{0}\notag\\
        &=\Bigg\{\frac{N-1}{2}\,\partial_b^2 \log\Upsilon_b\bigg(\frac{Q}{2}+\ii\,m\bigg)-\frac{N-1}{2}\,\partial_b^2\log\Upsilon_b^\prime(0)\label{B2def}\\[2mm]
  &\quad-\frac{1}{2}\partial_b^2 \bigg[\sum_{u<v=1}^{N} \log H_{\mathrm{v}}\bigg(\sqrt{\frac{\lambda}{8\pi^2N}}\,\,a_{uv};b\bigg)\bigg] \!+\!\frac{1}{2}\partial_b^2 \bigg[\sum_{u\neq v=1}^{N} \log H_{\mathrm{h}}\bigg(\sqrt{\frac{\lambda}{8\pi^2N}}\,\,a_{uv};b,m\bigg)\bigg]\Bigg\}_{\substack{m=0 \\ b=1}}~.\notag
\end{align}
After applying again the results of Appendix\,\ref{App:Hfunctions}, drastic simplifications occur and the above expression reduces to
\begin{align}
  \mathcal{B}_2^\NN&=(N^2-1)(1+\gamma) 
 +\sum_{n=1}^\infty\sum_{k=0}^{2n} (-1)^{n+k}\, \frac{(2n+1)!\,\zeta(2n+1)}{(2n-k)!\,k!} \Big(\frac{\lambda}{8\pi^2N}\Big)^{n}\tr a^{2n-k}\, \tr a^{k}~.
 \label{B2explicit}
\end{align}
A direct comparison with (\ref{M2explicit}) immediately shows that
\begin{align}
    \mathcal{B}_2^\NN=\mathcal{M}_2^\NN~.
    \label{relation1}
\end{align}
The remaining coefficients in (\ref{SN2*exp}) can likewise be expressed as power series of $\tr a$, leading to a uniform description of the expansion. Explicitly, we find 
\begin{subequations}
\begin{align}
\mathcal{M}_4^\NN
    &=-\frac{(N^2-1)}{2}\,\zeta(3)-\frac{1}{12}\sum_{n=1}^\infty\sum_{k=0}^{2n} (-1)^{n+k}\, \frac{(2n+3)!\,\zeta(2n+3)}{(2n-k)!\,k!} \Big(\frac{\lambda}{8\pi^2N}\Big)^{n}\tr a^{2n-k}\, \tr a^{k}~,
    \label{M4final}\\[1mm]
    \mathcal{C}^\NN&=(N^2-1)\Big(\frac{1}{3}-\zeta(3)\Big)\notag\\
  &\quad~+\frac{1}{6}
    \sum_{n=1}^\infty\sum_{k=0}^{2n}(-1)^{n+k}\frac{(4n^2+4n)\,(2n+1)!\,\zeta(2n+1)}{(2n-k)!\,k!}
    \,\Big(\frac{\lambda}{8\pi^2N}\Big)^{n}\,\tr a^{2n-k}\,\tr a^k\notag\\
    &\quad~-\frac{1}{6}
    \sum_{n=1}^\infty\sum_{k=0}^{2n}(-1)^{n+k}\frac{(2n+3)!\,\zeta(2n+3)}{(2n-k)!\,k!}\,\Big(\frac{\lambda}{8\pi^2N}\Big)^{n}\,
    \tr a^{2n-k}\,\tr a^k~,\label{Cfinal}\\[1mm]
    \mathcal{B}_4^\NN&=-\frac{N^2-1}{2}\Big(\zeta(3)-\frac{5\,\gamma}{2}-\frac{19}{6}\Big)\notag\\
&\qquad+\frac{1}{12}\sum_{n=1}^\infty\sum_{k=0}^{2n} (-1)^{n+k}\frac{(8n^2+8n+15)(2n+1)!\,\zeta(2n+1)}{(2n-k)!\,k!}\Big(\frac{\lambda}{8\pi^2N}\Big)^{n} \tr a^{2n-k}\,\tr a^k\notag\\
&\qquad-\frac{1}{12}\sum_{n=1}^\infty\sum_{k=0}^{2n} (-1)^{n+k}\frac{(2n+3)!\,\zeta(2n+3)}{(2n-k)!\,k!}\Big(\frac{\lambda}{8\pi^2N}\Big)^{n} \tr a^{2n-k}\,\tr a^k~.\label{B4final}
\end{align}%
\end{subequations}
From these expressions one can verify the following interesting identities among these quantities
\begin{align}
    \mathcal{B}_4^\NN-\frac{5}{4}\,\mathcal{B}_2^\NN+\mathcal{M}_4^\NN&=\mathcal{C}^\NN~,
    \label{relation2}\\
    \mathcal{C}^\NN-2\,\mathcal{M}_4^\NN&=\frac{4\,c^\NN}{3}+\frac{2}{3}\big(2\lambda\partial_\lambda+\lambda^2\partial_\lambda^2\big)\mathcal{M}_2^\NN \, , 
    \label{relation3}
\end{align}
where $c^\NN=(N^2-1)/4$ is the central charge of $\mathcal{N}=4$ SYM.

The expansion (\ref{SN2*exp}) of the effective action translates in the following expansion for $\log \mathcal{Z}^\NS$:
\begin{align}
    \log \mathcal{Z}^\NS&=-m^2\,\big\langle\mathcal{M}_2^\NN\big\rangle_0-\left[(b-1)^2-(b-1)^3\right]\,\big\langle\mathcal{B}_2^\NN\big\rangle_0\notag\\
    &\quad\,-m^4\Big[\big\langle\mathcal{M}_4^\NN\big\rangle_0-\frac{1}{2}\big\langle(\mathcal{M}_2^\NN)^{2}\big\rangle_0+\frac{1}{2}
    \big\langle\mathcal{M}_2^\NN\big\rangle_0^2\Big]\notag\\
    &\quad\,-m^2(b-1)^2\Big[\big\langle\mathcal{C}^\NN\big\rangle_0-\big\langle\mathcal{M}_2^\NN\,\mathcal{B}_2^\NN\big\rangle_0+\big\langle\mathcal{M}_2^\NN\big\rangle_0\,\big\langle\mathcal{B}_2^\NN\big\rangle_0\Big]\notag\\
    &\quad\,-(b-1)^4\Big[\big\langle\mathcal{B}_4^\NN\big\rangle_0-\frac{1}{2}\big\langle(\mathcal{B}_2^\NN)^{2}\big\rangle_0+\frac{1}{2}\big\langle\mathcal{B}_2^\NN\big\rangle_0^2\Big)+\ldots \, ,
    \label{logZexp}
\end{align}
where $\langle \, \cdot \,\rangle_0$ denotes the Gaussian matrix-model average in $\mathcal{N}=4$ SYM, 
\begin{align} \label{eq:gaussian}
\big\langle f(a) \big\rangle_0 :=  \int d a \, \rme^{- \tr a^2 } f(a) ~.
  \end{align}
Using (\ref{logZexp}), the identities (\ref{relation1}), (\ref{relation2}) and (\ref{relation3}) become
\begin{subequations}
    \begin{align}
        \partial_b^2 \log \mathcal{Z}^\NS\big|_{0}&=\partial_m^2 \log \mathcal{Z}^\NS\big|_{0}~,\label{relation1a}\\
        \Big(\partial_b^4-15\,\partial_b^2+\partial_m^4\Big)\log \mathcal{Z}^\NS\big|_{0}&=6\,\partial_m^2\partial_b^2\log \mathcal{Z}^\NS\big|_{0}~,\label{relation2a}\\
        \Big(3\,\partial_m^2\partial_b^2-\partial_m^4\Big)\log \mathcal{Z}^\NS\big|_{0}&=-16\,c^\NN+4\big(2\lambda\partial_\lambda+\lambda^2\partial_\lambda^2\big)\partial_m^2\log \mathcal{Z}^\NS\big|_{0}~.\label{relation3a}
    \end{align}
    \label{relationsN4}%
\end{subequations}
These relations perfectly match those proposed in \cite{Chester:2020vyz}\,\footnote{Note that these relations are valid even after including the instanton contributions \cite{Chester:2020vyz}. In the case of non-vanishing $\theta$-angle, the differential operator $(2\lambda\partial_\lambda+\lambda^2\partial_\lambda^2)$ in the third relation (\ref{relation3a}) should be promoted to the SL(2, $\mathbb{Z}$)-invariant hyperbolic Laplacian $\Delta_\tau=4\tau_2^2\partial_\tau\partial_{\bar\tau}$.} and show that the derivatives with respect to $b$ and $m$ are not independent. It is also worth noting that the first two relations, \eqref{relation1a} and \eqref{relation2a}, can be understood as the result of a supersymmetry enhancement when $m=\pm \frac{1}{2}\big(b-{1}/{b}\big)$ \cite{Minahan:2020wtz}.  

\subsubsection*{\texorpdfstring{Large-$N$}{} limit} 
We now briefly comment on the large-$N$ expansion of the operators introduced above.
A convenient way to organize this expansion is to perform a change of basis \cite{Billo:2024ftq}.
Rather than working directly with the traces $\tr a^k$, we introduce a new set of operators $\mathcal{P}_k$ defined through
\begin{align}
\label{changebasis}
\tr a^k = \Big(\frac{N}{2}\Big)^{\frac{k}{2}}\sum_{\ell=0}^{\lfloor\frac{k-1}{2}\rfloor}\sqrt{k-2\ell}\,\frac{k!}{\ell!\,(k-\ell)!}\,\,\mathcal{P}_{k-2\ell}   + 
\big\langle \tr a^k \big\rangle_0 \, ,
\end{align}
where the Gaussian averages are given by
\begin{align}
\big\langle \tr a^{2n+1} \big\rangle_0 &= 0~, \notag \\
\big\langle \tr a^{2n} \big\rangle_0 &= \frac{N^{n+1}}{2^n}\,\frac{(2n)!}{n!\,(n+1)!}-\frac{N^{n-1}}{2^{n+1}}\,\frac{(2n)!}{n!\,(n-1)!}\,\Big(1-\frac{n-1}{6}\Big)+O\big(N^{n-3}\big)~.
\label{tracesVEV}
\end{align}
The operators $\mathcal{P}_k$, which are proportional to those introduced in \cite{Rodriguez-Gomez:2016cem} in terms of Chebyshev polynomials, enjoy the important property of being orthonormal in the planar limit of the Gaussian model:
\begin{align}
    \big\langle \mathcal{P}_k\,\mathcal{P}_\ell\big\rangle_0=\delta_{k,\ell}+O\big(N^{-2}\big)~.
\end{align}
Under this change of basis, the double-trace and even single-trace terms reorganize schematically as
\begin{subequations}
\begin{align}
&N^{-n}\,\tr a^{2n-2\ell}\,\tr a^{2\ell}\,\longrightarrow\, N^{2}\,\alpha_0+N\,\alpha_1\,\mathcal{P}+\big(\alpha_2\,\mathcal{P}\,\mathcal{P}+\alpha_3\big)
+O\big(N^{-1}\big)~,\\
&N^{-n}\,\tr a^{2n-2\ell-1}\,\tr a^{2\ell+1}\,\longrightarrow\,
\beta \,\mathcal{P}\,\mathcal{P}~,\\
&N^{-n}\,\tr a^{2n}\,\longrightarrow\, N\,\gamma_0+\gamma_1\,\mathcal{P}+O\big( N^{-1} \big)~,
\end{align}
\label{largeN}%
\end{subequations}
with $\alpha_\#$, $\beta$ and $\gamma_\#$ being $N$-independent coefficients.
Consequently, in the planar limit any operator of the type considered here scales as $N^2$ and admits the following large-$N$ expansion:
\begin{align}
    \mathcal{O} =\sum_{g=0}^\infty N^{2-g}\,c_g(\mathcal{O}) \, , 
    \label{OlargeN}
\end{align}
where $\mathcal{O}$ stands for $\mathcal{M}_{2}^\NN$, $\mathcal{B}_{2}^\NN$, $\mathcal{M}_{4}^\NN$, $\mathcal{C}^\NN$ or $\mathcal{B}_{4}^\NN$. 

Using this notation, it is straightforward to verify from \eqref{M2a} and \eqref{tracesVEV} that the leading term of $\mathcal{M}_{2}^\NN$ is
given by 
\begin{align}
\label{eq:M2}
c_0\big(\mathcal{M}_{2}^\NN\big)&=(1+\gamma)+\sum_{n=1}^\infty\sum_{\ell=0}^{n}(-1)^n\,\frac{(2n+1)!\,\zeta(2n+1)}{(n-\ell)!\,(n-\ell+1)!\,\ell!\,(\ell+1)!}\Big(\frac{\lambda}{16\pi^2}\Big)^{n}\notag\\
   &=(1+\gamma)+\frac{8}{\pi}\sum_{n=1}^\infty(-1)^n\, \frac{\Gamma\big(n+\frac{3}{2}\big)^2\,\zeta(2n+1)}{\Gamma(n+2)\,\Gamma(n+3)}\,\Big(\frac{\lambda}{\pi^2}\Big)^n~,
\end{align}
which can be resummed to 
\begin{align}
c_0\big(\mathcal{M}_{2}^\NN\big)&=(1+\gamma)+\int_0^\infty\!dw\,\frac{w}{\sinh(w)^2}\bigg[\frac{4\pi^2}{w^2\lambda}J_1\Big(\frac{\sqrt{\lambda}\, w}{\pi} \Big)^2-1\bigg]~.
\label{c0exact}
\end{align}
Analogous exact expressions can be obtained for the leading terms of the other operators, as well as for their sub-leading contributions in the $1/N$ expansion. In particular, for $g$ odd one finds that $c_g(\cO)$ is linear in $\cP_k$ and thus its vacuum expectation value vanishes. Therefore, the large-$N$ expansion of $\langle \cO\rangle_0$ proceeds in powers of $1/N^2$, rather than $1/N$, and takes the form
\begin{align}
    \big\langle \mathcal{O}\big\rangle_0 =\sum_{h=0}^\infty N^{2-2h}\,\big\langle c_{2h}(\mathcal{O})\big\rangle_0~. 
    \label{OlargeNvev}
\end{align}
This structure is consistent with the fact that these expectation values compute integrated correlators that are holographically dual to closed string amplitudes.
Finally, since $c_0(\cO)$ does not depend on $\cP$, one simply has
\begin{align}
    \big\langle c_0\big(\cO\big)\big\rangle_0=c_0\big(\cO\big)~.
\end{align}

\subsection{\texorpdfstring{$\mathbf{D}$}{}-theory}
 We now extend our analysis to the \textbf{D}-theory, concentrating on the terms that depend on the mass $m$ of the anti-symmetric hypermultiplets and the squashing parameter $b$ in the expansion (\ref{SDSexp}), setting the fundamental mass $\mu$ to zero. The coefficient of $m^2$ is
\begin{align}
    \mathcal{M}^\D_{2,\mathrm{A}}&=N(N-1)(1+\gamma)+\sum_{n=1}^\infty\sum_{k=0}^{2n}(-1)^n\,\frac{(2n+1)!\,\zeta(2n+1)}{k!\,(2n-k)!
}\,\Big(\frac{\lambda}{8\pi^2N}\Big)^{n}\,\tr a^{2n-k}\,\tr a^{k}\notag \\
    &\qquad\qquad\qquad\qquad\qquad -\sum_{n=1}^\infty(-1)^n\,(2n+1)\,\zeta(2n+1)\,\Big(\frac{\lambda}{2\pi^2N}\Big)^{n}\,\tr a^{2n}~.\label{M2aDtr}
\end{align}
The first line closely resembles the expression (\ref{M2explicit}), although it is not identical. In contrast, the second line introduces a qualitatively novel structure involving single-trace terms of even degree. This difference arises because we are considering massive hypermultiplets in the anti-symmetric representation of SU($N$).
A similar analysis —though algebraically more involved— yields the coefficient
\begin{align}
    \mathcal{B}^\D_{2}&=\,(N^2-1)(1+\gamma) +\sum_{n=1}^\infty\sum_{\ell=0}^{n}(-1)^{n}\frac{(2n+1)!\,\zeta(2n+1)}{(2\ell)!(2n-2\ell)!}\Big(\frac{\lambda}{8\pi^2N}\Big)^{n}\,\tr a^{2n-2\ell}\,\tr a^{2\ell} \notag\\ 
    &~+\frac{1}{3}\sum_{n=1}^\infty\sum_{\ell=1}^{n-1}(-1)^{n}\frac{(2n+2)!\Big[4n\,\zeta(2n+1)\!+\!(2n+3)\,\zeta(2n+3)\Big]}{(2\ell+1)!\,(2n-2\ell+1)!}\Big(\frac{\lambda}{8\pi^2N}\Big)^{n+1}\tr a^{2n-2\ell+1}\,\tr a^{2\ell+1} \notag \\
     &~-\frac{4}{3}\sum_{n=1}^\infty (-1)^n\,\big(4^n-1)\Big[2n\,\zeta(2n+1)-(2n+3)\,\zeta(2n+3)\Big]\Big(\frac{\lambda}{8\pi^2N}\Big)^{n+1}\tr a^{2n+2}~.\label{B2Dtr}
\end{align}
As before, both double-trace and even single-trace contributions appear. Unlike the $\mathcal{N}=4$ SYM case, the operators $\mathcal{B}^\D_{2}$ and $\mathcal{M}^\D_{2,\mathrm{A}}$ differ significantly, and thus no simple relation analogous to (\ref{relation1}) can be established.

The other coefficients in the expansion of the effective action can also be derived within this formalism. Although the derivations are straightforward, they are algebraically lengthy. For completeness, here we write the explicit forms of $\mathcal{M}^\D_{4,\mathrm{A}}$ and
$\mathcal{C}^\D_{\mathrm{}{A}}$:
\begin{subequations}
    \begin{align}
\mathcal{M}^\D_{4,\mathrm{A}}&=-\frac{N(N-1)}{2}\zeta(3)-\frac{1}{12}\sum_{n=1}^\infty\sum_{k=0}^{2n}(-1)^n\,\frac{(2n+3)!\,\zeta(2n+3)}{k!\,(2n-k)!}\Big(\frac{\lambda}{8\pi^2N}\Big)^{n}\,\tr a^{2n-k}\,\tr a^{k}\notag \\
& \,+\frac{1}{12}\sum_{n=1}^\infty(-1)^n\,(2n+3)(2n+2)(2n+1)\,\zeta(2n+3)
\Big(\frac{\lambda}{2\pi^2N}\Big)^{n}\,\tr a^{2n}~,\label{M4aDtr}\\[2mm]
\mathcal{C}^\D_{\mathrm{A}}
&=N(N-1)\Big(\frac{1}{3}-\zeta(3)\Big)\notag\\
&~+\frac{1}{6}\sum_{n=1}^{\infty}\sum_{k=0}^{2n}(-1)^n \,\frac{(2n+2)!\Big[2n\,\zeta(2n+1)-(2n+3)\,\zeta(2n+3)\Big]}{k!(2n-k)!}\,\Big(\frac{\lambda}{8\pi^2N}\Big)^n\,\tr a^{2n-k}\,\tr a^{k} \notag \\
&~ -\frac{1}{6}\sum_{n=0}^{\infty}(-1)^n (2n+2)(2n+1)\Big[2n\,\zeta(2n+1)-(2n+3)\,\zeta(2n+3)\Big]\Big(\frac{\lambda}{2\pi^2N}\Big)^n\,\tr a^{2n}~.\label{CaDtr}
\end{align}
\end{subequations}
The expression for $\mathcal{B}^\D_{4}$, which is considerably longer, is reported in Appendix~\ref{App:B4}. In all cases, both double-trace and  single-trace structures occur.
It is evident from these formulas that these operators of the \textbf{D}-theory are significantly more complicated than those of $\cN=4$ SYM. However, we now show that at leading order in the ’t Hooft planar limit they greatly simplify and in fact match exactly with those in $\cN=4$ SYM.

\subsubsection*{\texorpdfstring{Large-$N$}{} limit}
Adopting the basis introduced in (\ref{changebasis}), one readily observes that, in the ’t Hooft limit, all operators defined above scale as
$N^2$ and admit a large-$N$ expansion with the same structure as in (\ref{OlargeN}):
\begin{align}
    \mathcal{O} =\sum_{g=0}^\infty N^{2-g}\,c_g(\mathcal{O}) ~, 
    \label{OlargeN2}
\end{align}
where $\mathcal{O}$ stands for any of the operators $\mathcal{M}_{2,\mathrm{A}}^\D$, $\mathcal{B}_{2}^\D$, $\mathcal{M}_{4,\mathrm{A}}^\D$, $\mathcal{C}_{\mathrm{A}}^\D$ or $\mathcal{B}_{4}^\D$.

Consider for example $\mathcal{M}_{2,\mathrm{A}}^\D$. Using (\ref{M2aDtr}) together with (\ref{tracesVEV}), one finds that the leading term $c_0\big(\mathcal{M}_{2,\mathrm{A}}^\D\big)$ coincides precisely with (\ref{c0exact}), namely
\begin{align}
   c_0\big(\mathcal{M}_{2,\mathrm{A}}^\D\big)=c_0\big(\mathcal{M}_{2}^\NN\big)~ . 
\end{align}
Hence, quite remarkably, in the planar limit the operator $\mathcal{M}_{2,\mathrm{A}}^\D$ becomes identical to its $\mathcal{N}=4$ counterpart,
differing only at the sub-leading orders in $1/N$. Explicitly,
 \begin{align}
     \mathcal{M}_{2,\mathrm{A}}^\D = \mathcal{M}_{2}^\NN +O (N)~. 
 \end{align}
An analogous pattern holds for all other operators:
\begin{align}
 \mathcal{B}_{2,\mathrm{A}}^\D &=
\mathcal{B}_{2}^\NN + O(N) ~, \qquad    \mathcal{M}_{4,\mathrm{A}}^\D = \mathcal{M}_{4}^\NN + O(N) ~, \notag\\     
\mathcal{C}_{4}^\D &= \mathcal{C}_{4}^\NN + O(N) ~,      \qquad\mathcal{B}_{4}^\D = \mathcal{B}_{4}^\NN + O(N) ~.
\end{align}

Recalling that the vacuum expectation value of any matrix operator $f(a)$ in the \textbf{D}-theory is defined as
\begin{align}
\big\langle f(a) \big\rangle := \frac{\big\langle  \rme^{ - S^\D }\,f(a) \big\rangle_0\phantom{\Big|}}{\big\langle  \rme^{ - S^\D} \big\rangle_0\phantom{\Big|}}~,
\end{align}
it follows immediately that
\begin{align}
    \big\langle c_0(\cO)\big\rangle =c_0(\cO)
\end{align}
since the leading coefficients are $\cP$-independent constants (see, for example, \eqref{eq:M2}). By contrast, the subleading terms
$c_g(\cO)$, including those with odd $g$, acquire a non-vanishing vacuum expectation value. 
This occurs because the operators $\cO$ in the \textbf{D}-theory contain explicit 
single-trace contributions. Consequently,
\begin{align}
    \big\langle \mathcal{O}\big\rangle =\sum_{g=0}^\infty N^{2-g}\,\big\langle c_{g}(\mathcal{O})\big\rangle~.
\end{align}
Thus, in contrast with (\ref{OlargeNvev}), the large-$N$ expansion of $\big\langle \mathcal{O}\big\rangle$ involves both even and odd powers of $1/N$. This is consistent with the fact that these expectation values are holographically dual to quantities in the theory involving open strings due to D7-branes.

Using relations such as
\begin{align}
    \partial_m^2\log \mathcal{Z}^\DS\big|_{\D}=-2\,\big\langle\mathcal{M}_{2,\mathrm{A}}^\D\big\rangle~,\quad \partial_b^2\log \mathcal{Z}^\DS\big|_{\D}=-2\,\big\langle\mathcal{B}_{2}^\D\big\rangle~,\quad\mbox{etc.}
\end{align}
together with the equality of the leading coefficients $c_0\big(\cO\big)$ in the \textbf{D}-theory and in $\cN=4$ SYM,
one obtains the following relations:
\begin{subequations}
    \begin{align}
        \partial_m^2 \log \mathcal{Z}^\DS\big|_{\D}&=\partial_m^2 \log \mathcal{Z}^\NS\big|_{0}+ O(N)~, \label{relation0a1}
        \\
        \partial_m^4 \log \mathcal{Z}^\DS\big|_{\D}&=\partial_m^4  \log \mathcal{Z}^\NS\big|_{0}+ O(N)~,\label{relation2aD} \\  \partial_m^2\partial_b^2 \log \mathcal{Z}^\DS\big|_{\D}&=\partial_m^2\partial_b^2 \log \mathcal{Z}^\NS\big|_{0} + O(N)~.\label{relation3aD} \\
           \Big(\partial_b^4-15\,\partial_b^2\Big)\log \mathcal{Z}^\DS\big|_{\D}&=\Big(\partial_b^4-15\,\partial_b^2\Big)\log \mathcal{Z}^\NS\big|_{0}+ O(N)~.\label{relation4aD}
    \end{align}
    \label{relationsD}%
\end{subequations}
These relations demonstrate that the identities (\ref{relationsN4}), which hold exactly in $\cN=4$ SYM, continue to be valid in the 
\textbf{D}-theory only at leading order in the large-$N$ expansion.
This observation is consistent with the planar equivalence of the \textbf{D}-theory with $\mathcal{N}=4$ SYM, provided one considers only operators that in the holographic dual description correspond to modes of the closed string sector. This equivalence can also be understood from the superstring amplitude viewpoint, since at leading order in the large-$N$ limit, the four-graviton scattering occurs entirely in the bulk and is not affected by the presence of the D7-branes.

\section{Integrated correlators with open and closed string modes}
\label{secn:integratedmixed}
We now analyze the response of the \textbf{D}-theory when the four fundamental hypermultiplets are assigned a mass $\mu$. This scenario differs qualitatively from the cases previously considered, since such deformations correspond to open-string excitations in the holographic dual description \cite{Behan:2023fqq,Billo:2024ftq,Chester:2025ssu}. Within the matrix model, this response is captured by the coefficients in the expansion of the effective action that involve powers of $\mu$. The quadratic and quartic terms in $\mu$ were already studied in detail in \cite{Behan:2023fqq,Billo:2024ftq}, providing integrated results that can be used as constraints for the four-gluon scattering process in AdS. Here, instead, we focus on the mixed structures proportional to $\mu^2 m^2$ and $\mu^2 (b-1)^2$. These terms are associated with integrated correlators of the \textbf{D}-theory that, in the holographic dual, correspond to mixed scattering amplitudes involving both open and closed strings.
The $\mu^2 m^2$ structure was recently analyzed in \cite{Chester:2025ssu} for a conformal $\mathcal{N}=2$ SYM theory with gauge group Sp($N$). In this work, we extend that analysis to the \textbf{D}-theory, also including the effects of the squashing deformation.

From the expansion (\ref{SDSexp}), one finds
\begin{equation}
    \begin{aligned}
    \log \mathcal{Z}^\DS &=\log \mathcal{Z}^\D+\ldots+\mu^2 m^2\Big[\big\langle\mathcal{M}_{2,\mathrm{F}}^\D\,\,\mathcal{M}_{2,\mathrm{A}}^\D\big\rangle-\big\langle\mathcal{M}_{2,\mathrm{F}}^\D\big\rangle\,\big\langle\mathcal{M}_{2,\mathrm{A}}^\D\big\rangle
    \Big] \\[1mm]
    &\quad -\mu^2(b-1)^2\Big[\big\langle\mathcal{C}_{\mathrm{F}}^\D\big\rangle-\big\langle\mathcal{M}_{2,\mathrm{F}}^\D\,\,\mathcal{B}_2^\D\big\rangle+\big\langle\mathcal{M}_{2,\mathrm{F}}^\D\big\rangle\,\big\langle\mathcal{B}_2^\D\big\rangle\Big] +\ldots\, ,
    \label{logZexpD}
\end{aligned}
\end{equation}
where we have displayed only the terms that are relevant to our present discussion. These will be studied in detail in the following.

\subsection{\texorpdfstring{The $\mu$/$m$ mixed derivative: $\partial_\mu^2\,\partial_m^2\log\mathcal{Z}^\DS$}{}}
We first analyze the mixed derivative
\begin{align}
\partial_\mu^2\,\partial_m^2\log\mathcal{Z}^\DS\big|_{\D}=4\Big[\big\langle\mathcal{M}_{2,\mathrm{F}}^\D\,\,\mathcal{M}_{2,\mathrm{A}}^\D\big\rangle-\big\langle\mathcal{M}_{2,\mathrm{F}}^\D\big\rangle\,\big\langle\mathcal{M}_{2,\mathrm{A}}^\D\big\rangle \Big] \, ,
\label{intcorr1}
\end{align}
where $\mathcal{M}_{2,\mathrm{A}}^\D$ is defined in (\ref{M2aDtr}) and $\mathcal{M}_{2,\mathrm{F}}^\D$ is
\begin{align}
    \mathcal{M}_{2,\mathrm{F}}^\D=4\biggl[N(1+\gamma)+\sum_{n=1}^\infty(-1)^n\,(2n+1)\,\zeta(2n+1)\Big(\frac{\lambda}{8\pi^2N}\Big)^{n}\,\tr a^{2n}\biggr]~.\label{M2fDtr}
\end{align}
A direct evaluation of the correlators in the right-hand side of (\ref{intcorr1}) is challenging. However, upon performing a large-$N$ expansion of the operators $\mathcal{M}_{2,\mathrm{F}}^\D$ and $\mathcal{M}_{2,\mathrm{A}}^\D$, the computation becomes tractable and can be carried out systematically order by order in $1/N$. In what follows, we demonstrate this procedure first using the full Lie-algebra approach, and then independently using the topological recursion.

\subsubsection*{Full Lie-algebra approach}
As shown in (\ref{OlargeN}), the term $\mathcal{M}_{2,\mathrm{A}}^\D$ admits the expansion
\begin{align}
\mathcal{M}_{2,\mathrm{A}}^\D=N^2\,c_0\big(\mathcal{M}_{2,\mathrm{A}}^\D\big)+N\,c_1\big(\mathcal{M}_{2,\mathrm{A}}^\D\big)+c_2\big(\mathcal{M}_{2,\mathrm{A}}^\D\big)+O(1/N)\, , 
\label{M2Aexpanded}
\end{align}
where $c_0\big(\mathcal{M}_{2,\mathrm{A}}^\D\big)$ is given in (\ref{c0exact}).
The sub-leading coefficients can be expressed in terms of the $\mathcal{P}$-operators after performing in (\ref{M2aDtr}) the change of basis (\ref{changebasis}) and resumming the perturbative series in $\lambda$ using Bessel functions, as detailed in \cite{Billo:2024ftq}.
 The first two coefficients are\,\footnote{The explicit expressions of the $\mathcal{P}$-independent parts of $c_1$ and $c_2$ are not required since they cancel in the correlators relevant to our analysis.}
\begin{subequations}
    \begin{align}
    c_1\big(\mathcal{M}_{2,\mathrm{A}}^\D\big)&=\frac{8\,\pi}{\sqrt{\lambda}}\sum_{k=1}^\infty(-1)^k\sqrt{2k}\,\mathsf{M}_{1,2k}^{(1)}\,\mathcal{P}_{2k}+\mathcal{P}\text{-independent terms}~,\label{c1M2A}\\
    c_2\big(\mathcal{M}_{2,\mathrm{A}}^\D\big)&=\!\sum_{n,m=1}^{\infty}\!\!(-1)^{n+m}\Bigl(\sqrt{2n}\sqrt{2m}\,\mathsf{M}_{2n,2m}^{(2)}\mathcal{P}_{2n}\mathcal{P}_{2m}\!-\!\sqrt{2n+1}\sqrt{2m+1}\,\mathsf{M}_{2n+1,2m+1}^{(2)}\mathcal{P}_{2n+1}\mathcal{P}_{2m+1}\Bigr)\notag\\
    &\quad-\sum_{k=1}^\infty(-1)^k\,\sqrt{2k}\,\widehat{\mathsf{Z}}_{2k}^{(2)}\,\mathcal{P}_{2k}+\mathcal{P}\text{-independent terms}~,\label{c2M2A}
\end{align}
\end{subequations}
where we defined the integral representations:
    \begin{align}
\mathsf{M}_{n,m}^{(p)}&=\int_0^{\infty}\! \frac{dw}{4w}\,\frac{ (2w)^{\,p}}{ \sinh^2w}\,J_{n}\bigg(\frac{\sqrt{\lambda}\,w}{\pi}\bigg)\,J_{m}\bigg(\frac{\sqrt{\lambda}\,w}{\pi}\bigg)~,
    \label{Mkp}\\[2mm]
  \widehat{\mathsf{Z}}_n^{(p)}&=\int_0^{\infty}\! \frac{dw}{4w}\,\frac{ (2w)^{\,p}}{\sinh^2w}\,J_{n}\bigg(\frac{2
  \sqrt{\lambda}\,w}{\pi}\bigg)~,
    \label{Zhatkp}    
\end{align}
which are valid at arbitrary positive coupling $\lambda$.

Applying the same procedure to (\ref{M2fDtr}), we note a crucial difference: the leading contribution of $\mathcal{M}_{2,\mathrm{F}}^\D$ scales as $N$, rather than $N^2$. This is consistent with the fact that the derivative with respect to fundamental mass $\mu$ is related to open string modes. Consequently, its large-$N$ expansion takes the form
\begin{align}
\mathcal{M}_{2,\mathrm{F}}^\D=N\,c_0\big(\mathcal{M}_{2,\mathrm{F}}^\D\big)+c_1\big(\mathcal{M}_{2,\mathrm{F}}^\D\big)+O(1/N)\, ,
\label{M2Fexpanded}
\end{align}
where
\begin{subequations}
    \begin{align}
       c_0\big(\mathcal{M}_{2,\mathrm{F}}^\D\big)&=4(1+\gamma)+2\sum_{n=1}^\infty(-1)^n\frac{(2n+2)!\,\zeta(2n+1)}{\Gamma(n+2)^2}\Big(\frac{\lambda}{16\pi^2}\Big)^n\notag\\
       &=4(1+\gamma)+\int_0^\infty\!dw\,\frac{4w}{\sinh^2w}\bigg[\frac{2\pi}{w\sqrt{\lambda}}\,J_1\bigg(\frac{\sqrt{\lambda}\,w}{\pi} \bigg)-1\bigg]
       ~,\label{c0M2F}\\[2mm]
       c_1\big(\mathcal{M}_{2,\mathrm{F}}^\D\big)&=4\,\sum_{k=1}^\infty(-1)^k\,\sqrt{2k}\,\,\mathsf{Z}_{2k}^{(2)}\,\mathcal{P}_{2k}~,\label{c1M2F}
    \end{align}
\end{subequations}
with
\begin{align}
    \mathsf{Z}_n^{(p)}&=\int_0^{\infty}\! \frac{dw}{4w}\,\frac{ (2w)^{\,p}}{\sinh^2w}\,J_{n}\bigg(\frac{
  \sqrt{\lambda}\,w}{\pi}\bigg)~.
    \label{Zkp}  
\end{align}
Substituting the large-$N$ expansions (\ref{M2Aexpanded}) and (\ref{M2Fexpanded}) into (\ref{intcorr1}), we easily realize that the terms $N^2\,c_0\big(\mathcal{M}_{2,\mathrm{A}}^\D\big)$
and $N\,c_0\big(\mathcal{M}_{2,\mathrm{F}}^\D\big)$ can be dropped since, being $\mathcal{P}$-independent, they cancel in the combination in the right-hand side of (\ref{intcorr1}). Thus, the mixed derivative simplifies to
\begin{align}
    \partial_\mu^2\,\partial_m^2\log\mathcal{Z}^\DS\big|_{\D}\,&=4N\,\Big[ \big\langle c_1\big(\mathcal{M}_{2,\mathrm{F}}^\D\big)\,c_1\big(\mathcal{M}_{2,\mathrm{A}}^\D\big)\big\rangle-\big\langle c_1\big(\mathcal{M}_{2,\mathrm{F}}^\D\big)\big\rangle\,\big\langle\,c_1\big(\mathcal{M}_{2,\mathrm{A}}^\D\big)\big\rangle\Big]\\
    &\quad+4\Big[\big\langle c_1\big(\mathcal{M}_{2,\mathrm{F}}^\D\big)\,c_2\big(\mathcal{M}_{2,\mathrm{A}}^\D\big)\big\rangle-\,\big\langle c_1\big(\mathcal{M}_{2,\mathrm{F}}^\D\big)\big\rangle\,\big\langle c_2\big(\mathcal{M}_{2,\mathrm{A}}^\D\big)\big\rangle\Big]+O(1/N)~.\notag
\end{align}
From the previously derived expressions we obtain
\begin{subequations}
    \begin{align}
    &\big\langle c_1\big(\mathcal{M}_{2,\mathrm{F}}^\D\big)\,c_1\big(\mathcal{M}_{2,\mathrm{A}}^\D\big)\big\rangle-\big\langle c_1\big(\mathcal{M}_{2,\mathrm{F}}^\D\big)\big\rangle\,\big\langle\,c_1\big(\mathcal{M}_{2,\mathrm{A}}^\D\big)\big\rangle\notag\\
    &\qquad
   =\frac{32\,\pi}{\sqrt{\lambda}}\sum_{k,\ell=1}^\infty(-1)^{k+\ell}\,\sqrt{2k}\,\sqrt{2\ell}\,\,\mathsf{Z}_{2k}^{(2)}\,\mathsf{M}_{1,2\ell}^{(1)}\,\,\big\langle  \mathcal{P}_{2k}\,\mathcal{P}_{2\ell}\big\rangle^c~,\\[2mm]
   &\big\langle c_1\big(\mathcal{M}_{2,\mathrm{F}}^\D\big)\,c_2\big(\mathcal{M}_{2,\mathrm{A}}^\D\big)\big\rangle-\,\big\langle c_1\big(\mathcal{M}_{2,\mathrm{F}}^\D\big)\big\rangle\,\big\langle c_2\big(\mathcal{M}_{2,\mathrm{A}}^\D\big)\big\rangle=\notag\\
    &\qquad=4\!\!\sum_{k,\ell,m=1}^\infty(-1)^{k+\ell+m}\,\sqrt{2k}\,\,\mathsf{Z}_{2k}^{(2)}\,\sqrt{2\ell}\,\sqrt{2m}\,\,\mathsf{M}_{2\ell,2m}^{(2)}\Big[\big\langle\mathcal{P}_{2k}\,\mathcal{P}_{2\ell}\,\mathcal{P}_{2m}\big\rangle-\big\langle\mathcal{P}_{2k}\big\rangle \big\langle\mathcal{P}_{2\ell}\,\mathcal{P}_{2m}\big\rangle\Big]\notag \\
    &\qquad-4\!\!\sum_{k,\ell,m=1}^\infty(-1)^{k+\ell+m}\,\sqrt{2k}\,\,\mathsf{Z}_{2k}^{(2)}\,\sqrt{2\ell+1}\,\sqrt{2m+1}\,\,\mathsf{M}_{2\ell+1,2m+1}^{(2)}\,\big\langle\mathcal{P}_{2k}\,\mathcal{P}_{2\ell+1}\,\mathcal{P}_{2m+1}\big\rangle^c\notag\\
    &\qquad-4\,\sum_{k,\ell=1}^\infty(-1)^{k+\ell}\,\sqrt{2k}\,\sqrt{2\ell}\,\,\mathsf{Z}_{2k}^{(2)}\,\widehat{\mathsf{Z}}_{2\ell}^{(2)}\,\big\langle  \mathcal{P}_{2k}\,\mathcal{P}_{2\ell}\big\rangle^c \, , 
\end{align}%
\end{subequations}
where $\langle\,\cdot\,\rangle^c$ denotes a connected correlator.
These formulas show that everything is reduced to the calculation of the expectation values of products of $\mathcal{P}$-operators in the matrix model of the \textbf{D}-theory. Such expectation values have been computed in \cite{Billo:2024ftq} and here we report them in the large-$N$ expansion up to the order needed for our purposes:
\begin{subequations}
    \begin{align}
\big\langle  \mathcal{P}_{2k}\,\mathcal{P}_{2\ell}\big\rangle^c &=\,\delta_{k,\ell}+
    \frac{\sqrt{2k}\,\sqrt{2\ell}\,\,{\mathsf{Y}}}{N}+O(1/N^2)~, \label{P2nP2mcD} \\
\big\langle  \mathcal{P}_{2k}\,\mathcal{P}_{2\ell}\,\mathcal{P}_{2m}\big\rangle-\big\langle  \mathcal{P}_{2k}\big\rangle\big\langle\mathcal{P}_{2\ell}\,\mathcal{P}_{2m}\big\rangle&=\, \delta_{k,\ell}\,\mathsf{Y}_{2m}+\delta_{k,m}\,\mathsf{Y}_{2\ell} +O(1/N)~, \label{PPPevenD} \\[1mm]
\big\langle  \mathcal{P}_{2k}\,\mathcal{P}_{2m+1}\,\mathcal{P}_{2n+1}\big\rangle^c&=0+O(1/N)~, \label{PPPoddD}
\end{align}
\label{vevsP}%
\end{subequations}
where\,\footnote{We point out that in \cite{Billo:2024ftq} the integral representations for $\mathsf{Y}_{2k}$ and $\mathsf{Y}$ are written with a different kernel, but they are identical to those presented here.}
\begin{align}
\mathsf{Y}_{2k} &=(-1)^{k+1}2\sqrt{2k}\,\Big(\,\widehat{\mathsf{Z}}_{2k}^{(0)}-4\mathsf{Z}_{2k}^{(0)}\Big)~,\label{Y2k}\\
\mathsf{Y}&=\sum_{k=1}^\infty\sqrt{2k}\,\mathsf{Y}_{2k}=\frac{\sqrt{\lambda}}{\pi}
\,\Big(\,\widehat{\mathsf{Z}}_{1}^{(1)}-2\mathsf{Z}_{1}^{(1)}\Big)~.
\label{Y}
\end{align}
Using these results, it is immediate to obtain
\begin{align}
    \partial_\mu^2\,\partial_m^2\log\mathcal{Z}^\DS\big|_{\D}\,=\sum_{g=0}^\infty N^{1-g}\,\mathcal{F}_g^\D \, , 
    \label{dmudmlargeN}
\end{align}
where all coefficients $\mathcal{F}_g^\D$ can be written in terms of the integral kernels introduced above. After some algebra, one obtains the following expressions for the first two coefficients:
\begin{subequations}
\begin{align}
    \mathcal{F}_0^\D&=\frac{128\,\pi}{\sqrt{\lambda}}\,\sum_{k=1}^\infty(2k)\,\mathsf{Z}_{2k}^{(2)}\,\mathsf{M}_{1,2k}^{(1)}~,\label{F0}\\
    \mathcal{F}_1^\D&=\frac{128\,\pi}{\sqrt{\lambda}}\,\textsf{Y}\,\sum_{k,\ell=1}^\infty(-1)^{k+\ell}\,(2k)(2\ell)\,\mathsf{Z}_{2k}^{(2)}\,\mathsf{M}_{1,2\ell}^{(1)}-16\,\sum_{k=1}^\infty(2k)\,\mathsf{Z}_{2k}^{(2)}\,\widehat{\mathsf{Z}}_{2k}^{(2)}\notag\\
    &\quad+32\sum_{k,\ell=1}^\infty(-1)^\ell(2k)\,\sqrt{2\ell}\,\,\mathsf{Z}_{2k}^{(2)}\,\mathsf{M}_{2k,2\ell}^{(2)}\,\textsf{Y}_{2\ell}~.\label{F1}
\end{align}
\label{eq:dmu2m2F0andF1}%
\end{subequations}
These results are exact in the ’t Hooft coupling $\lambda$.
The higher-order terms $\mathcal{F}_g^\D$ ($g\geq 2$) can also be systematically computed in this way, but their derivation becomes increasingly intricate. Substituting the integral definitions in terms of Bessel functions, one can prove that
\begin{align}
    \mathcal{F}_0^\D=\frac{1}{2}F_1~,
    \label{F0F1}
\end{align}
with $F_1$ defined in Eq.\,(B.11) of \cite{Chester:2025ssu}
and
\begin{align}
    32\sum_{k,\ell=1}^\infty(-1)^\ell(2k)\,\sqrt{2\ell}\,\,\mathsf{Z}_{2k}^{(2)}\,\mathsf{M}_{2k,2\ell}^{(2)}\,\textsf{Y}_{2\ell}=-\frac{1}{2}F^{\mathrm{hard}}~, 
\end{align}
where $F^{\mathrm{hard}}_2$ is given in Eq.\,(B.18) of the published version of \cite{Chester:2025ssu}.

\subsubsection*{Topological recursion approach}
It is instructive to rederive the large-$N$ results using the method of topological recursion \cite{Eynard:2004mh,Eynard:2008we}, which has been employed in related contexts, for example in \cite{Chester:2019pvm,Behan:2023fqq,Chester:2025ssu}. This provides both an alternative derivation and an independent check of our results.

To illustrate the method, we first rewrite the matrix-model action of the \textbf{D}-theory, originally given in (\ref{Sd}), as
\begin{align} 
\label{eq:actionD}
S^\D = \int_0^\infty \!\!\frac{d\omega}{4\omega\sinh^2 \omega}\, \bigg  \{6N \!-\! \Big[(f(2\ii \omega) {-} f(-2\ii \omega)\Big]^2
+ \Big[ f(4\ii \omega)  - 4 f(2\ii \omega) +(\omega \rightarrow -w) \Big] \bigg \}~,
\end{align}
with
\begin{align} \label{eq:fx}
    f(x)= \tr \exp\bigg(\sqrt{\frac{\lambda}{8\pi^2N}}\,a\,x\bigg)~.
\end{align}
Expanding the integrand of (\ref{eq:actionD}) in powers of $\omega$ and evaluating the resulting integrals reproduces the full series in (\ref{Sd}), where the double- and single-trace terms arise, respectively, from the quadratic and linear contributions in $f$. A similar rewriting applies to the expansion coefficients of the $\mathbf{D}^*$-theory action (\ref{SDSexp}). For example, one finds
\begin{subequations}
    \begin{align}
\mathcal{M}_{2,\mathrm{F}}^\D &=4N(1+\gamma) -\!\int_0^\infty\!\! \frac{2\,\omega \,d\omega}{\sinh^2 \omega} \,\Big[2N - f(2\ii \omega) - f(-2\ii \omega) \Big]~,\\[2mm]
\mathcal{M}_{2,\mathrm{A}}^\D 
&=N(N{-}1)(1{+}\gamma)-\!\int_0^\infty\!\!\frac{\omega\,d\omega}{2\sinh^2 \omega}\,\Big  \{ 2N(N{-}1) + \Big [ f(4\ii \omega)  {-} f(2\ii \omega)^2+(\omega \rightarrow -\omega) \Big] \Big \}~,
\end{align}
\label{eq:ZfZaS0}%
\end{subequations}
which agree precisely with (\ref{M2fDtr}) and (\ref{M2aDtr}). Substituting these expressions into \eqref{intcorr1} gives
\begin{align}
\partial_\mu^2\,\partial_m^2\log\mathcal{Z}^\DS\big|_{\D}&=\,\iint_{0}^\infty\!\frac{4\,\omega\,\nu\,d\omega\,d\nu}{\sinh^2\omega\,\sinh^2\nu}\,\Bigg\{\bigg[\big\langle f(2\ii\nu)^2\,f(2\ii\omega)\big\rangle-\big\langle f(2\ii\nu)^2\big\rangle\,\big\langle f(2\ii\omega)\big\rangle
\label{d2mud2mtop1}\\
&\qquad-\big\langle f(4\ii\nu)\,f(2\ii\omega)\big\rangle+\big\langle f(4\ii\nu)\big\rangle\,\big\langle f(2\ii\omega)\big\rangle+(\omega\rightarrow-\omega)\bigg]+(\nu\rightarrow-\nu)\Bigg\}~.
\notag
\end{align}
Thus, the relevant quantities are the correlators
\begin{align}
    \big\langle \prod_{i=1}^n f(z_i)\big\rangle =
    \frac{\displaystyle{\big\langle  \rme^{ - S^\D }\,\prod_{i=1}^n f(z_i) \big\rangle_0\phantom{\Big|}}}{\displaystyle{\big\langle  \rme^{ - S^\D} \big\rangle_0\phantom{\Big|}}}~.
\end{align}
Expanding the exponential factors and expressing $S^\D$ in terms of $f$ via (\ref{eq:actionD}), one finds that the computation reduces to evaluating free correlators of the following form
\begin{align} \label{eq:fzi0}
    \big\langle \prod_{i=1}^n f(z_i)\big\rangle_0 \, ,
\end{align}
for various values of $n$. 
These Gaussian correlators decompose naturally into connected components. The latter can be systematically obtained through topological recursion \cite{Chester:2019pvm}, which organizes them as a genus expansion in the large-$N$ limit:
\begin{align}
    \label{W-funs}
    \big\langle \prod_{i=1}^n f(z_i) \big\rangle_0^c = \sum_{g=0}^\infty N^{2-2g-n}\, W_g^n(z_1,\cdots,z_n)~. 
\end{align} 
Explicit expressions for the first few genus-0 coefficients $W_0^n$ in terms of Bessel functions are provided in Appendix~\ref{app:resolvent}.

Applying this procedure to (\ref{d2mud2mtop1}) allows for a systematic generation of the large-$N$ expansion, reproducing the results in \eqref{eq:dmu2m2F0andF1}. At the planar level, the action $S^\D$ itself does not contribute, and the dominant terms arise from products of free connected correlators with the maximal number of factors. Consequently,
\begin{align}
     \mathcal{F}_0^\D &=
     \iint_0^\infty \!\frac{8 \,\omega\, \nu\, d \omega \,d \nu}{\sinh^2\omega \,\sinh^2 \nu} \,\bigg\{\Big[ W_0^1(2 \ii\nu)\,W_0^2(2\ii\nu,2\ii\omega) + (\omega\rightarrow -\omega)\Big]+(\nu\rightarrow -\nu)\Big]\bigg\}\notag\\[1mm]
     &= -\iint_0^\infty \!\frac{32\,\omega^2\,\nu\,d \omega \,d\nu}{\sinh^2w \sinh^2 \nu}\, \frac{J_1\big(\frac{\sqrt{\lambda}\,\nu}{\pi }\big)\Big[
   \nu\, J_0\big(\frac{\sqrt{\lambda}\,\nu}{\pi}\big)
   \,J_1\big(\frac{\sqrt{\lambda}\,\omega}{\pi}\big)-\omega \,J_1\big(\frac{
   \sqrt{\lambda}\,\nu}{\pi}\big) \,J_0\big(\frac{\sqrt{\lambda}\,\omega}{\pi}\big)\Big]}{\nu^2-\omega^2}~.\label{eq:F1intnew}
\end{align}
Using the Bessel kernel identity
\begin{equation} \label{eq:BJ-id}
   \frac{x\,J_0\big(\frac{\sqrt{\lambda}\,x}{\pi}\big)\, J_1\big(\frac{\sqrt{\lambda}\,y}{\pi}\big) - y\, J_1\big(\frac{\sqrt{\lambda}\,x}{\pi}\big) \,J_0\big(\frac{\sqrt{\lambda}\,y}{\pi}\big) }{y^2 -x^2}= \frac{4\pi}{\sqrt{\lambda}}\sum_{k=1}^\infty \frac{k}{x\,y} \,J_{2k}\Big(\frac{\sqrt{\lambda}\,x}{\pi}\Big)\, J_{2k}\Big(\frac{\sqrt{\lambda}\,y}{\pi}\Big)~,
\end{equation}
one verifies that \eqref{eq:F1intnew} precisely reproduces $\mathcal{F}_0^\D$ in \eqref{F0}. Extending the analysis to the sub-leading order reveals that only the single-trace part of the action $S^\D$ is required, considerably simplifying the derivation. Proceeding in this way, we have found complete agreement with $\mathcal{F}_1^\D$ in \eqref{F1}, thereby establishing the full equivalence between the Lie-algebra approach and the topological recursion approach.

\subsection{\texorpdfstring{The $\mu$/$b$ mixed derivative: $\partial_\mu^2\,\partial_b^2\log\mathcal{Z}^\DS$}{}}

We now consider the $\mu^2(b-1)^2$ term in the effective action (\ref{logZexpD}), corresponding to
\begin{align}
\partial_\mu^2\partial_b^2\log \mathcal{Z}^\DS\big|_{\D}=-4\Big[\big\langle\mathcal{C}_{\mathrm{F}}^\D\big\rangle-\big\langle\mathcal{M}_{2,\mathrm{F}}^\D\,\,\mathcal{B}_2^\D\big\rangle+\big\langle\mathcal{M}_{2,\mathrm{F}}^\D\big\rangle\,\big\langle\mathcal{B}_2^\D\big\rangle\Big]~.
\label{intcorr2}
\end{align}
Here, the operators $\mathcal{M}_{2,\mathrm{F}}^\D$ and $\mathcal{B}_2^\D$ are defined, respectively, in (\ref{M2fDtr}) and (\ref{B2Dtr}), while $\mathcal{C}_{\mathrm{F}}^\D$ is given by
\begin{align}
\mathcal{C}_{\mathrm{F}}^\D&=
4N\Big(\frac{1}{3}-\zeta(3)\Big)+\frac{2}{3}\sum_{n=1}^{\infty}(-1)^n \Big[2n(2n+2)(2n+1)\,\zeta(2n+1)
\notag\\&\qquad\qquad\qquad\qquad
-(2n+3)(2n+2)(2n+1)\,\zeta(2n+3)\Big]\Big(\frac{\lambda}{8\pi^2N}\Big)^n\,\tr a^{2n}~.\label{CfDtr}
\end{align}
To extract explicit results, we proceed again via a large-$N$ expansion, beginning with the full Lie-algebra approach.

\subsubsection*{Full Lie-algebra approach}

After performing the change of basis (\ref{changebasis}) and resumming the perturbative series as discussed in previous sections, the operator (\ref{CfDtr}) can be rewritten as
\begin{align}
 \mathcal{C}_{\mathrm{F}}^\D=N\,c_0\big(\mathcal{C}_{\mathrm{F}}^\D\big)+c_1\big(\mathcal{C}_{\mathrm{F}}^\D\big)+O(1/N) \, ,
\label{CFexpanded}
\end{align}
where
\begin{align}
    c_0\big(\mathcal{C}_{\mathrm{F}}^\D\big)&=-\frac{2\sqrt{\lambda}}{3\pi}\,\mathsf{Z}^{(3)}_{1}-\frac{8\pi}{3\sqrt{\lambda}}\mathsf{Z}^{(3)}_{1}+\frac{4}{3}~,\\[2mm]
    c_1\big(\mathcal{C}_{\mathrm{F}}^\D\big)&=\frac{2}{3}\sum_{k=1}^\infty\,(-1)^k\sqrt{2k}\,\bigg[4k(k+1)\mathsf{Z}^{(2)}_{2k}-\Big(1+\frac{\lambda}{4\pi^2}\Big)\mathsf{Z}^{(4)}_{2k}-\frac{\sqrt{\lambda}}{\pi}\,\mathsf{Z}^{(3)}_{2k+1} \bigg]\,\mathcal{P}_{2k}~.
\end{align}
As is typical for operators associated with derivatives with respect to the fundamental mass $\mu$, $\mathcal{C}_{\mathrm{F}}^\D$ scales as $N$ in the planar limit. In contrast, the operator $\mathcal{B}_{2}^\D$ exhibits the following large-$N$ expansion (cf. (\ref{OlargeN})):
\begin{align}
\mathcal{B}_{2}^\D=N^2\,c_0\big(\mathcal{B}_{2}^\D\big)+N\,c_1\big(\mathcal{B}_{2}^\D\big)+c_2\big(\mathcal{B}_{2}^\D\big)+O(1/N) \, ,
\label{B2expanded}
\end{align}
where 
\begin{subequations}
    \begin{align}
        c_0\big(\mathcal{B}_{2}^\D\big)&=c_0\big(\mathcal{M}_{2,\mathrm{A}}^\D\big)~,\\
        c_1\big(\mathcal{B}_{2}^\D\big)&=c_1\big(\mathcal{M}_{2,\mathrm{A}}^\D\big)+\mathcal{P}\text{-independent terms} \, , 
        \label{c1B2}
    \end{align}
    \label{c01B2}%
\end{subequations}
with $c_0\big(\mathcal{M}_{2,\mathrm{A}}^\D\big)$ and $c_1\big(\mathcal{M}_{2,\mathrm{A}}^\D\big)$ given in (\ref{c0exact}) and (\ref{c1M2A}). Notice that in the combination appearing in the right-hand side of (\ref{intcorr2}), all $\mathcal{P}$-independent components of $\mathcal{B}_2^\D$ cancel. This explains why we do not need to write them explicitly. 

The coefficient $c_2\big(\mathcal{B}_{2}^\D\big)$ is more intricate and its relation to $c_2\big(\mathcal{M}_{2,\mathrm{A}}^\D\big)$, defined in (\ref{c2M2A}), is
\begin{align}
c_2\big(\mathcal{B}_{2}^\D\big)&= c_2\big(\mathcal{M}_{2,\mathrm{A}}^\D\big)
\notag \\
&\quad+\frac{1}{3}\sum_{n,m=1}^{\infty}(-1)^{n+m}\sqrt{2n+1}\,\sqrt{2m+1}\,\Bigg[\frac{\lambda}{\pi^2}\,\mathsf{M}^{(2)}_{2n+2,2m+2}\notag\\
&\qquad\qquad\qquad+\Big(4-\frac{\lambda}{\pi^2}\Big)\mathsf{M}_{2n+1,2m+1}^{(2)}-\frac{4}{\pi}m\,\mathsf{M}^{(1)}_{2n+2,2m+1} -\frac{4}{\pi} n\,\mathsf{M}^{(1)}_{2n+1,2m+2}\notag\\
&\qquad\qquad\qquad+8(n+m)(n+m+1)\,\mathsf{M}^{(0)}_{2n+1,2m+1}\Bigg]\mathcal{P}_{2n+1}\mathcal{P}_{2m+1}\notag \\
&\quad+\frac{1}{3}\sum_{k=1}^\infty\,(-1)^k\;\sqrt{2k}\,\Bigg[16k(1-k)\,\mathsf{Z}^{(0)}_{2k}+\bigg(4+\frac{\lambda}{\pi^2}\bigg)\,\mathsf{Z}^{(2)}_{2k}-\frac{4\sqrt{\lambda}}{\pi}\,\mathsf{Z}^{(1)}_{2k+1}
\notag\\
&\qquad\qquad\qquad-4k(1-k)\,\widehat{\mathsf{Z}}^{(0)}_{2k}+\bigg(2-\frac{\lambda}{\pi^2}\bigg)\,\widehat{\mathsf{Z}}^{(2)}_{2k}+\frac{2\sqrt{\lambda}}{\pi}\,\widehat{\mathsf{Z}}^{(1)}_{2k+1}\Bigg] \,\mathcal{P}_{2k}\, ,
\label{B2DP}
\end{align}
up to $\mathcal{P}$-independent terms.

As we already remarked, all $\mathcal{P}$-independent components of $\mathcal{B}_2^\D$ cancel. Since the leading $N^2$-term belongs to this class, the mixed derivative (\ref{intcorr2}) scales as $N$ and admits an expansion analogous to (\ref{dmudmlargeN}):
\begin{align}
    \partial_\mu^2\,\partial_b^2\log\mathcal{Z}^\DS\big|_{\D}\,=\sum_{g=0}^\infty N^{1-g}\,\widetilde{\mathcal{F}}_g^\D~.
    \label{dmudblargeN}
\end{align}
The coefficients $\widetilde{\mathcal{F}}_g^\D$ can be expressed in terms of the integral kernels introduced above and take the form
\begin{align}
    \widetilde{\mathcal{F}}_g^\D=\mathcal{F}_g^\D+\Delta_g^\D~.
    \label{FFtilde}
\end{align}
After an explicit but lengthy computation, the first two corrections are found to be
\begin{subequations}
    \begin{align}
        \Delta_0^\D&=
\frac{8\sqrt{\lambda}}{3\pi}\,\mathsf{Z}^{(3)}_{1}+\frac{32\pi}{3\sqrt{\lambda}}\mathsf{Z}^{(3)}_{1}-\frac{16}{3}~, \label{Delta0}\\
\Delta_1^\D&=
\frac{32\sqrt{\lambda}}{3\pi}\sum_{k=1}^\infty(2k)\,\mathsf{Z}^{(2)}_{2k}\Big(\widehat{\mathsf{Z}}^{(1)}_{2k+1}-2\,\mathsf{Z}^{(1)}_{2k+1} \Big)-\frac{16\sqrt{\lambda}}{3\pi}\sum_{k=1}^\infty (2k)\,\mathsf{Z}^{(3)}_{2k+1}\Big(\widehat{\mathsf{Z}}^{(0)}_{2k}-4\,\mathsf{Z}^{(0)}_{2k} \Big)\notag \\
&\quad+\frac{32}{3}\sum_{k=1}^\infty (2k)^3\,\mathsf{Z}^{(2)}_{2k}\Big(\widehat{\mathsf{Z}}^{(0)}_{2k}-4\mathsf{Z}^{(0)}_{2k} \Big) 
+\frac{32}{3}\,\sum_{k=1}^\infty (2k)\,\mathsf{Z}_{2k}^{(2)}\,\Big(\widehat{\mathsf{Z}}^{(2)}_{2k}+2\,\mathsf{Z}^{(2)}_{2k}\Big)\label{Delta1}\\
&\quad-\frac{16 \lambda}{3\pi^2}\sum_{k=1}^\infty (2k)\,\mathsf{Z}_{2k}^{(2)}\Big(\widehat{\mathsf{Z}}^{(2)}_{2k}-\mathsf{Z}^{(2)}_{2k}\Big) -\frac{16}{3}\Big(1+\frac{\lambda}{4\pi^2}\Big)\sum_{k=1}^\infty (2k)\,\mathsf{Z}_{2k}^{(4)}\Big(\widehat{\mathsf{Z}}^{(0)}_{2k}-4\,\mathsf{Z}^{(0)}_{2k}\Big) \notag~.
    \end{align}
\end{subequations}
We emphasize that these results are exact in $\lambda$. The expressions (especially $\Delta_1^\D$) may look complicated, but we note that only $\mathsf{Z}_{n}^{(p)}$ and $\widehat{\mathsf{Z}}^{(p)}_{n}$  appear, but not $\mathsf{M}_{n,m}^{(p)}$. This has important consequences and will lead to a huge simplicity in the strong-coupling regime, as we will demonstrate in Section\,\ref{secn:strongcoupling}.

\subsubsection*{Topological recursion  approach}
We now perform an independent consistency check of the previous results using the topological recursion method. We first expand the functions $\log H_{\mathrm{v}}(x;b)$ and $ \log H_{\mathrm{h}}(x;b)$ around $b = 1$, finding
\begin{equation}
\begin{aligned}
     \partial_b^2 \log H_{\mathrm{v}}(x;b)\big|_{b=1} &= -4(1+\gamma) + 4\int_0^\infty \!\frac{d\omega}{\sinh^2\omega} \,\big[\,\omega +\coth\omega \,(\omega  \coth \omega -1)\big]\, \sin^2(\omega\,x )~,\notag\\
   \partial_b^2 \log H_{\mathrm{h}}(x;b,m)\big|_{b=1} &= -\int_0^\infty \!\frac{d\omega}{\sinh^4 \omega} \,\big[\sinh (2 \omega )-2 \omega \big]\, \sin ^2[\omega  (m+x)]~.
\end{aligned}
\end{equation}
We then use these results to rewrite $\mathcal{B}_2^\D$
and $\mathcal{C}_{\mathrm{F}}^\D$ as follows
\begin{align}
    \mathcal{B}_2^\D &= (N^2-1)(1+\gamma)+\int_0^\infty \frac{\big[\omega+\coth\omega \,(\omega \coth \omega-1)\big]\,d\omega}{2\sinh^2 \omega}\, \Big[f(2\ii\omega)f(-2\ii\omega) 
    -  N^2\Big]\notag\\
&+ \int_0^\infty \frac{\big[2 \omega-\sinh (2 \omega)\big]\,d\omega}{8\sinh^4 \omega} \, \Big[ f(4\ii \omega) -f(2\ii\omega)^2 -4f(2\ii\omega)+2N(N+1) + (\omega \rightarrow -\omega)\Big ]~,\\[2mm]
\mathcal{C}_{\mathrm{F}}^\D &= 4N\Big(\frac{1}{3}-\zeta(3)\Big) + \int_0^\infty \frac{\omega^2\big[2 \omega-\sinh (2 \omega)\big]\,d\omega}{\sinh^4 \omega}\, \Big[f(2\ii\omega) -N + (\omega \rightarrow -\omega)\Big]~.
\end{align}
These results provide all the necessary ingredients for the large-$N$ expansion of $\partial_\mu^2\partial_b^2\log \mathcal{Z}^\DS$, as written in \eqref{intcorr2}, within the framework of topological recursion. We find that this approach reproduces the results obtained through the full Lie-algebraic computation with exact agreement.
As an illustrative example, we focus on the leading large-$N$ contribution which is $O(N)$. In this case, by using \eqref{eq:MBD} we find
\begin{align}  \label{eq:M2F} 4\Big[\big\langle\mathcal{M}_{2,\mathrm{F}}^\D\,\,  \mathcal{B}_2^\D\big\rangle  & -\big\langle\mathcal{M}_{2,\mathrm{F}}^\D\big\rangle\,\big\langle\mathcal{B}_2^\D\big\rangle\Big]\bigg|_{O(N)} =  \int \frac{4\nu\big[\omega+\coth \omega\,(\omega \coth \omega-1)\big]\,d\omega\, d\nu }{\sinh^2 \omega \,\sinh^2 \nu} 
\\  
& \qquad\qquad\qquad\times \bigg\{ \Big[ W_0^2(2\ii\nu,-2\ii\omega) \,W_0^1(2\ii\omega) + (\omega \rightarrow - \omega) \Big] +(\nu \rightarrow - \nu)\bigg\} \notag\\
        -\int& \frac{2\nu\,\big[2\omega- \sinh^2 \omega\big]\,d\omega\, d\nu }{\sinh^4 \omega \,\sinh^2 \nu} \bigg\{\Big[ W_0^2(2\ii\nu,-2\ii\omega)\,W_0^1(-2\ii\omega) + (\omega \rightarrow -\omega)
        \Big] +(\nu \rightarrow - \nu)\bigg\}~, \nonumber
\end{align}
which reproduces $\mathcal{F}_0^\D$ in \eqref{eq:F1intnew} upon substituting the explicit expressions for $W_0^n$'s given in \eqref{eq:W-examples}.
For the leading $O(N)$ contribution to $\big\langle\mathcal{C}_{\mathrm{F}}^\D\big\rangle$, from \eqref{eq:CFD} we find
\begin{align} 
    -4 \big\langle\mathcal{C}_{\mathrm{F}}^\D\big\rangle\Big|_{O(N)} &=   16\zeta(3)-\frac{16}{3}+\int_0^\infty \frac{4\omega^2\,d\omega}{\sinh^4 \omega}\,\big[\sinh (2 \omega) - 2 \omega\big] \, \Big(W_0^1(2\ii\omega)-1+  (\omega \rightarrow -\omega)\Big)\notag\\
   &= \frac {16 \pi}{\sqrt{\lambda }} \int_0^\infty \frac{\omega\, d\omega}{\sinh^4 \omega}\, \big[\sinh (2 \omega)-2 \omega\big] \,  J_1\Big(\frac{\omega \sqrt{\lambda }}{\pi}\Big)~, \label{eq:CDF}
\end{align}
which, although written in a different form, can be verified to coincide with $\Delta_0^\D$ in \eqref{Delta0}. 

Once again, we see that the topological recursion and the full Lie-algebra approach lead to same final results.

\section{Strong-coupling universality}
\label{secn:strongcoupling}

In the previous section we derived the large-$N$ expansions
\begin{subequations}
    \begin{align}
    \partial_\mu^2\,\partial_m^2\log\mathcal{Z}^\DS\big|_{\D}\,&=N\,\mathcal{F}_0^\D+\mathcal{F}_1^\D+\ldots~,\\
    \partial_\mu^2\,\partial_b^2\log\mathcal{Z}^\DS\big|_{\D}\,&=N\,\widetilde{\mathcal{F}}_0^\D+\widetilde{\mathcal{F}}_1^\D+\ldots~,
\end{align}
\label{largeNexpansions}%
\end{subequations}
and obtained integral representations of $\mathcal{F}_{0,1}^\D$ and $\widetilde{\mathcal{F}}_{0,1}^\D$ in terms of Bessel functions valid for all values of $\lambda$. We now turn to the strong-coupling regime $\lambda\to\infty$.  

Before considering explicit cases, we will comment on the general structure of the strong-coupling expansion. To obtain it we will utilize the Mellin–Barnes representations of (products of) Bessel functions, which appear in the building blocks of the integrated correlators, namely $\widehat{\mathsf{Z}}^{(p)}_{n}$, $\mathsf{Z}_{n}^{(p)}$, and $\mathsf{M}_{n,m}^{(p)}$ (defined in \eqref{Zhatkp}, \eqref{Zkp}, and \eqref{Mkp}, respectively). We note that both $\widehat{\mathsf{Z}}^{(p)}_{n}$, $\mathsf{Z}_{n}^{(p)}$ contain a single Bessel function, whereas $\mathsf{M}_{n,m}^{(p)}$ is given by a product of two Bessel functions, and that their Mellin–Barnes representations take the following form,  
\begin{align}
\begin{split}
    J_{\nu}(x) &= \frac{1}{2\pi \ii} \int_{-\ii \, \infty}^{+ \ii\, \infty} \!dt~ \frac{\Gamma(-t)\, x^{\nu+2t}}{2^{\nu+2t}\, \Gamma(\nu+t+1)}~, \\
J_{\mu}(x) \,J_{\nu}(x) &= \frac{1}{2\pi \ii} \int_{-\ii \, \infty}^{+ \ii\, \infty} \!dt ~
\frac{\Gamma(-t)\,\Gamma(2t+\mu+\nu+1)\, x^{\mu+\nu+2t}}{2^{\mu+\nu+2t}\, \Gamma(\mu+t+1)\,\Gamma(\nu+t+1)\,\Gamma(\mu+\nu+t+1)}~.
\end{split}
\label{MellinJJ}
\end{align}
We will see that terms involving a single Bessel function contribute only a finite number of terms to the strong-coupling expansion (up to exponentially suppressed corrections). Consequently, the full asymptotic series of the integrated correlators at strong coupling is governed entirely by $\mathsf{M}_{n,m}^{(p)}$. This leads to a striking simplification in the strong-coupling regime, in sharp contrast to the intricate structures that appear at weak coupling, reported in Appendix \ref{App:WeakCoupling}.

\subsection{\texorpdfstring{Leading term}{}}
The leading term $\mathcal{F}_0^\D$, given in (\ref{F0}), coincides with $\frac{1}{2}F_1$, where $F_1$ is defined in Eq. (B.11) of \cite{Chester:2025ssu}. Although the strong-coupling expansion of $F_1$ was already established in that work, here we provide an independent derivation based on our method. This approach will prove instrumental for analyzing other quantities.

Employing the Mellin–Barnes representations of Bessel functions given in (\ref{MellinJJ}), together with the integral representation of the Riemann $\zeta$-function, we can rewrite (\ref{F0}), or equivalently (\ref{eq:F1intnew}), as follows
\begin{align}
    \mathcal{F}_0^\D&=32\iint_{-\ii\infty}^{+\ii\infty}\!\frac{ds\,ds'}{(2\pi\ii)^2}\,\sum_{k=1}^\infty\,(2k)
    \,\frac{\Gamma(-s)\,\Gamma(2s+2k+2)\,\zeta(2s+2k+1)}{\Gamma(s+2k+1)}\,\times\notag\\[2mm]
    &\qquad\qquad\times\frac{\Gamma(-s')\,\Gamma(2s'+2k+2)^2\,\zeta(2s'+2k+1)}{\Gamma(s'+2)\,\Gamma(s'+2k+1)\,\Gamma(s'+2k+2)}\,\bigg(\frac{\sqrt{\lambda}}{4\pi}\bigg)^{2s+2s'+4k}~.
    \label{F0D0}
\end{align}
It is worth noting that, for the two terms in (\ref{F0}), the $s'$-integral is associated to the Mellin–Barnes representation of $\mathsf{M}_{1,2k}^{(1)}$ and the $s$-integral to that of $\mathsf{Z}_{2k}^{(2)}$. This fact will become important in the strong-coupling expansion. 
After the shifts $s\to s-k+1$ and $s'\to s'-k+1$, the sum over $k$ in (\ref{F0D0}) can be evaluated as
\begin{align}
   & \sum_{k=1}^\infty\,(2k)\frac{\Gamma(-s+k-1)\,\Gamma(-s'+k-1)}{\Gamma(s+k+2)\,\Gamma(s'-k+3)\,\Gamma(s'+k+2)\,\Gamma(s'+k+3)}
   \notag\\[1mm]
   &\qquad=\frac{2\,\Gamma(-s)\,\Gamma(-s')\,\,{}_4F_3(2,-s,-s',-s'-1;s+3,s'+3,s'+4;-1)}{\Gamma(s+3)\,\Gamma(s'+2)\,\Gamma(s'+3)\,\Gamma(s'+4)}\notag\\[1mm]
   &\qquad=\frac{\Gamma(-s)\,\Gamma(-s')\,\,{}_3F_2(-s,-s',s'+3;2,s'+4;1)}{\Gamma(s+s'+3)\,\Gamma(s'+2)\,\Gamma(s'+4)}\, ,
\end{align}
where the last step follows from the following identity
\begin{equation}
    \begin{aligned}
    \, _4F_3(a,b,c,d;&a-b+1,a-c+1,a-d+1;-1)\label{identity4F3}\\[1mm]
    &=\frac{\Gamma (a-b+1) \Gamma (a-c+1) \,
   _3F_2\left(b,c,\frac{a}{2}-d+1;\frac{a}{2}+1,a-d+1;1\right)}{\Gamma (a+1) \Gamma (a-b-c+1)}~.
\end{aligned}
\end{equation}
This yields the compact representation
\begin{align}
    \mathcal{F}_0^\D=\iint_{-\ii\infty}^{+\ii\infty}\!\frac{ds\,ds'}{(2\pi\ii)^2}
\,\,\mathcal{G}_0(s,s') \, , 
\end{align}
with
\begin{align}
    \mathcal{G}_0(s,s')&=32\,
    \Gamma(2s+4)\,\zeta(2s+3)\,\Gamma(2s'+4)^2\,\zeta(2s'+3)~\times\notag\\[2mm]
    &\quad\times~\frac{\Gamma(-s)\,\Gamma(-s')\,\,{}_3F_2(-s,-s',s'+3;2,s'+4;1)~}{\Gamma(s+s'+3)\,\Gamma(s'+2)\,\Gamma(s'+4)}\bigg(\frac{\sqrt{\lambda}}{4\pi}\bigg)^{2s+2s'+4}~.
\end{align}
The strong-coupling expansion is obtained by closing the integration contours counter-clockwise in the $\mathfrak{Re}\, s<0$ and $\mathfrak{Re}\,s'<0$ planes, and summing the residues of all poles located at $s=-n$ with $n\in \mathbb{Z}_{>0}$.
Computing the residues yields
\begin{align}
\mathcal{F}_0^\D\,\underset{\lambda \rightarrow \infty}{\sim}\,\,
\sum_{n=1}^\infty  f_0^{(n)}
~,
\end{align}
where
\begin{align}
    f_0^{(n)} &=\int_{-\ii\infty}^{+\ii\infty}\!\frac{ds'}{2\pi\ii}\,\,\mathrm{Res}\Big[\,\mathcal{G}_0(s,s')\Big]_{s=-n}~ \notag\\[2mm]
     &=-16
    \int_{-\ii\infty}^{+\ii\infty}\!\frac{ds'}{2\pi\ii}\,
    \frac{\Gamma(n)\,\Gamma(2s'+4)^2\,\Gamma(-s')\,\zeta(2s'+3)}{\Gamma(s'+3-n)\,\Gamma(s'+2)\,\Gamma(s'+4)}\,\times\notag\\
    &\qquad\qquad\qquad\times \frac{B_{2n-2}\,(2n-3)\,\,{}_3F_2(n,-s',s'+3;2,s'+4;1)}{(2n-2)!}\,\bigg(\frac{\sqrt{\lambda}}{4\pi}\bigg)^{2s'+4-2n}\, ,
    \label{F1n}
\end{align}
with $B_{m}$ denoting Bernoulli numbers.

For $n=1$, exploiting the identity 
\begin{align}
   {}_3F_2(1,a,b;2,d;1) =\frac{(d-1) }{(a-1) (b-1)}\left(\frac{\Gamma (d-1) \Gamma (-a-b+d+1)}{\Gamma (d-a) \Gamma (d-b)}-1\right)~,
\label{identity3F2}
\end{align}
setting $s'=t-1$ and using the duplication formula of the $\Gamma$-functions, one finds
\begin{align}
    f_0^{(1)}
    =-16 \int_{-\ii\infty}^{+\ii\infty}\!\frac{dt}{2\pi\ii}\,
    \frac{\Gamma(-t)\,\Gamma(2t+2)\,\zeta(2t+1)}{\Gamma(t+2)}\,
    \bigg(\frac{2^{2t+1}\,\Gamma(t+\frac{3}{2})}{\sqrt{\pi}\,\Gamma(t+2)}-1\bigg)\bigg(\frac{\sqrt{\lambda}}{4\pi}\bigg)^t ~. 
    \label{F11a}
\end{align}
A comparison with the planar term $\mathcal{C}^{(0)}_{\mathcal{D}}$ of the integrated giant-graviton correlator in $\mathcal{N}=4$ SYM \cite{Brown:2024tru} (see Eq.\,(5.27) of that reference and appendix \ref{app:GiantG}) reveals the remarkably simple relation
\begin{align}
    f_0^{(1)}= 4\, \mathcal{C}^{(0)}_{\mathcal{D}}~.
\end{align}
Expanding (\ref{F11a}) at strong-coupling as discussed in \cite{Brown:2024tru} gives
\begin{align}
    f_0^{(1)}\underset{\lambda \rightarrow \infty}{\sim} \,&\,8-\frac{16\pi^2}{3\,\lambda}-\sum_{n=1}^\infty\frac{64\,n\,\Gamma(n-\frac{1}{2})^2\,\Gamma(n+\frac{1}{2})\,\zeta(2n+1)}{\pi^{3/2}\,\Gamma(n)\,\lambda^{n+1/2}}~.
    \label{F11strong}
\end{align}

For $n=2$, we obtain
\begin{align}
    f_0^{(2)}&=-\frac{4}{3}
    \int_{-\ii\infty}^{+\ii\infty}\!\frac{ds'}{2\pi\ii}\,
    \frac{\Gamma(2s'+4)^2\,\Gamma(-s')\,\zeta(2s'+3)\,\,{}_3F_2(2,-s',s'+3;2,s'+4;1)}{\Gamma(s'+1)\,\Gamma(s'+2)\,\Gamma(s'+4)}\,
    \bigg(\frac{\sqrt{\lambda}}{4\pi}\bigg)^{2s'}\notag\\
    &=-\frac{4}{3}
    \int_{-\ii\infty}^{+\ii\infty}\!\frac{ds'}{2\pi\ii}\,
    \frac{\Gamma(2s'+4)\,\Gamma(-s')\,\zeta(2s'+3)}{\Gamma(s'+2)}\,
    \bigg(\frac{\sqrt{\lambda}}{4\pi}\bigg)^{2s'}\notag\\
    &=-\frac{16\pi}{3\sqrt{\lambda}}\,Z_1^{(3)} \, , 
\end{align}
where in the second step we exploited the properties of the hypergeometric functions while in the last step we introduced the quantity $Z_1^{(3)}$ defined in (\ref{Zkp}). Using the asymptotic behavior $Z_1^{(3)}\underset{\lambda \rightarrow \infty}{\sim} \,\frac{2\pi}{\sqrt{\lambda}}$ established in \cite{Billo:2024ftq}, it follows that
\begin{align}
   f_0^{(2)}\underset{\lambda \rightarrow \infty}{\sim} -\frac{32\pi^2}{3\lambda}~.
   \label{F12strong}
\end{align}

For $n\geq3$, one finds
\begin{align}
f_0^{(n)}&=\frac{16(2n-3)\,B_{2n-2}}{(2n-2)!}\!
    \int_{-\ii\infty}^{+\ii\infty}\!\frac{ds'}{2\pi\ii}\,
    \frac{\Gamma(-s'-3+n)\,\Gamma(2s'+4)\,\zeta(2s'+3)}{\Gamma(s'+3-n)}
    \bigg(\frac{\sqrt{\lambda}}{4\pi}\bigg)^{2s'+4-2n} ~.
\end{align}
Since the integrand has no poles for $s'<0$, all these terms vanish
in the strong-coupling limit:
\begin{align}
   f_0^{(n)}\underset{\lambda \rightarrow \infty}{\sim} 0 \qquad
   \text{for all $n\geq 3$}~.
   \label{F1nstrong}
\end{align}

Combining all contributions, we arrive at
\begin{align}
    \mathcal{F}_0^\D\underset{\lambda \rightarrow \infty}{\sim} \,&\,8-\frac{16\pi^2}{\lambda}-\sum_{n=1}^\infty\frac{64\,n\,\Gamma(n-\frac{1}{2})^2\,\Gamma(n+\frac{1}{2})\,\zeta(2n+1)}{\pi^{3/2}\,\Gamma(n)\,\lambda^{n+1/2}} \, ,
    \label{SCF0D}
\end{align}
in complete agreement, up to an overall factor of $1/2$, with the independent derivation of \cite{Chester:2025ssu}.

Comparing with the strong-coupling expansion of the planar part of the integrated giant-graviton correlator of $\mathcal{N}=4$ SYM computed in \cite{Brown:2024tru} and displayed in \eqref{eq:GG_strong}, we may write
\begin{align}
\mathcal{F}_0^\D-4\,\mathcal{C}_{\mathcal{D}}^{(0)}\underset{\lambda \rightarrow \infty}{\sim} \,-\frac{32\pi^2}{3\lambda}~,
    \label{univ1}
\end{align}
demonstrating that the difference between the two observables reduces to a single $1/\lambda$ correction at strong coupling. Instead, $\mathcal{F}_0^\D$ and $\mathcal{C}_{\mathcal{D}}^{(0)}$ have a completely different behavior at weak coupling, as one can easily see considering \eqref{C0Weak} and \eqref{WeakF0}. 

The above analysis shows that the $s$-integral (associated with $\mathsf{Z}_{n}^{(p)}$) only contributes a finite number of terms in the strong coupling expansion, as we stated earlier; on the contrary, the $s'$-integral (associated with $\mathsf{M}_{n,m}^{(p)}$) is responsible for the infinite number terms of the strong-coupling asymptotic series. 

\subsection{\texorpdfstring{Sub-leading term}{}}
The sub-leading coefficient $\mathcal{F}_1^\D$, given in (\ref{F1}) decomposes as
\begin{align}
    \mathcal{F}_1^\D = A_1+A_2+A_3 \, ,
    \label{F1Dcomponents}
\end{align}
with 
\begin{subequations}
\begin{align}
    A_1&=\frac{128\,\pi}{\sqrt{\lambda}}\,\textsf{Y}\,\sum_{k,\ell=1}^\infty(-1)^{k+\ell}\,(2k)(2\ell)\,\mathsf{Z}_{2k}^{(2)}\,\mathsf{M}_{1,2\ell}^{(1)}~,\label{A1}\\
    A_2&=-16\,\sum_{k=1}^\infty(2k)\,\mathsf{Z}_{2k}^{(2)}\,\widehat{\mathsf{Z}}_{2k}^{(2)}~,\label{A2}\\
    A_3&=32\sum_{k,\ell=1}^\infty(-1)^\ell(2k)\,\sqrt{2\ell}\,\,\mathsf{Z}_{2k}^{(2)}\,\mathsf{M}_{2k,2\ell}^{(2)}\,\textsf{Y}_{2\ell}~.\label{A3}
\end{align}
\label{A123}%
\end{subequations}
The techniques used for $\mathcal{F}_0^\D$ apply to $A_{1,2,3}$ as well. Details are given in Appendix \ref{app:strong}, and here we summarize the results.

For $A_1$, which factorizes completely, we find
\begin{align}
 A_1   \underset{\lambda \rightarrow \infty}{\sim}\, &~\frac{64\pi}{\sqrt{\lambda}}\bigg(\frac{1}{4}-\frac{\log (2)}{2\pi^2}\,\lambda\bigg)
 \bigg(\frac{\sqrt{\lambda}}{8\pi}-\frac{1}{4\pi}+\sum_{n=1}^\infty\frac{(n-\frac{1}{2})\,\Gamma(n-\frac{1}{2})^2\,\Gamma(n+\frac{3}{2})\,\zeta(2n+1)}{2\pi^{5/2}\,\Gamma(n)\,\lambda^n}\bigg) ~.
 \label{A1strong}
\end{align}
For $A_2$, the large-$\lambda$ behavior is logarithmic:
\begin{align}
A_2 \underset{\lambda \rightarrow \infty}{\sim} ~-4\log\Big(\frac{\lambda}{4\pi^2}\Big)-8\gamma+11\,\zeta(3)-\frac{44}{3}~.
\label{A2strong}
\end{align}
The structure of $A_3$ is more involved and relates, up to a factor
$-1/2$, to the quantity $F_2^{\mathrm{hard}}$ introduced in \cite{Chester:2025ssu} where semi-numerical methods were used to find the first terms of its strong-coupling expansion. Our approach, instead, enables us to derive this expansion in a fully analytic form\,\footnote{Some details of the derivation are given in Appendix \ref{app:strong1}.}:
\begin{align}
    A_3\,&\underset{\lambda \rightarrow \infty}{\sim}\,\frac{4\log(2)}{\pi^2}\,\lambda-\frac{8\log(2)}{\pi^2}\,\sqrt{\lambda}-2 \log\Big(\frac{\lambda}{\pi^2}\Big)+k+\frac{4}{\sqrt{\lambda}}
    \label{A3strong}\\
    &\qquad+\sum_{n=1}^\infty\frac{8\,\Gamma(n+\frac{1}{2})^3\,\zeta(2n+1)}{\pi^{3/2}\,\Gamma(n)\,\lambda^{n+1/2}}-\log(2)
    \sum_{n=1}^\infty\frac{16 (n+\frac{1}{2})\,\Gamma(n-\frac{1}{2})^2\,\Gamma(n+\frac{3}{2})\,\zeta(2n+1)}{\pi^{7/2}\,\Gamma(n)\,\lambda^{n-1/2}} \, , \notag
\end{align}
where
\begin{align}
    k=-\frac{53}{15}-4(1+\gamma)+8\log(2)+\frac{\zeta(3)}{10}~.
    \label{constantK}
\end{align}
The first few terms of this expansion explicitly read
\begin{align}
    A_3&~\underset{\lambda \rightarrow \infty}{\sim}
    \frac{4\log(2)}{\pi^2}\,\lambda-\frac{8\log(2)}{\pi^2}\,\sqrt{\lambda}
-2\log\Big(\frac{\lambda}{\pi^2}\Big)+k+\bigg(4-\frac{18\,\zeta(3)\log(2)}{\pi^2}\bigg)\frac{1}{\lambda^{1/2}} \label{A3strongexpl}\\
&\qquad+\bigg(2\,\zeta(3)-\frac{75\,\zeta(5)\log(2)}{2\pi^2}\bigg)\frac{1}{\lambda^{3/2}}+\bigg(\frac{27\,\zeta(5)}{4}-\frac{6615\,\zeta(7)\log(2)}{32\pi^2}\bigg)\frac{1}{\lambda^{5/2}}+O\big(\lambda^{-7/2}\big)~.
\notag
\end{align}
The first line above agrees, up to an overall factor $-1/2$, with Eq.\,(B.19) of \cite{Chester:2025ssu}. 
This comparison also allows us to fix the constant $c_{F_2}^{\mathrm{hard}}$ defined in \cite{Chester:2025ssu}, yielding
\begin{align}
    c_{F_2}^{\mathrm{hard}}=-2k=\frac{106}{15}+8(1+\gamma)-16\log(2)-\frac{\zeta(3)}{5}~\simeq ~ 8.35363~.
\end{align}
The numerical estimate $c_{F_2}^{\mathrm{hard}}=8.3\pm0.1$ of \cite{Chester:2025ssu} is in good agreement with the exact value.

Summing all contributions, the asymptotic form of $\mathcal{F}_1^\D$ is
\begin{align}
    \mathcal{F}_1^\D~\underset{\lambda \rightarrow \infty}{\sim}&
    -6\log\Big(\frac{\lambda}{\pi^2}\Big) +k_{\mathcal{F}_1^\D}-\log(2)\sum_{n=1}^\infty \frac{32\,n\,\Gamma(n-\frac{1}{2})^2\,\Gamma(n+\frac{3}{2})\,\zeta(2n+1)}{\pi^{7/2}\,\Gamma(n)\,\lambda^{n-1/2}}\notag\\
    &\quad+\sum_{n=1}^\infty \frac{16\,n\,\Gamma(n-\frac{1}{2})\,\Gamma(n+\frac{1}{2})^2\,\zeta(2n+1)}{\pi^{3/2}\,\Gamma(n)\,\lambda^{n+1/2}}\, ,
    \label{F1strongfinal}
\end{align}
where the constant term is
\begin{align}
    k_{\mathcal{F}_1^\D}=\frac{111 \,\zeta (3)}{10}-\frac{101}{5}-12 \gamma +16 \log (2)~.
    \label{kF1D}
\end{align}
Once again, we see that $A_1$ and $A_3$, which contain $\mathsf{M}_{n,m}^{(p)}$, produce infinite series in $1/\lambda$, whereas $A_2$, which is given in terms of $\mathsf{Z}_{m}^{(p)}$ and $\widehat{\mathsf{Z}}_{m}^{(p)}$, truncates in the large-$\lambda$ expansion. 

Strikingly, the last line of (\ref{F1strongfinal}) matches twice the asymptotic series appearing in the sub-leading contribution $\mathcal{C}_\mathcal{D}^{(1)}$ to the giant-graviton correlator at strong coupling computed in \cite{Brown:2024tru} and given in \eqref{eq:GG_1_strong}. Therefore, taking into account also the $\log(\lambda)$-terms of $\mathcal{C}_\mathcal{D}^{(1)}$, we can write
\begin{align}
    \mathcal{F}_1^\D-2\,\mathcal{C}_\mathcal{D}^{(1)}~\underset{\lambda \rightarrow \infty}{\sim} -4\log\Big(\frac{\lambda}{\pi^2}\Big)+\text{constant terms}+\cdots~,
    \label{F1C1}
\end{align}
where the ellipsis denotes a tail of $\log(2)$-dependent contributions. As shown in the next section, these can be absorbed into a redefinition of the coupling, leaving a difference that contains only a $\log(\lambda)$ term and $\lambda$-independent constants. This parallels the behavior observed at the planar level in (\ref{univ1}), reinforcing the picture of a remarkable underlying universality between very different scattering processes. 

\subsection{\texorpdfstring{Strong-coupling expansion for $\partial_\mu^2\,\partial_b^2\log\mathcal{Z}^\DS$ and summary}{}}
The strong-coupling expansions of $\widetilde{\mathcal{F}}_0^\D$ and $\widetilde{\mathcal{F}}_1^\D$ appearing in $\partial_\mu^2\,\partial_b^2\log\mathcal{Z}^\DS\big|_{\D}$ can be obtained from the relation (\ref{FFtilde}), namely
\begin{align} \label{eq:Ftitle}
\widetilde{\mathcal{F}}_g^\D=\mathcal{F}_g^\D+\Delta_g^\D~.
\end{align}
Since we have computed the strong coupling expansions for $\mathcal{F}_g^\D$ for $g=0, 1$ in the previous subsection, we now just need to analyze the behavior of the differences $\Delta_0^\D$ and $\Delta_1^\D$ as $\lambda\to\infty$. 

For the genus-zero term, $\Delta_0^\D$, as given in (\ref{Delta0}), one may exploit results from \cite{Billo:2024ftq} to find
\begin{align}
    \Delta_0^\D~\underset{\lambda \rightarrow \infty}{\sim} \frac{64\pi^2}{3\lambda}~.\label{Delta0strong}
\end{align}
The derivation of the expansion for $\Delta_1^\D$ for $\lambda\to\infty$ is much more involved (details are provided in Appendix \ref{app:strong2}) but the final expression is very simple:
\begin{align}
    \Delta_1^\D~\underset{\lambda \rightarrow \infty}{\sim} 8\log\Big(\frac{\lambda}{\pi^2}\Big)-\frac{76\,\zeta(3)}{5}+\frac{152}{5}+16\gamma-\frac{80\log(2)}{3}~.\label{Delta1strong}
\end{align}
We note that both $\Delta_0^\D$ and $ \Delta_1^\D$
truncate in the large-$\lambda$ expansion
and that the ratio between the coefficients of the $1/\lambda$ and $\log(\lambda)$ terms remains fixed at $8\pi^2/3$, as for 
$\mathcal{F}_0^\D$ and $\mathcal{F}_1^\D$. This fact will be important for the $\mathrm{SL}(2, \mathbb{Z})$ completion of these integrated correlators.

In summary, we find that, despite being distinct functions, the mixed derivatives $\partial_\mu^2\,\partial_m^2\log\mathcal{Z}^\DS$ and
$\partial_\mu^2\,\partial_b^2\log\mathcal{Z}^\DS$ exhibit striking similarities in the large--$N$ limit, both among themselves and with the giant-graviton correlators of $\mathcal{N}=4$ SYM. At strong coupling, the differences between these observables simplify drastically, and all these integrated correlators are captured by universal expressions like \eqref{eq:Ft0} for the leading large-$N$ order and \eqref{eq:Ft1} for the sub-leading order. 
As emphasized earlier, this remarkable behavior can be understood from the fact that the differences among these observables contain only the building blocks $\mathsf{Z}_{m}^{(p)}$ and $\widehat{\mathsf{Z}}_{m}^{(p)}$. These quantities truncate in the large-$\lambda$ expansion, even though they give rise to infinite series in the small-$\lambda$ regime. Furthermore, the remaining structure $\mathsf{M}_{n,m}^{(p)}$, which does produce an infinite expansion in powers of $1/\lambda$, appears only linearly in all the quantities we have studied. This leads to the fact  that the coefficient of each order in the $1/\lambda$ expansion contains only one Riemann $\zeta$-value, in contrast to the products of many 
Riemann $\zeta$-values that appear in the weak-coupling expansion, as shown in Appendix \ref{App:WeakCoupling}. This remarkably simple structure of the strong-coupling regime enables us to propose an SL(2, $\mathbb{Z}$)-invariant completion
of our results, as discussed in the next section.

\section{The ``very strong-coupling'' limit and modularity}
\label{secn:verystrong}
We now turn to a regime distinct from that analyzed in the previous sections, namely the limit of large $N$ at fixed YM coupling $\gym$, rather than fixed ’t Hooft coupling $\lambda$. In this regime, often referred to as the ``very strong-coupling limit'', instanton contributions become significant. 
When these effects are taken into account, the resulting expressions change qualitatively in structure, and typically acquire nontrivial transformation properties under the modular group $\mathrm{SL}(2,\mathbb{Z})$.

In this context, however, we expect the observables discussed in Section\,\ref{secn:strongcoupling} to exhibit modular invariance. Indeed, the mixed derivative $\partial_\mu^2\,\partial_m^2 \log \mathcal{Z}^\DS\big|_{\D}$ encodes the terms in the effective action quadratic in the masses of both the fundamental and anti-symmetric hypermultiplets.
In the string theory realization, the latter correspond to open strings that begin and end on D3-branes, crossing the orientifold plane. Since these states are associated with D3-branes, we naturally expect their mass-dependent contributions to be modular invariant. By contrast, the four fundamental hypermultiplets arise from open strings ending on D7-branes
and their masses transform nontrivially under $\mathrm{SL}(2,\mathbb{Z})$ \cite{Seiberg:1994aj} (see also \cite{Billo:2010mg}). 
Nevertheless, the quadratic combination of such masses is invariant under modular transformations, implying that its coefficient must also be modular invariant. 
Similar considerations apply to the other mixed derivative $\partial_\mu^2\,\partial_b^2 \log \mathcal{Z}^\DS\big|_{\D}$. Another way to see this comes from the holographic scattering amplitude perspective, where the SL$(2, \mathbb{Z})$ transformations permute the color factors of gluons. Since the scattering amplitudes corresponding to $\partial_\mu^2\,\partial_m^2 \log \mathcal{Z}^\DS\big|_{\D}$ and $\partial_\mu^2\,\partial_b^2 \log \mathcal{Z}^\DS\big|_{\D}$ involve two gluons and two gravitons, their overall color factor is trivial, and therefore they remain invariant under SL$(2, \mathbb{Z})$ \footnote{Conversely, observables involving four $\mu$-derivatives, corresponding to four-gluon amplitudes, in general transform non-trivially under SL$(2,\mathbb{Z})$. In the \textbf{D}-theory, these four-gluon amplitudes carry U(4) color factors. Since U(4) is embedded into SO(8), the associated color factors are permuted by SL$(2,\mathbb{Z})$ transformations, in direct analogy with the $\mathcal{N}=2$ Sp($N$) theory studied in \cite{Behan:2023fqq}, where the gluons transform under SO(8). Consequently, the modular properties of the integrated four-gluon amplitudes in the \textbf{D}-theory can be obtained following the same reasoning.}.
This pattern closely parallels the one found in the $\mathcal{N}=2$ Sp($N$) theory studied in \cite{Chester:2025ssu}. We can therefore follow the same approach and promote the strong-coupling results obtained in the zero-instanton sector to an $\mathrm{SL}(2,\mathbb{Z})$-invariant completion. As we will see that the results can be expressed in terms of non-holomorphic Eisenstein series as in \cite{Chester:2019jas, Dorigoni:2021guq, Paul:2022piq, Brown:2024tru,Chester:2025ssu}. 

To illustrate this proposal in some detail, we consider the large-$N$ strong-coupling expansion of the mixed derivative
$\partial_\mu^2\,\partial_m^2\log\mathcal{Z}^\DS$, 
which for convenience we rewrite explicitly:
\begin{align}
    \partial_\mu^2\,\partial_m^2\log\mathcal{Z}^\DS  \big|_{\D}~& \simeq~
    N\bigg[8-\frac{16\pi^2}{\lambda}-\sum_{n=1}^\infty\frac{64\,n\,\Gamma(n-\frac{1}{2})^2\,\Gamma(n+\frac{1}{2})\,\zeta(2n+1)}{\pi^{3/2}\,\Gamma(n)\,\lambda^{n+1/2}}\bigg]\notag\\
    &\, +\bigg[-6\log\Big(\frac{\lambda}{\pi^2}\Big) +k_{\mathcal{F}_1^\D}+\sum_{n=1}^\infty \frac{16\,n\,\Gamma(n-\frac{1}{2})\,\Gamma(n+\frac{1}{2})^2\,\zeta(2n+1)}{\pi^{3/2}\,\Gamma(n)\,\lambda^{n+1/2}}\notag\\
    &\, -\log(2)\sum_{n=1}^\infty \frac{32\,n\,\Gamma(n-\frac{1}{2})^2\,\Gamma(n+\frac{3}{2})\,\zeta(2n+1)}{\pi^{7/2}\,\Gamma(n)\,\lambda^{n-1/2}} \bigg]+O(N^{-1}) \, , 
    \label{d2md2mufinal}
\end{align}
with the constant $k_{\mathcal{F}_1^\D}$ given in (\ref{kF1D}).

A useful observation is that the $\log(2)$-dependent terms in \eqref{d2md2mufinal} can be absorbed by redefining the coupling as \cite{Beccaria:2022kxy, Billo:2024ftq}
\begin{align}
    \frac{1}{\lambda}\to\frac{1}{\lambda^\prime}=\frac{1}{\lambda}+\frac{\log(2)}{2\pi^2N}\qquad\text{with}\qquad \lambda^\prime=\frac{8\pi N}{\tau_2}~.
    \label{lambdaprime}
\end{align}
Expressing the result in terms of $\tau_2\sim 1/g_{_{\rm YM}}^2$, we have
\begin{align}
 \partial_\mu^2\,\partial_m^2\log\mathcal{Z}^\DS\big|_{\D}~&=\,\text{constant terms}-6\biggl[\frac{\pi\tau_2}{3}+2\gamma-\log(4\pi\tau_2)+O(\tau_2^{-1})\biggr]\notag\\
     &\quad-\biggl[\frac{\sqrt{2}\,\zeta(3)\,\tau_2^{3/2}}{\pi^{3/2}}+O(\tau_2^{-1/2})\biggr]\frac{1}{N^{1/2}}
     \notag\\&\quad
     -\biggl[\frac{3\,\zeta(5)\,\tau_2^{5/2}}{16\sqrt{2}\,\pi^{5/2}}-\frac{\zeta(3)\,\tau_2^{3/2}}{4\sqrt{2}\,\pi^{3/2}}+O(\tau_2^{1/2})\biggr]\frac{1}{N^{3/2}}
     \label{d2md2muexpanded}\\
     &\quad-\biggl[\frac{405\,\zeta(7)\,\tau_2^{7/2}}{4096\sqrt{2}\,\pi^{7/2}}-\frac{9\,\zeta(5)\,\tau_2^{5/2}}{128\sqrt{2}\,\pi^{5/2}}+O(\tau_2^{3/2})\biggr]\frac{1}{N^{5/2}}
      \notag\\
     &\quad-\biggl[\frac{7875\,\zeta(9)\,\tau_2^{9/2}}{65536\sqrt{2}\,\pi^{9/2}}-\frac{2025\,\zeta(7)\,\tau_2^{7/2}}{32768\sqrt{2}\,\pi^{7/2}}+O(\tau_2^{5/2})\biggr]\frac{1}{N^{7/2}}
     +O(N^{-9/2})~,\notag
\end{align}
where the ``constant terms" are coupling-independent contributions. 
In each square bracket, the terms denoted as $O(\tau_2^{\#})$ represent further contributions involving either integer powers of $\log(2)/\tau_2$ or half-integer powers of $\tau_2$ with Riemann $\zeta$-valued coefficients. Upon including the higher-order corrections $\mathcal{F}_{2}^\D, \mathcal{F}_{3}^\D, \ldots$ in the $1/N$ expansion of \eqref{d2md2mufinal}, we expect all $\log(2)$-term to cancel and the $\zeta$-valued contributions to be fixed\,\footnote{Such expectations were explicitly verified in the Sp($N$) theory studied in \cite{Chester:2025ssu} up to order $1/N$.}.

Since the correlators under consideration are modular invariant, their perturbative expansion should be completed by modular invariant functions.  We will \emph{assume} that the large-$N$ (finite-$\tau_2$) expansion, as given in \eqref{d2md2muexpanded}, is completed by the non-holomorphic Eisenstein series, precisely as in \cite{Chester:2019jas, Brown:2024tru,Chester:2025ssu}. The non-holomorphic Eisenstein series of index $s$ is a modular invariant function defined by the lattice sum
\begin{align}
    E(s;\tau,\bar\tau)=\sum_{(m,n)\not=(0,0)}\frac{\tau_2^s}{\pi^s\,|m+n\tau|^{2s}} \, ,
\end{align}
with $\tau=\tau_1+\ii\,\tau_2$, and admits a Fourier expansion of the type 
\begin{align}
    E(s;\tau,\bar\tau)=\sum_{k \in \mathbb{Z}}  \mathcal{E}_k(s;\tau_2)\,\rme^{2k\pi\ii\tau_1}\, ,
\end{align}
where the 0-instanton term is given by\,\footnote{For $s=1$ we have chosen to regularize the divergence by simply subtracting the pole $\frac{1}{s-1}$. More explicitly, we have defined $\mathcal{E}_0(1,\tau_2)$ as $\lim_{s\to1}\Big[\mathcal{E}_0(s,\tau_2)-\frac{1}{s-1}\Big]$; see also \cite{Brown:2024tru}.}
\begin{align}
    \mathcal{E}_0(s;\tau_2)=\begin{cases} \displaystyle{\frac{\pi\tau_2}{3}+2\gamma-\log(4\pi\tau_2)} & \text{for}~s=1~, \\ \\\displaystyle{\frac{2\,\zeta(2s)\,\tau_2^s}{\pi^s}}+\frac{2\sqrt{\pi}\,\zeta(2s-1)\,\Gamma\big(s-\frac{1}{2}\big)\,\tau_2^{1-s}}{\pi^s\,\Gamma(s)} & \text{for}~s\not=1~. \end{cases}
\end{align}
Notably, in the $O(N^0)$ contribution of \eqref{d2md2muexpanded} the coefficients precisely reproduce $\mathcal{E}_0(1;\tau_2)$, including the scheme-dependent constants. At sub-leading orders in $1/N$, we can similarly identify the perturbative part of $\mathcal{E}_0(s;\tau_2)$ for $s\not=1$. Therefore, following the same arguments of \cite{Chester:2019jas, Brown:2024tru,Chester:2025ssu}, we propose the modular invariant completion of our results via the substitution rule
\begin{align}
\begin{split}
    \frac{\pi\tau_2}{3}+2\gamma-\log(4\pi\tau_2)&\to E(1;\tau,\bar\tau)~,\\\frac{2\,\zeta(2s)\,\tau_2^s}{\pi^s}&\to E(s;\tau,\bar\tau)\quad\textsf{for}~s\not=1~.
\end{split}
\label{rule}
\end{align}
Applying this prescription to (\ref{d2md2muexpanded}) leads to
\begin{align}
    \partial_\mu^2\,\partial_m^2\log\mathcal{Z}^\DS\big|_{\D}~&=\,\text{constant terms}-6\,E(1;\tau,\bar{\tau})-\frac{E\big(\frac{3}{2};\tau,\bar{\tau}\big)}{(2N)^{1/2}}\notag\\[2mm]
    &\quad-\frac{3\,E\big(\frac{5}{2};\tau,\bar{\tau}\big)-4\,E\big(\frac{3}{2};\tau,\bar{\tau}\big)}{16 \, (2N)^{3/2}}
    -\frac{405\,E\big(\frac{7}{2};\tau,\bar{\tau}\big)-288\,E\big(\frac{5}{2};\tau,\bar{\tau}\big)+\ldots}{2048 \,(2N)^{5/2}}\notag\\[2mm]
    &\quad-\frac{7875\,E\big(\frac{9}{2};\tau,\bar{\tau}\big)-4050\,E\big(\frac{7}{2};\tau,\bar{\tau}\big)+\ldots}{16384 \,(2N)^{7/2}}+O(N^{-9/2}) \, , \label{d2md2muexpandedE}
\end{align}
where the ellipses denote additional contributions determined by the higher-order terms in the large-$N$ expansion.
Several remarks are in order. First, the fact that in the 
$O(N^0)$-term of \eqref{d2md2muexpanded} the $\tau_2$- and $\log(\tau_2)$-dependent contributions, originating respectively from $\mathcal{F}_0^\D$ and $\mathcal{F}_1^\D$, combine with the exact relative coefficient of the perturbative part of $E(1;\tau,\bar\tau)$ is highly nontrivial, given the very different structure of $\mathcal{F}_0^\D$ and $\mathcal{F}_1^\D$. This agreement can be regarded as strong evidence in favor of our proposal. Moreover, also all $\gamma$-dependent terms can be nicely absorbed into $E(1;\tau,\bar\tau)$.
Second, from superstring theory we expect the appearance of the Eisenstein series $E(1;\tau,\bar\tau)$ and $E(\frac{3}{2};\tau,\bar\tau)$, since they correspond to the coefficients of the higher-derivative terms $R^2F^2$ and $D^2R^2F^2$, arising from the low-energy expansion of flat-space superstring amplitudes in the presence of D-branes. These coefficients are fixed to be $E(1;\tau,\bar\tau)$ and $E(\frac{3}{2};\tau,\bar\tau)$ by the supersymmetry and modular invariance, following the arguments of \cite{Bachas:1999um, Green:2000ke, Kiritsis:2000zi, Lin:2015ixa}.\footnote{Note for the (unintegrated) superstring amplitudes, only the first two higher-derivative terms are determined by supersymmetry. This can be achieved by deriving differential equations obeyed by the coefficients of $R^2F^2$ and $ D^2R^2F^2$ using supersymmetry as in \cite{Lin:2015ixa}, whose modular invariant solutions are given by $E(1;\tau,\bar\tau)$ and $E(\frac{3}{2};\tau,\bar\tau)$.}  Third, the prescription (\ref{rule}) has precise implications for the strong-coupling behavior of the sub-leading coefficients $\mathcal{F}_{2}^\D, \mathcal{F}_{3}^\D, \ldots$ of the large-$N$ expansion. For example, consider the term proportional to $E(\frac{3}{2};\tau,\bar\tau)$ in the first line of (\ref{d2md2muexpandedE}). Its zero-mode part contains both a $\zeta(3)\,\tau_2^{3/2}$-contribution, matching the $\zeta(3)/{\lambda'}^{3/2}$-term found in $\mathcal{F}_0^\D$, and a $1/\sqrt{\tau_2}$-term. The latter predicts the existence of a $\sqrt{\lambda'}$-contribution with a specific coefficient in the strong-coupling expansion of $\mathcal{F}_2^\D$ at order $1/N$. Similar considerations apply to the higher Eisenstein series. Therefore, a further verification of our proposal would require computing also the higher-order corrections in the $1/N$ expansion and analyzing their strong-coupling behavior, as well as the non-perturbative instanton sectors. However, such calculations are technically very demanding and beyond the scope of this paper\,\footnote{The difficulty stems from the necessity of computing connected correlators of the $\mathcal{P}$-operators in the \textbf{D}-theory at higher orders in $1/N$ as compared to (\ref{vevsP}). This calculation involves the double-trace part of the action $S^\D$, significantly increasing the technical difficulties. This complexity is absent in the Sp($N$) theory studied in \cite{Chester:2025ssu} where the matrix-model action contains only single-trace terms (see Appendix\,\ref{App:Sptheory}).}. We should therefore regard the SL$(2, \mathbb{Z})$ completion in (\ref{d2md2muexpandedE}) as a well-motivated proposal, supported by the cases studied in \cite{Chester:2019jas, Brown:2024tru} and fully analogous to those in \cite{Chester:2025ssu}. 

Applying the same prescription to the mixed derivative    $\partial_\mu^2\,\partial_b^2\log\mathcal{Z}^\DS$, we find
\begin{align}
\partial_\mu^2\,\partial_b^2\log\mathcal{Z}^\DS\big|_{\D}~&=\, {\rm constant  \,\,\, terms}+2\,E(1;\tau,\bar{\tau})-\frac{E\big(\frac{3}{2};\tau,\bar{\tau}\big)}{(2N)^{1/2}}\notag\\[2mm]
    &\quad-\frac{3\,E\big(\frac{5}{2};\tau,\bar{\tau}\big)-4\,E\big(\frac{3}{2};\tau,\bar{\tau}\big)}{16\,(2N)^{3/2}}
    -\frac{405\,E\big(\frac{7}{2};\tau,\bar{\tau}\big)-288\,E\big(\frac{5}{2};\tau,\bar{\tau}\big)+\ldots}{2028 \, (2N)^{5/2}}\notag\\[2mm]
    &\quad-\frac{7875\,E\big(\frac{9}{2};\tau,\bar{\tau}\big)-4050\,E\big(\frac{7}{2};\tau,\bar{\tau}\big)+\ldots}{16384 \,(2N)^{7/2}}+O(N^{-9/2})~.\label{d2bd2muexpandedE}
\end{align}
This expression closely parallels \eqref{d2md2muexpandedE}: the same combinations of Eisenstein series appear at successive orders, with only differences confined in the constant terms and the coefficient of $E(1;\tau,\bar\tau)$.
This fact strongly suggests  a form of universality among distinct $\mathcal{N}=2$ observables in the limit $N\to\infty$ at fixed gauge coupling.

Even more striking is the similarity to the integrated giant-graviton correlator $\mathcal{G}$ in $\mathcal{N}=4$ SYM. The first terms of its large-$N$ expansion at fixed $\tau_2$ were obtained in \cite{Brown:2024tru}, but additional contributions can be generated straightforwardly, yielding\,\footnote{In this case, being a $\cN=4$ observable, we have used the relation $\lambda = 4\pi N/\tau_2$.}
\begin{align}
 \label{GGexpandedE}
    \mathcal{G}&= \, {\rm constant \,\,\, terms} -E(1;\tau,\bar\tau)-\frac{E(\frac{3}{2};\tau,\bar\tau)}{2\,N^{1/2}}-\frac{3\,E\big(\frac{5}{2};\tau,\bar{\tau}\big)-4\,E\big(\frac{3}{2};\tau,\bar{\tau}\big)}{32\,N^{3/2}}
   \\[2mm]
     &\quad
    -\frac{405\,E\big(\frac{7}{2};\tau,\bar{\tau}\big)-288\,E\big(\frac{5}{2};\tau,\bar{\tau}\big)+\ldots}{4096\,N^{5/2}}-\frac{7875\,E\big(\frac{9}{2};\tau,\bar{\tau}\big)-4050\,E\big(\frac{7}{2};\tau,\bar{\tau}\big)+\ldots}{32768\,N^{7/2}}
     +O(N^{-9/2})~. \nonumber
\end{align}
We see that the very same combinations of Eisenstein series found in (\ref{d2md2muexpandedE}) and (\ref{d2bd2muexpandedE}) reappear again in $\mathcal{G}$ at successive orders. A closer inspection of the coefficients reveals a simple correspondence: twice the integrated giant-graviton correlator $\mathcal{G}$ in $\mathcal{N}=4$ SYM with gauge group SU($2N$) matches the mixed derivatives in the SU($N$) $\mathbf{D}$-theory\,\footnote{This correspondence is natural given that the SU($N$) $\mathbf{D}$-theory can be realized via orbifold/orientifold projections of SU($2N$) $\mathcal{N}=4$ SYM.}, up to constant terms and the coefficient of $E(1;\tau,\bar\tau)$.
This again highlights a universality across different observables in the large-$N$ regime at fixed $\tau_2$.

Finally, using the results in Appendix \ref{App:Sptheory}, one may also promote the strong-coupling expansions of integrated correlators in the $\mathcal{N}=2$ Sp$(N)$ theory to be SL$(2, \mathbb{Z})$ invariant. In particular, we find once again that the results may be expressed in terms of linear combinations of non-holomorphic Eisenstein series. The case of the mixed derivative $\partial_\mu^2\,\partial_m^2\log\widetilde{\mathcal{Z}}^{*}$ has been worked out in \cite{Chester:2025ssu} and is given in Eq.\,(1.3) of that reference. Here, we provide the expression for $\partial_\mu^2\,\partial_b^2\log\widetilde{\mathcal{Z}}^{*}$, which reads: 
\begin{align}
\partial_\mu^2\,\partial_b^2\log\widetilde{\mathcal{Z}}^{*}\Big|_{\substack{m,\mu=0 \\ b=1}}~&= \, {\rm constant \,\,\, terms}+
    4\,E(1;\tau,\bar{\tau})-\frac{2\, E(\frac{3}{2};\tau,\bar\tau)}{(2N)^{1/2}}\notag\\[2mm]
     &\quad-\frac{3\,E\big(\frac{5}{2};\tau,\bar{\tau}\big)-8\,E\big(\frac{3}{2};\tau,\bar{\tau}\big)}{8\,(2N)^{3/2}}-\frac{405\,E\big(\frac{7}{2};\tau,\bar{\tau}\big)-576\,E\big(\frac{5}{2};\tau,\bar{\tau}\big)+\ldots}{1204\,(2N)^{5/2}}\notag\\[2mm]
     &\quad-\frac{7875\,E\big(\frac{9}{2};\tau,\bar{\tau}\big)-8100\,E\big(\frac{7}{2};\tau,\bar{\tau}\big)+\ldots}{8192\,(2N)^{7/2}}
     +O(N^{-9/2})~.\label{d2bd2muexpandedESp}
\end{align}
It is easy to see that it takes the same form as $\partial_\mu^2\,\partial_m^2\log\widetilde{\mathcal{Z}}^{*}$ in \cite{Chester:2025ssu} (except for the constant terms and the coefficient of $E(1;\tau,\bar{\tau})$). We also note that, unlike the integrated correlators in the $\mathbf{D}$-theory, it matches with the integrated giant gravitons of SU$(2N)$ $\mathcal{N}=4$ SYM only for the Eisenstein series with highest index at each order in the $1/N$-expansion, whereas the coefficients of the Eisenstein series with next-to-highest index differ by a factor of $2$. This difference is in agreement with \cite{Alday:2021vfb, Dorigoni:2022zcr}, where it was shown that integrated correlators in $\mathcal{N}=4$ SYM with gauge groups SU$(2N)$ and Sp$(N)$ only match for the Eisenstein series with highest index in the large-$N$ expansion. We therefore expect that the result \eqref{d2bd2muexpandedESp} should be compared with the integrated giant-graviton correlators in $\mathcal{N}=4$ SYM with Sp$(N)$ gauge group.

\section{Conclusions and outlook}
\label{sec:concl}

In this paper we studied the leading and sub-leading orders of the large-$N$ expansion of various mixed derivatives of the partition function  in the matrix model of a special $\mathcal{N}=2$ theory, called $\mathbf{D}^*$ theory. These quantities correspond to several classes of integrated correlators, which holographically correspond to scattering amplitudes of gluons and gravitons in AdS space, in the presence of D7-branes. The most notable cases are $\partial_\mu^2\,\partial_m^2\log\mathcal{Z}^\DS$ and $\partial_\mu^2\,\partial_b^2\log\mathcal{Z}^\DS$, corresponding to the integrated correlators of four moment map operators and two moment map and two stress energy tensor operators, respectively. In the dual perspective they both describe mixed scattering amplitudes of two gluons and two gravitons. Even though their weak-coupling expansions take very different expressions, their strong-coupling expansions are in fact governed  by exactly the same asymptotic series. Such common behavior becomes even more surprising by finding that the integrated correlators of gluons and gravitons  in a different $\mathcal{N}=2$ theory \cite{Chester:2025ssu} and  the integrated giant-graviton correlators in $\mathcal{N}=4$ SYM \cite{Brown:2024tru, Brown:2025huy} are given by precisely the same asymptotic series in the strong-coupling expansion. These results point to a remarkable universal property of the integrated correlators for graviton scattering in the presence of D-branes. 

As we argued, the integrated correlators considered here should be modular invariant, and exhibit a modular completion governed by non-holomorphic Eisenstein series of half-integer indices and $E(1; \tau, \bar{\tau})$. This is, once again, exactly in agreement with the case of giant-graviton correlators in $\mathcal{N}=4$ SYM, where the ${\rm SL}(2, \mathbb{Z})$ completion by the non-holomorphic Eisenstein series is better understood and expected \cite{Chester:2019jas, Dorigoni:2021guq, Collier:2022emf, Paul:2022piq}. The precise matching of perturbative contributions to the zero-mode of $E(1;\tau,\bar\tau)$, together with the recurrence of the same Eisenstein series combinations across different observables, provides strong evidence in support of this prescription. This is also in agreement with expectations from flat space superstring amplitudes in the presence of D-branes. 

The close parallel with the integrated giant-graviton correlator in $\mathcal{N}=4$ SYM highlights an unexpected universality: different supersymmetric theories, disparate quantities (with or without the presence of heavy operators) and realized in different gauge-theory contexts, seem to share the same modular structures in their large-$N$ expansions. This universality strongly suggests the presence of an underlying $\mathrm{SL}(2,\mathbb{Z})$-invariant framework that organizes sub-leading corrections across a wide class of observables. Our results and these observations have also opened up many future directions.  

One natural question is to consider the higher order terms of integrated correlators in the large-$N$ expansion. As we have commented, the computation becomes more involved beyond the large-$N$ orders considered in this paper, due to the double-trace contribution in the action $S^{\mathbf{D}}$, which will start to play a role. These results will be important for exploring whether the strong-coupling universality persists at higher orders in the large-$N$ expansion. Furthermore, the higher order results will provide further checks on the ${\rm SL}(2, \mathbb{Z})$ completion we proposed. Relating to further understanding the ${\rm SL}(2, \mathbb{Z})$ completion, it will be of interest to compute explicitly the non-perturbative instanton contributions and compare with the predictions from the non-holomorphic Eisenstein series. 

It is also important to understand the origin of the strong-coupling universality and to study its impact on the (unintegrated) correlators. As we emphasized, all these observables have the nice holographic interpretation as gravitons (with additional gluons)  scattering  off D-branes. It appears that being modular invariant, which we discussed above, also plays an important role.  As a comparison, one may consider the integrated correlators dual to scattering amplitude of two gravitons in the presence of D1-branes, studied in \cite{Pufu:2023vwo,Billo:2023ncz,Billo:2024kri, Dorigoni:2024vrb}, which is not modular invariant, and one finds the strong-coupling expansion is not governed by the same asymptotic series.

Provided that all the integrated correlators we considered here do take very different forms in the weak-coupling expansion, the universal strong-coupling asymptotic series must be accompanied with exponentially suppressed terms. These non-perturbative terms should account for the differences of these observables in the strong-coupling regime. It would be interesting to analyze those terms explicitly and to study their resurgence properties. Similar non-perturbative analyses have already been carried out in \cite{Brown:2025huy} for the leading large-$N$ terms of the integrated giant-graviton correlators and for integrated correlators in the Sp$(N)$ $\mathcal{N}=2$ theory considered in \cite{Chester:2025ssu}. 

\vskip 1cm
\noindent {\large {\bf Acknowledgments}}
\vskip 0.2cm
We sincerely thank Marco Bill\`o for collaboration and many fruitful and enlightening discussions throughout the whole project. We also thank Augustus Brown, Daniele Dorigoni, Alba Grassi, Cristoforo Iossa, Alessandro Pini, Xin Wang and Deliang Zhong for insightful discussions and exchange of ideas.

This research is partially supported by the INFN project ST\&FI ``String Theory \& Fundamental Interactions''.   ZD is supported by the Swiss National Science Foundation (SNSF) through a Postdoctoral Fellowship (TMPFP2\_234009). FG is  supported by the Italian Ministry of University and Research (MUR) under the FIS grant BootBeyond (CUP: D53C24005470001). CW is  supported by a Royal Society University Research Fellowship,  URF$\backslash$R$\backslash$221015. ZD, FG and CW are also supported by a STFC Consolidated Grant, ST$\backslash$T000686$\backslash$1 ``Amplitudes, strings \& duality".

\vskip 1cm

\appendix

\section{\texorpdfstring{Integrated constraints from mass deformations of $\cN=4$ SYM}{}}
\label{app:A}
We review the main results for the integrated correlators of half-BPS four-point functions coming from the mass deformation of $\cN=4$ SYM. These provide concrete examples of the supersymmetry preserving integration measures of integrated correlators, namely the LHS of the general formula in \eqref{eq:int_corr_schem}.

We consider the four-point function of $\mathbf{20^\prime}$ scalar operator, namely the top component of the $\cN=4$ stress-tensor multiplet. 
$\cO_{\mathbf{20^\prime}}$ is  the following gauge invariant combinations of the six scalar fields $\Phi^I$,
\begin{equation}\label{eq:O20prime}
 \cO_{\mathbf{20^\prime}}(x, Y)  = \frac{1}{2} Y_{I_1}Y_{I_2} \Tr\left( \Phi^{I_1}(x)\Phi^{I_2}(x) \right)~,
\end{equation}
where $Y_I$, with $I=1,\dots 6$, is the $SO(6)$ R-symmetry null vector. The four-point function of $\cO_{\mathbf{20^\prime}}$ reads \cite{Eden:2000bk, Nirschl:2004pa}:
\begin{equation}\label{eq:4pt2222}
    \langle \cO_{\mathbf{20^\prime}}(x_1,Y_1)\dots \cO_{\mathbf{20^\prime}}(x_4,Y_4)\rangle = \cT_{\rm free}(x_i,Y_i) + R_4(x_i,Y_i) ~ \cT(u,v;\tau) ~,
\end{equation}
where $\cT_{\rm free}$ is the free theory result obtained via Wick contractions, and $R_4$ is the following prefactor
\begin{equation}
    R_4(x_i,Y_i) = \frac{(z-\alpha)(z-\bar \alpha)(\bar z-\alpha)(\bar z-\bar \alpha)}{z {\bar z} (1-z)(1-\bar z)} \,\frac{(Y_1\cdot Y_3)^2}{x_{13}^4} \frac{(Y_2\cdot Y_4)^2}{x_{24}^4}~,
\end{equation}
with the R-symmetry cross-ratios
\begin{equation}
    \alpha\, \bar{\alpha}= \frac{Y_1\cdot Y_2 Y_3\cdot Y_4}{Y_1\cdot Y_3 Y_2\cdot Y_4}~, \quad \quad (1-\alpha)(1- \bar{\alpha})= \frac{Y_1\cdot Y_4 Y_2\cdot Y_3 }{Y_1\cdot Y_3 Y_2\cdot Y_4}~,
\end{equation}
and spacetime cross-ratios:
\begin{equation}
    \label{eq:zzb_def}
   u =z\, \bar{z}= \frac{x_{12}^2 x_{34}^2}{x_{13}^2 x_{24}^2} ~, \quad \quad v=(1-z)(1- \bar{z})= \frac{x_{14}^2 x_{23}^2}{x_{13}^2 x_{24}^2} ~. 
\end{equation}
The relevant function containing all the dynamical properties is the so-called reduced correlator $\cT(u,v; \tau)$. In particular, one can define two integrated correlators constraining $\cT$, coming from two or four mass derivatives. Hence, we define:
\begin{equation}\label{eq:measure_2}
    \cI_2[\cT] = -\frac{2}{\pi} \int_0^{\infty} dr \int^{\pi}_0 d \theta \,\frac{r^3 \sin^2 \theta}{u^2}\, \cT(u,v;\tau) ~, 
\end{equation}
and:
\begin{equation}\label{eq:measure_4}
     \cI_4[\cT] = \frac{1}{\pi} \int_0^{\infty} dr \int^{\pi}_0 d \theta \,\frac{r^3 \sin^2 \theta}{u^2} (1+u+v) \,\bar D_{1111}(u,v)\,\cT(u,v;\tau) \, , 
\end{equation}
where $u=1-2r \cos\theta  +r^2$, $v=r^2$ and $\bar D_{1111}(u,v)$ is the following box integral:
\begin{equation}
 \bar D_{1111}(u,v) = \frac{1}{z-\bar z} \left( \log z\bar z \, \log \frac{1-z}{1-\bar z} + 2\text{Li}_2(z) -2\text{Li}_2(\bar z)  \right)~.
\end{equation}
As written in the main text around \eqref{eq:int_corr_schem}, the integrated correlators \eqref{eq:measure_2} and \eqref{eq:measure_4} can be computed via supersymmetric localisation (where $m$ is the mass deformation of $\cN=4$ SYM) as follows:
\begin{equation}
I_2[\cT] = \tau_2^2 \partial_{\tau}\partial_{\bar\tau}\partial^2_m\log \cZ_{\cN=2^*}\big|_{m=0}~, ~~~~~ I_4[\cT] =\partial^4_m \log \cZ_{\cN=2^*}\big|_{m=0}~.
\end{equation}
As explained in Section\,\ref{secn:General}, since the forms of the integration measures displayed in \eqref{eq:measure_2} and \eqref{eq:measure_4} follow from $\cN=2$ supersymmetric Ward identities, they can be employed to constrain classes of four-point functions for general $\cN=2$ SCFTs in the presence of mass deformations.
We expect similar expressions for the integration measures coming from the squashing deformation. We leave the explicit derivation of such integration measures for future work.

\section{\texorpdfstring{Exact results for integrated giant-graviton  correlators in $\mathcal{N}=4$ SYM}{}}
\label{app:GiantG}
We recall here the results from \cite{Brown:2024tru}, where the integrated correlator in presence of maximal determinant operators - namely $\cD(x, Y) = \det_N\, Y\cdot \phi(x)$, dual to giant-graviton D3 branes - was computed using supersymmetric localisation. Such integrated correlator can be computed in the mass-deformed matrix model as follows:
\begin{align}\label{eq:GG_def}
    \cC_{\cD} (\tau;N) = \frac{\partial_{\cD} \partial_{\cD} \partial_m^2 \log \cZ^{\cN=2^*}(\tau, \tau'; m) \, {\vert}_{\tau', m=0}}{\partial_{\cD} \partial_{\cD} \log \cZ^{\cN=2^*}(\tau, \tau'; m) \, {\vert}_{\tau', m=0}} \, ,
\end{align}
and can be performed by re-expressing the $\vev{\cD\cD\cO_2\cO_2}$
integrated correlator as an infinite sum over protected three-point functions. This method allows to obtain exact results in the 't Hooft coupling $\lambda$ at the first orders in the topological expansion:
\begin{align}\label{eq:GG_genus}
       \mathcal{C}_{\mathcal{D}}(\lambda; N) = \sum_{g=0}^{\infty} N^{1-g} \, \cC^{(g)}_{\cD}(\lambda) \, , 
\end{align}
where the exact results for the first two orders read \cite{Brown:2024tru}:
\begin{equation}
\begin{aligned}\label{eq:CD_exact}
     \mathcal C^{(0)}_{\cD} (\lambda) &=-  \int_0^{\infty} \frac{8w\; dw}{\sinh(w)^2}   \frac{1}{v} \left(J_0(v)-1\right)
   J_1(v) \, , \\
   \mathcal C^{(1)}_{\cD} (\lambda)  &= \int_0^{\infty} \frac{2w\; dw}{\sinh(w)^2}  \left[ J_1(v) \left( J_1(v)- \frac{v}{2} \right)  - \left(J_0(v)-1\right)^2\right]\, , 
\end{aligned}
\end{equation}
with $v=w \sqrt{\lambda }/\pi$. 
In the holographic  interpretation, these results provide exact constraints for the scattering process of two gravitons off D3-branes (wrapping an $S^3$ inside of $S^5$), where the direct calculation of the correlators
becomes challenging\,\footnote{See \cite{Jiang:2019xdz, Chen:2025yxg} for the results of the first orders of the giant-graviton correlator in the planar limit at weak and strong coupling expansions, which have also been shown to match with the localisation computations.}.
Expanding such results at weak coupling yields the following perturbative expansions:
\begin{equation}
    \begin{aligned} 
    \mathcal{C}^{(0)}_{\cD} (\lambda) &=4   \sum_{\ell=1}^{\infty}  (-1)^{\ell+1} {\zeta (2 \ell{+}1)} \left[\binom{2 \ell+1}{\ell}^2-\binom{2 \ell+1}{\ell}\right] \bigg( \frac{\lambda}{16 \pi^2}\bigg)^{\ell} \, , \label{C0Weak}\\ 
    \mathcal{C}_{\cD}^{(1)}(\lambda) &=2 \sum_{\ell=1}^\infty (-1)^{\ell+1} \zeta (2 \ell{+}1) \, (\ell{+}1) \left[ \binom{2 \ell+1}{\ell}^2 -(\ell{+}2) \binom{2\ell+1}{\ell}\right]\bigg(\frac{\lambda}{16 \pi^2} \bigg)^\ell ~.
\end{aligned}
\end{equation}
Similarly one can expand in the strong-coupling regime, where the first two orders in the topological expansion \eqref{eq:GG_genus} read:\footnote{At genus-0 order in the expansion \eqref{eq:GG_genus}, the result has recently been extended to giant gravitons with general dimension $\alpha N$ (sub-determinant operators with $0<\alpha<1$) and the $AdS$ giant gravitons with dimension $\beta N$ (symmetric Schur polynomial operators with $\beta>0$) \cite{Brown:2025huy}. In this reference, it was discovered that the integrated ($AdS$) giant-graviton correlators in the planar limit enjoy the same strong coupling universality as discussed in this paper. In particular, it was found that all these integrated correlators share the same universal asymptotic series at strong coupling as given in \eqref{eq:GG_strong}; the dimension dependence (i.e. $\alpha$ or $\beta$) only appears in the leading coupling-independent factor (namely the supergravity regime in the strong-coupling expansion).}
\begin{align} \label{eq:GG_strong}
\mathcal{C}_{\cD}^{(0)}(\lambda) &\sim   2 - \frac{4\pi^2}{3 \lambda } - \sum_{n=1}^{\infty}   \frac{16 n \zeta (2 n{+}1) \Gamma
   \left(n-\frac{1}{2}\right)^2 \Gamma \left(n+\frac{1}{2}\right)}{  \lambda ^{  n+\frac{1}{2}}\, \pi ^{3/2}\, \Gamma (n)}  \, , \\
   \mathcal{C}_{\mathcal{D}}^{(1)}(\lambda) &\sim -2 \gamma -2  -\log \bigg(\frac{\lambda}{16\pi^2 }\bigg) + \sum_{n=1}^{\infty}  \frac{8 n  \zeta (2 n{+}1) \Gamma \left(n-\frac{1}{2}\right)
   \Gamma \left(n+\frac{1}{2}\right)^2}{ \lambda ^{  n+\frac{1}{2}}\, \pi ^{3/2} \Gamma (n)}  \, ,\label{eq:GG_1_strong}
\end{align}
where $\gamma$ is the Euler-Mascheroni constant. Remarkably, despite the very different set up, the integrated giant-graviton  correlator in the strong-coupling expansion is governed by the same asymptotic series as integrated correlators in the $\mathbf{D}$-theory discussed in the main text, as commented around \eqref{univ1} and \eqref{F1C1}, as well as around \eqref{GGexpandedE} after considering the SL$(2, \mathbb{Z})$ completion. 

\section{The functions \texorpdfstring{$\Upsilon_b(x)$}{}, \texorpdfstring{$H_{\mathrm{v}}(x;b)$}{} and \texorpdfstring{$H_{\mathrm{h}}(x;b,m)$}{}}
\label{App:Hfunctions}
Here we collect some properties of the functions $H_{\mathrm{v}}(x;b)$ and $H_{\mathrm{h}}(x;b,m)$ defined in (\ref{Hfunctions}) in terms of the function $\Upsilon_b(x)$ introduced in \cite{Hama:2012bg,Zamolodchikov:1995aa}. In particular, we provide the expansions of their logarithms around the point $(m=0,b=1)$. 

The basic ingredient is the following integral representation of the logarithm of the $\Upsilon_b$ function
\cite{Nakayama:2004vk}
\begin{align}
    \log \Upsilon_b(x)=\int_0^\infty\!\frac{d\omega}{\omega}\,\left[\rme^{-2\omega}\bigg(\frac{Q}{2}-x\bigg)^2-\frac{\sinh^2{\left(\omega\left(\frac{Q}{2}-x\right)\right)}}{\sinh(b\,\omega)\,\sinh(\omega/b)}\right]~.
    \label{integral}
\end{align}
The integrand can be easily expanded in powers of $x$ and $(b-1)$, and the corresponding integrals over $\omega$ can be explicitly evaluated. Proceeding in this way and using the definitions (\ref{Hfunctions}), one finds
\begin{align}
    \log H_{\mathrm{v}}(x;b)&=-(b-1)^2\Big\{2(1+\gamma)-\sum_{n=0}^\infty (-1)^n\,\frac{4x^{2n+2}}{3}\,\big[n\,\zeta(2n+1)+(2n+3)\,\zeta(2n+3)\big]\Big\}\notag\\
    &\quad+(b-1)^3\Big\{2(1+\gamma)-\sum_{n=0}^\infty (-1)^n\,\frac{4x^{2n+2}}{3}\,\big[n\,\zeta(2n+1)+(2n+3)\,\zeta(2n+3)\big]\Big\}\notag\\
    &\quad\,+(b-1)^4 \Big\{\zeta (3)-\frac{5 \gamma }{2}-\frac{19}{6}+\sum_{n=0}^\infty (-1)^n\,\frac{x^{2n+2}}{45}\big[3n(4n^2+4n+17)\,\zeta(2n+1) \notag \\
& \qquad\qquad\qquad+5(2n+3)(8n^2+24n+31)\zeta(2n+3)\notag\\
&\qquad\qquad\qquad-4(2n+5)(2n+4)(2n+3)\,\zeta(2n+5)\big]\Big\}+O\Big((b-1)^5\Big)~.
\label{logHvexpanded}
\end{align}
Similarly, one gets
\begin{align}
  \log H_{\mathrm{h}}(x;b,m)=&\,m^2\Big[(1+\gamma)+\sum_{n=1}^\infty (-1)^n\,x^{2n}\,(2n+1)\,\zeta(2n+1)\Big]\notag\\
  &+(b-1)^2\sum_{n=0}^\infty (-1)^n\,\frac{x^{2n+2}}{3}\,\Big[2n\,\zeta(2n+1)-(2n+3)\,\zeta(2n+3)\Big]\notag\\
  &-(b-1)^3\sum_{n=0}^\infty (-1)^n\,\frac{x^{2n+2}}{3}\,\Big[2n\,\zeta(2n+1)-(2n+3)\,\zeta(2n+3)\Big]\notag\\
  &-m^4\sum_{n=0}^\infty (-1)^n \,\frac{x^{2n}}{12}\,(2n+3)(2n+2)(2n+1)\,\zeta(2n+3)\notag\\
  &+m^2(b-1)^2\sum_{n=0}^\infty (-1)^n\,\frac{x^{2n}}{6}\,\Big[(4n^2+4n)(2n+1)\,\zeta(2n+1)\notag
  \\
  &\qquad-(2n+3)(2n+2)(2n+1)\,\zeta(2n+3)\Big]\notag\\
  &+(b-1)^4\sum_{n=0}^\infty (-1)^n\,\frac{x^{2n+2}}{180}\Big[6n(4n^2+4n+17)\,\zeta(2n+1) \notag \\
& \qquad-\,5(2n+3)(8n^2+24n+31)\,\zeta(2n+3)\notag\\
&\qquad+7(2n+5)(2n+4)(2n+3)\,\zeta(2n+5)\Big]+ O\Big((b-1)^5, m^6\Big)~.
\label{logHhexpanded}
\end{align}
These expansions are somehow related to those appearing in Appendix\,B of \cite{Mitev:2015oty}.

Using the integral representation (\ref{integral}), it is possible to derive the following results
\begin{subequations}
    \begin{align}
    \partial_m^2 \log\Upsilon_b\bigg(\frac{Q}{2}+\ii\,m\bigg)\bigg|_{\substack{m=0 \\ b=1}}&=2(1+\gamma)~,\label{d2mlogYb}\\[1mm] \partial_m^4 \log\Upsilon_b\bigg(\frac{Q}{2}+\ii\,m\bigg)\bigg|_{\substack{m=0 \\ b=1}}&=-12\,\zeta(3)~,\label{d4mlogYb}\\[1mm]
    \partial_b^2 \log\Upsilon_b\bigg(\frac{Q}{2}+\ii\,m\bigg)\bigg|_{\substack{m=0 \\ b=1}}&=\partial_b^4 \log\Upsilon_b\bigg(\frac{Q}{2}+\ii\,m\bigg)\bigg|_{\substack{m=0 \\ b=1}}=0~,\label{d2bd4blogYb}\\[1mm]
    \partial_m^2\partial_b^2 \log\Upsilon_b\bigg(\frac{Q}{2}+\ii\,m\bigg)\bigg|_{\substack{m=0 \\ b=1}}&=\frac{4}{3}-4\,\zeta(3)~.\label{d2md2blogYb}
\end{align}
\end{subequations}
and \cite{Chester:2020vyz}
\begin{align}
    \partial_b^2\log \Upsilon_b^\prime(0)\bigg|_{b=1}&=-2(1+\gamma)~,\qquad
    \partial_b^4\log \Upsilon_b^\prime(0)\bigg|_{b=1}=12\,\zeta(3)-30\,\gamma-38~.
    \label{d24blogYprime}
\end{align}

\section{The term \texorpdfstring{$\mathcal{B}^\D_{4}$}{}}
\label{App:B4}
Here we give the explicit expression of $\mathcal{B}^\D_{4}$ appearing as the coefficient of $(b-1)^4$ in the expansion of the effective action of the matrix model of the $\mathbf{D}^*$-theory, see (\ref{SDSexp}). This is
\begin{align}
    \mathcal{B}^\D_{4}&=\frac{1}{24}\partial_b^4 S^\DS\big|_{\D}\notag\\
        &=-\frac{N^2-1}{2}\Big(\zeta(3)-\frac{5}{2}\gamma-\frac{19}{6}\Big) \notag\\
        &\quad+\frac{1}{12}\sum_{n=1}^\infty\sum_{k=0}^n(-1)^n\frac{(2n)!}{(2k)!(2n-2k)!}\,\bigg[(2n+1)(8n^2+8n+15)\,\zeta(2n+1) \notag \\
     & \qquad\qquad\qquad\qquad-(2n+3)(2n+2)(2n+1)\,\zeta(2n+3)\bigg]\Big(\frac{\lambda}{8\pi^2N}\Big)^{n} \tr a^{2n-2k}\,\tr a^{2k}\notag \\
    &\quad+\frac{1}{180}\sum_{n=0}^\infty\sum_{k=1}^{n-1}(-1)^n\frac{(2n+2)!}{(2k+1)!(2n-2k+1)!}\,\bigg[\frac{12n}{15}(4n^2+4n+17)\,\zeta(2n+1)\notag\\
    &\qquad\qquad\qquad\qquad+5(2n+3)(8n^2+24n+31)\,\zeta(2n+3)\notag\\
    &\qquad\qquad\qquad\qquad-(2n+5)(2n+4)(2n+3)\,\zeta(2n+5)\bigg]\Big(\frac{\lambda}{8\pi^2N}\Big)^{n+1}\tr a^{2n-2k+1}\,\tr a^{2k+1}\notag \\ 
    &\quad-\frac{1}{180}\sum_{n=0}^\infty (-1)^n\bigg[6n(4n^2+4n+17)\,\zeta(2n+1)-5(2n+3)(8n^2+24n+31)\,\zeta(2n+3)\notag \\
    &\qquad\qquad\qquad\qquad+7(2n+5)(2n+4)(2n+3)\zeta(2n+5)\Big]\Big(\frac{\lambda}{2\pi^2N}\Big)^{n+1}\,\mathrm{tr}a^{2n+2} \notag \\
    &\quad+\frac{1}{45}\sum_{n=0}^\infty (-1)^n\bigg[6n(4n^2+4n+17)\,\zeta(2n+1)-5(2n+3)(8n^2+24n+31)\,\zeta(2n+3)\notag \\
    &\qquad\qquad\qquad\qquad+7(2n+5)(2n+4)(2n+3)\zeta(2n+5)\Big]\Big(\frac{\lambda}{8\pi^2N}\Big)^{n+1}\,\mathrm{tr}a^{2n+2}~.
\end{align}

\section{Weak-coupling results} 
\label{App:WeakCoupling}
In this Appendix we provide the first few terms in the weak-coupling expansion of the quantities 
$\mathcal{F}_0^\D$, $\mathcal{F}_1^\D$, $ \Delta_0^\D $ and $\Delta_1^\D$ defined, respectively, in \eqref{F0}, \eqref{F1}, \eqref{Delta0} and \eqref{Delta1}. By comparing them with their respective strong-coupling counterparts, reported in \eqref{SCF0D}, \eqref{F1strongfinal}, \eqref{Delta0strong} and \eqref{Delta1strong}, we can clearly realize how much more involved these weak-coupling expressions are. We get 
\begin{subequations}
\begin{align}
    \mathcal{F}_0^\D \underset{\lambda \to 0}{\sim} &\frac{9  \zeta (3)^2}{4 \pi ^4}\lambda ^2 -\frac{105  \zeta (3) \zeta (5)}{32 \pi ^6} \lambda ^3 +\frac{105  \left(5 \zeta (5)^2+12 \zeta (3) \zeta (7)\right)}{512 \pi ^8}\lambda ^4 \nonumber  \\ 
    & -\frac{21 (65 \zeta (5) \zeta (7)+99 \zeta (3) \zeta (9))}{1024
   \pi ^{10}} \lambda^5 + O(\lambda^6) \, ,  \label{WeakF0}\\
  \mathcal{F}_1^{\D} \underset{\lambda \to 0}{\sim} &-\frac{9  \zeta (3)^2}{2 \pi ^4} \lambda ^2+ \frac{75  \zeta (3) \zeta (5)}{8 \pi ^6}\lambda ^3 -\frac{9  \left(72 \zeta (3)^3+595 \zeta (7) \zeta (3)+200 \zeta (5)^2\right)}{512 \pi ^8}\lambda ^4 \nonumber \\
    & +\frac{45 \left(90 \zeta (5) \zeta (3)^2+273 \zeta (9) \zeta (3)+140 \zeta (5) \zeta (7)\right)}{1024 \pi ^{10}} \lambda ^5 + O(\lambda^6) \, , \label{WeakF1} \\
    \Delta_0^{\D} \underset{\lambda \to 0}{\sim} & \frac{16 (3 \zeta (3) -1)}{3} -\frac{2   (5 \zeta (5) - 2 \zeta (3))}{\pi ^2}\lambda +\frac{5  (7 \zeta (7)-4 \zeta (5))}{8 \pi ^4}  \lambda ^2 - \frac{35 (3 \zeta (9) - 2 \zeta (7))}{64 \pi ^6} \lambda ^3 \nonumber \\ 
    & + \, \frac{105  (11 \zeta (11)-8 \zeta (9))}{2048 \pi ^8} \lambda ^4 - \frac{231 (13 \zeta (13)- 10 \zeta (11) )}{16384 \pi ^{10}} \lambda ^5 + O(\lambda^6) \, , \label{WeakDelta0} \\
    \Delta_1^\D \underset{\lambda \to 0}{\sim} &\frac{9  \zeta (3)^2}{2 \pi ^4}\lambda ^2-\frac{3  \left(4 \zeta (3)^2+15 \zeta (5) \zeta (3)\right)}{8 \pi ^6} \lambda ^3+ \frac{15  \left(40 \zeta (5)^2+128 \zeta (3) \zeta (5)+189 \zeta (3) \zeta (7)\right)}{512 \pi ^8} \lambda ^4\nonumber \\
    &- \frac{9  \left(400 \zeta (5)^2+385 \zeta (7) \zeta (5)+882 \zeta (3) \zeta (7)+1596 \zeta (3) \zeta (9)\right)}{2048 \pi ^{10}}\lambda ^5 + O(\lambda^6)  \label{WeakDelta1}\;.
\end{align}
\label{WeakExp}%
\end{subequations}
Although the higher-order terms can be obtained straightforwardly, they become increasingly more complicated. The most striking feature of these weak-coupling results, in contrast to their strong-coupling counterparts, is the appearance of products of many Riemann $\zeta$-values. 

\section{Details on the strong-coupling expansion}
\label{app:strong}
In this Appendix we provide the technical details underlying the strong-coupling expansions presented in Section\,\ref{secn:strongcoupling}.

\subsection{Strong-coupling behavior of \texorpdfstring{$\mathcal{F}_1^\D$}{}}
\label{app:strong1}
Our analysis focuses on the three contributions $A_1, A_2$ and $A_3$ defined in \eqref{F1Dcomponents} and \eqref{A123}, and we systematically derive their behavior in the limit $\lambda \to \infty$.
\subsubsection*{\texorpdfstring{$\bullet \quad A_1$}{}}
From its definition in \eqref{A1}, the quantity $A_1$ factorizes into three independent components.
The first factor, denoted by \textsf{Y}, was analyzed in \cite{Billo:2024ftq}, where its strong-coupling limit was obtained as
\begin{align}
    \textsf{Y} \underset{\lambda \to \infty}{\sim}-\frac{\log(2)}{2\pi^2}\,\lambda + \frac{1}{4}~.
    \label{ScY}
\end{align}
The second factor is given by
\begin{align}
    \sum_{k=1}^{\infty}(-1)^k\;(2k)\; \textsf{Z}_{2k}^{(2)}~,
\end{align}
and can be simplified using Eq.\,(3.25) of \cite{Billo:2024ftq}. Its strong-coupling expansion can then be read directly from the first term in Eq.\, (A.6) of that reference:
\begin{align}
  \sum_{k=1}^{\infty}(-1)^k\;(2k)\; \textsf{Z}_{2k}^{(2)}\;    \underset{\lambda \to \infty}{\sim} -\frac{1}{2} ~.
  \label{StrongsumZ2k2}
\end{align}
Finally, employing \eqref{Mkp}, performing the sum over $\ell$, and introducing the Mellin–Barnes representation of the Bessel functions, the third factor of $A_1$ can be rewritten as
\begin{align}
 \sum_{\ell=1}^{\infty}(-1)^{\ell}\; (2\ell)\; \textsf{M}_{1, 2\ell}^{(1)}\, = \, -\frac{\sqrt{\lambda}}{4\pi}\int_{-i \infty}^{i \infty} \frac{ds}{2\pi i}\frac{\Gamma(-s)\Gamma(2s+3)\Gamma(2s+4)\zeta(2s+3)}{\Gamma^2(s+2)\Gamma(s+3)}\; \bigg(\frac{\sqrt{\lambda}}{4\pi} \bigg)^{2s+2}~.
\end{align}
For $\lambda \to \infty$, the contour can be closed counter-clockwise, picking up residues at the poles along the negative real axis of $s$. This yields
\begin{align}
  \sum_{\ell=1}^{\infty}(-1)^{\ell}\; (2\ell)\; \textsf{M}_{1, 2\ell}^{(1)}\;  \underset{\lambda \to \infty}{\sim} -\frac{\sqrt{\lambda}}{8\pi} +\frac{1}{4\pi} -\sum_{n=1}^\infty \frac{(n-\frac{1}{2})\,\Gamma(n-\frac{1}{2})^2\,\Gamma(n+\frac{3}{2})\,\zeta(2n+1)}{2\pi^{5/2}\,\Gamma(n)\,\lambda^n}~.
  \label{StrongsumM}
\end{align}
Combining \eqref{ScY}, \eqref{StrongsumZ2k2}, and \eqref{StrongsumM}, one finds the strong-coupling expansion of $A_1$ as quoted in \eqref{A1strong}. 

\subsubsection*{\texorpdfstring{$\bullet \quad A_2$}{}}
\label{app:strongA2}
The expression for $A_2$, given in \eqref{A2}, is structurally similar to the third quantity analyzed in Appendix A of \cite{Billo:2024ftq}. Following the same procedure as in that work, we obtain
\begin{align}
A_2 =  -4 &\iint_{-\ii \infty}^{+\ii \infty} \frac{ds\,ds'}{(2 \pi \ii)^2} \,\Big(\frac{\sqrt{\lambda}}{2\pi} \Big)^{2s+2s^\prime+4}\,\;2^{-2s}\;\,\zeta(2s+3)\,\zeta(2s^\prime+3) \nonumber \\
& \qquad\qquad\times \frac{\Gamma(-s)\,\Gamma(-s^\prime)\,\Gamma(2s+4)\,\Gamma(2s^\prime+4)}{(s+s^\prime+2)\,\Gamma(s+2)\,\Gamma(s^\prime+2)} ~.
\label{ZZ'sc} 
\end{align}
Closing both contours counter-clockwise yields two types of contributions:  
($i$) residues at $(s=-1,\,s^\prime=-1)$, giving
\begin{align}
    -4\,\log\!\Big(\frac{\lambda}{\pi^2}\Big) -8\gamma + 8\log(2)-8~,
    \label{Icontr}
\end{align}
and ($ii$) residues at $(s=-s^\prime-2,\,s^\prime=-n)$ with $n=1,2,\ldots$, producing
\begin{align}
    8\log(2) - 8\sum_{n=1}^{\infty}\frac{\Gamma(2n+2)}{4^n \Gamma(2n-1)}\zeta(1-2n)\zeta(1+2n)~.
    \label{IIcontr}
\end{align}
The infinity sum in \eqref{IIcontr} is divergent but can be regularized using the functional equation
\begin{align}
    \zeta(1-2n)= 2 \frac{(-1)^n\Gamma(2n)}{(2\pi)^{2n}}\zeta(2n)\;,
    \label{funceq}
\end{align}
together with the integral representation
\begin{align}
    \zeta(n)= \frac{1}{\Gamma(n)} \int_0^{\infty}dx\; \frac{x^{n-1}}{e^x-1}~.
    \label{intzeta2}
\end{align}
Carrying out the computation yields
\begin{align}
&8\log(2)-8\sum_{n=1}^{\infty}\frac{\Gamma(2n+2)}{4^n \Gamma(2n-1)}\,\zeta(1-2n)\,\zeta(1+2n)\, \notag\\
    &\qquad=
    8\log(2)+\frac{1}{4\pi^3}\int_0^{\infty}dx \frac{x^2}{e^x-1}\int_0^{\infty}dy \frac{y}{e^y-1}  \left(12 \pi  \cos \left(\frac{x y}{4 \pi }\right)-x y \sin \left(\frac{x y}{4 \pi }\right)\right)\notag\\
    &
    \qquad=-\frac{20}{3}+11\,\zeta(3)~.
    \label{IIa}
\end{align}
Combining \eqref{Icontr} and \eqref{IIa}, the strong-coupling behavior of $A_2$ is
\begin{align}
A_2 \underset{\lambda \rightarrow \infty}{\sim} ~-4\log\Big(\frac{\lambda}{\pi^2}\Big)+8\log(2)-8\gamma+11\,\zeta(3)-\frac{44}{3}~,
\label{ZZstrongapp}
\end{align}
as reported in \eqref{A2strong}.

\subsubsection*{\texorpdfstring{$\bullet\quad A_3$}{}}
The derivation of the strong-coupling behavior of $A_3$ given in \eqref{A3} is considerably more involved. Substituting the definitions~\eqref{Mkp}, \eqref{Zkp}, and \eqref{Y2k} into \eqref{A3}, and employing the Mellin–Barnes representations of Bessel functions together with the integral form of the Riemann $\zeta$-function, one obtains
\begin{align}
    A_3&=64\iiint_{-\ii\infty}^{+\ii\infty}\!\frac{ds\,ds'\,ds''}{(2\pi\ii)^3}\,\sum_{k,n=1}^\infty\,(2k)\,(2n)
    \,\Big(\frac{\sqrt{\lambda}}{4\pi}\Big)^{2s+2s'+2s''+4k+4n}~\times\notag\\
    &\qquad\qquad\times\,\frac{\Gamma(-s)\,\Gamma(2s+2k+2)\,\zeta(2s+2k+1)}{\Gamma(s+2k+1)}\,\times\notag\\[2mm]
    &\qquad\qquad\times\frac{\Gamma(-s')\,\Gamma(2s'+2n+2k+1)\,\Gamma(2s'+2n+2k+2)\,\zeta(2s'+2n+2k+1)}{\Gamma(s'+2n+1)\,\Gamma(s'+2k+1)\,\Gamma(s'+2n+2k+1)}~\times\notag\\[2mm]
    &\qquad\qquad\times\,\big(4-4^{s''+n}\big)\,\frac{\Gamma(-s'')\,\Gamma(2s''+2n)\,\zeta(2s''+2n-1)}{\Gamma(s''+2n+1)}~.
\end{align}
After shifting the integration variables as $s\to s-k+1$, $s'\to s'-k-n+1$, and $s''\to s''-n+2$, the sum over $n$ can be performed. This yields
\begin{align}
   & \sum_{n=1}^\infty(2n)
    \,\frac{\Gamma(-s'+k+n-1)\,\Gamma(-s''+n-2)}{\Gamma(s'+n-k+2)\,\Gamma(s'+k-n+2)\,\Gamma(s'+n+k+2)\,\Gamma(s''+n+3)}\notag\\
    &\quad =\frac{2\,\Gamma(-s'+k)\,\Gamma(-s''-1)}{\Gamma(s'-k+3)\,\Gamma(s'+k+1)\,\Gamma(s'+k+3)\,\Gamma(s''+4)}~\times\notag\\[1mm]
    &\qquad\qquad\times\,{}_4F_{3}(2,-s'-k,-s'+k,-s''-1;s'-k+3,s'+k+3,s''+4;-1)\notag\\[2mm]
    &\quad=\frac{\Gamma(-s'+k)\,\Gamma(-s''-1)}{\Gamma(s'+k+1)\,\Gamma(2s'+3)\,\Gamma(s''+4)}\,\,{}_3F_{2}(-s'-k,-s'+k,s''+3;2,s''+4;1) \, ,
    \label{sum1}
\end{align}
where the last step follows from the identity (\ref{identity4F3}).
Consequently, we can write 
\begin{align}
     A_3
    &=\iiint_{-\ii\infty}^{+\ii\infty}\!\frac{ds\,ds'\,ds''}{(2\pi\ii)^3} ~\mathcal{A}(s,s',s'') \, , 
\end{align}
where
\begin{align}
    \mathcal{A}(s,s',s'')
    &=256\,\,\Big(\frac{\sqrt{\lambda}}{4\pi}\Big)^{2s+2s'+2s''+8}\big(1-4^{s''+1}\big)
    ~\times\label{F2H2}\\[2mm]
    &\times
    \frac{\Gamma(2s+4)\,\zeta(2s+3)\,\Gamma(2s'+4)\,\zeta(2s'+3)\,\Gamma(2s''+4)\,\zeta(2s''+3)\,\Gamma(-s''-1)}{\Gamma(s''+4)}~\times\notag\\[2mm]
    &\times \sum_{k=1}^\infty\,(2k)
    \,\frac{\Gamma(-s+k-1)\,\Gamma(-s'+k)}{\Gamma(s+k+2)\,\Gamma(s'+k+1)}\,\,{}_3F_{2}(-s'-k,-s'+k,s''+3;2,s''+4;1)~.\notag
\end{align}
The strong-coupling expansion of $A_3$ can now be obtained by closing the integration contours counter-clockwise in the half-planes $\mathfrak{Re}(s)<0$, $\mathfrak{Re}(s')<0$, $\mathfrak{Re}(s'')<0$, and summing over the residues. Owing to the structure of the integrand, it is convenient to first consider the $s''$-integration, since $s''$ does not mix with the summation index $k$. At strong coupling we may thus write
\begin{align}
    A_3~ \underset{\lambda \to \infty}{\sim} ~\sum_{n=1}^{\infty} A_3^{(n)}\quad\textsf{with}\quad A_3^{(n)}  =\text{Res}\;\mathcal{A}(s,s',s'') \bigg|_{s''=-n}~.
    \label{A3expansion}
    \end{align}
    \paragraph{Residue at $s''=-1$:}  
The contribution from $s''=-1$ reads
\begin{align}
    A_3^{(1)}&=256\log(2)\!\iint_{-\ii\infty}^{+\ii\infty}\!\!\frac{ds\,ds'}{(2\pi\ii)^2}\,\Big(\frac{\sqrt{\lambda}}{4\pi}\Big)^{2s+2s'+6}\,\Gamma(2s+4)\,\zeta(2s+3)\,\Gamma(2s'+4)\,\zeta(2s'+3)
    \,\times\notag\\[2mm]
    &\qquad\quad\times\,\Gamma(2s'+3)\,
    \sum_{k=1}^\infty\,(2k)
    \,\frac{\Gamma(-s+k-1)\,\Gamma(-s'+k)}{\Gamma(s+k+2)\,\Gamma(s'+k+1)\,\Gamma(s'-k+3)\,\Gamma(s'+k+3)}~.
    \label{F2H-1a}
\end{align}
Applying the same method described in Section\,\ref{secn:strongcoupling} for $\mathcal{F}_0^{\textbf{D}}$, we obtain
\begin{align}
    A_3^{(1)} &\underset{\lambda \to \infty}{\sim} \frac{4\,\log(2)\,\lambda}{\pi^2}-\frac{8\,\log(2)\,\lambda^{1/2}}{\pi^2}-8\,\log(2)\notag\\
    &\qquad\qquad-\sum_{n=1}^\infty
    \frac{16\,\log(2)\, (n+\frac{1}{2})\,\Gamma(n-\frac{1}{2})^2\,\Gamma(n+\frac{3}{2})\,\zeta(2n+1)}{\pi^{7/2}\,\Gamma(n)\,\lambda^{n-1/2}}~,
    \label{A3first}
\end{align}
with intermediate steps reported in the ancillary  {\tt Mathematica}  file.

\paragraph{Residue at $s''=-2$:}  
The next contribution arises from $s''=-2$ and is
\begin{align}
    A_3^{(2)} &=16\iint_{-\ii\infty}^{+\ii\infty}\!\frac{ds\,ds'}{(2\pi\ii)^2}\,\Big(\frac{\sqrt{\lambda}}{4\pi}\Big)^{2s+2s'+4}\,\,\Gamma(2s+4)\,\zeta(2s+3)\,\Gamma(2s'+4)\,\zeta(2s'+3)\,\times\notag\\
    &\qquad\qquad\times\,\frac{\Gamma(-s)\,\Gamma(-s')\,\Gamma(2s'+3)}{\Gamma(s+s'+3)\,\Gamma(s'+1)\,\Gamma(s'+3)}\,
    {}_3F_2(-s,-s',s'+2;2,s'+3;1)
    \notag\\[3mm]
    &\quad-16\iint_{-\ii\infty}^{+\ii\infty}\!\frac{ds\,ds'}{(2\pi\ii)^2}\,\Big(\frac{\sqrt{\lambda}}{4\pi}\Big)^{2s+2s'+4}\,\,\Gamma(2s+4)\,\zeta(2s+3)\,\Gamma(2s'+4)\,\zeta(2s'+3)\,\times\notag\\
    &\qquad\qquad\times\,\frac{\Gamma(-s)\,\Gamma(-s')}{(s+s'+2)\,\Gamma(s+2)\,\Gamma(s'+2)}~.
    \label{F2H2b}
\end{align}
 The first two integrals are evaluated using the same method as for $\mathcal{F}_0^{\textbf{D}}$, while for the last two integrals we follow the strategy described before for $A_2$. Altogether, one finds
\begin{align}
A_3^{(2)}&~\underset{\lambda \rightarrow \infty}{\sim}~
    - 2 \log\Big(\frac{\lambda}{\pi^2}\Big)-4(1+\gamma)+ 8 \log 2 -\frac{10}{3} +4\,\zeta(3)+\sum_{n=1}^\infty\frac{8\,(n-1)\,\Gamma(n-\frac{1}{2})^3\,\zeta(2n-1)}{\pi^{3/2}\,\Gamma(n)\,\lambda^{n-1/2}}~.
    \label{A3second}
\end{align}
Again the details can be found in the ancillary  {\tt Mathematica}  file.

\paragraph{Residues at $s''=-3,-4,\ldots$:}  All such residues 
can be written as
\begin{align}
A_3^{(3+\ell)}
    &=
  \frac{128\,(1-2^{-2\ell-4})\,\zeta(-2\ell-3) }{(2\ell+2)!}\,\iint_{-\ii\infty}^{+\ii\infty}\!\frac{ds\,ds'}{(2\pi\ii)^2}\,\Big(\frac{\sqrt{\lambda}}{4\pi}\Big)^{2s+2s'+2-2\ell}
    ~\times\notag\\[2mm]
    &\qquad\quad\times~
    \frac{\Gamma(2s+4)\,\zeta(2s+3)\,\Gamma(2s'+4)\,\zeta(2s'+3)\,\Gamma(-s)\,\Gamma(-s'+1+\ell)}{(s+s'+1-\ell)\,\Gamma(s+2)\,\Gamma(s'+1-\ell)}
    \label{F2H-3a}
\end{align}
for $\ell\geq0$.
Only the poles at $(s=-s'-1+\ell,\,s'=-m)$ with $m=1,2,\ldots$ contribute yielding constant terms:
\begin{align}
A_3^{(3+\ell)}&~\underset{\lambda \rightarrow \infty}{\sim}~c_\ell~.
    \label{F2H31}
\end{align}
The sum over all such constants is
\begin{align}
\sum_{\ell=0}^\infty \,c_\ell =  \sum_{\ell=0}^\infty\,\sum_{m=1}^\infty\frac{(64 - 4^{1-\ell})\,(2m-3)\,(2\ell+3)\,\Gamma(2\ell {+} 2m {+}2)\,\zeta(2\ell{+} 2m {+}1)}{ (2m-2)!\,(2\ell+4)!}\,B_{2\ell+4}\,B_{2m-2} \, , 
\end{align}
where $B_n$ are Bernoulli numbers. Using the integral representation of the Riemann $\zeta$-function, both sums can be performed independently, yielding
\begin{align}
    \sum_{\ell=0}^\infty c_\ell=\int_0^\infty \frac{w^3 \, dw}{\sinh^2(w)} \, \text{sech}^4\Big(\frac{w}{2}\Big) =-\frac{1}{5}+8\log(2)-\frac{39\,\zeta(3)}{10}~.
    \label{A3others}
\end{align}

Collecting \eqref{A3first}, \eqref{A3second}, and \eqref{A3others}, we arrive at the strong-coupling expansion
\begin{align}
    A_3\,&\underset{\lambda \rightarrow \infty}{\sim}\,\frac{4\log(2)}{\pi^2}\,\lambda-\frac{8\log(2)}{\pi^2}\,\sqrt{\lambda}-2 \log\Big(\frac{\lambda}{\pi^2}\Big)+k+\frac{4}{\sqrt{\lambda}}
    \label{A3strongapp}\\
    &\qquad+\sum_{n=1}^\infty\frac{8\,\Gamma(n+\frac{1}{2})^3\,\zeta(2n+1)}{\pi^{3/2}\,\Gamma(n)\,\lambda^{n+1/2}}-\log(2)
    \sum_{n=1}^\infty\frac{16 (n+\frac{1}{2})\,\Gamma(n-\frac{1}{2})^2\,\Gamma(n+\frac{3}{2})\,\zeta(2n+1)}{\pi^{7/2}\,\Gamma(n)\,\lambda^{n-1/2}} \, ,\notag
\end{align}
where
\begin{align}
    k=-\frac{53}{15}-4(1+\gamma)+8\log(2)+\frac{\zeta(3)}{10} \, ,
    \label{constantKapp}
\end{align}
in agreement with (\ref{A3strong}) and (\ref{constantK}).

\subsection{Strong-coupling behavior of \texorpdfstring{$\Delta_1^\D$}{}}
\label{app:strong2}

We now consider the quantity $\Delta_1^\D$, defined in \eqref{Delta1}, and write it as the sum of three terms:
\begin{align}
    \Delta_1^\D = B_1+B_2+ B_3 \, ,
\end{align}
with
\begin{subequations}
    \begin{align}
   B_1=& \frac{32}{3}\sum_{k=1}^\infty (2k)^3\,\mathsf{Z}^{(2)}_{2k}\Big(\widehat{\mathsf{Z}}^{(0)}_{2k}-4\mathsf{Z}^{(0)}_{2k} \Big)~, \label{Q1}\\
     B_2=& \frac{32\sqrt{\lambda}}{3\pi}\sum_{k=1}^\infty(2k)\,\mathsf{Z}^{(2)}_{2k}\Big(\widehat{\mathsf{Z}}^{(1)}_{2k+1}-2\,\mathsf{Z}^{(1)}_{2k+1} \Big)-\frac{16\sqrt{\lambda}}{3\pi}\sum_{k=1}^\infty (2k)\,\mathsf{Z}^{(3)}_{2k+1}\Big(\widehat{\mathsf{Z}}^{(0)}_{2k}-4\,\mathsf{Z}^{(0)}_{2k} \Big)~,\label{Q2} \\
     B_3=& \frac{32}{3}\,\sum_{k=1}^\infty (2k)\,\mathsf{Z}_{2k}^{(2)}\,\Big(\widehat{\mathsf{Z}}^{(2)}_{2k}+2\,\mathsf{Z}^{(2)}_{2k}\Big) -\frac{16 \lambda}{3\pi^2}\sum_{k=1}^\infty (2k)\,\mathsf{Z}_{2k}^{(2)}\Big(\widehat{\mathsf{Z}}^{(2)}_{2k}-\mathsf{Z}^{(2)}_{2k}\Big) \notag \\
    & -\frac{16}{3}\Big(1+\frac{\lambda}{4\pi^2}\Big)\sum_{k=1}^\infty (2k)\,\mathsf{Z}_{2k}^{(4)}\Big(\widehat{\mathsf{Z}}^{(0)}_{2k}-4\,\mathsf{Z}^{(0)}_{2k}\Big)~. \label{Q3}
\end{align}
\label{B123}%
\end{subequations}
We now derive the strong-coupling behavior of each of these terms.

\subsubsection*{\texorpdfstring{$\bullet \quad B_1$}{}}
Using the Mellin-Barnes representation of the Bessel functions and the integral representation of the Riemann $\zeta$-function, we can rewrite $B_1$ as
\begin{align}
   B_1
   = &\,  \frac{256}{3}\sum_{k=1}^{\infty} k^3   \iint_{-i \infty}^{i \infty}\frac{ds\,ds'}{(2\pi \ii)^2}\, \frac{\Gamma(-s+k-1)\Gamma(-s'+k-2)}{\Gamma(s+k+2) \Gamma(s'+k+3)}\Gamma(2s+4)\zeta(2s+3)\nonumber \,\times\\ & \qquad\qquad\qquad \,\times\Gamma(2s'+4)\zeta(2s'+3)  \Big( \frac{\sqrt{\lambda}}{4\pi}\Big)^{2s'+2s+6}\,\big(4^{s'+2} - 4 \big)~.
\end{align}
To proceed we adopt the same strategy as before: we first perform the sum and then close the integration contours counter-clockwise in the half-planes $\mathfrak{Re}(s)<0$ and $\mathfrak{Re}(s')<0$. Summing up the residues and adopting the same regularization procedure discussed above, we obtain
\begin{align}
 B_1   \underset{\lambda \to \infty}{\sim} -\frac{2   \zeta (3)}{\pi ^2} \lambda+ \frac{52 \,\zeta (3)}{15}+\frac{8}{45}~.
\label{SCQ1}
\end{align}
The details on the intermediate steps can be found in the ancillary {\tt Mathematica} file.

\subsubsection*{\texorpdfstring{$\bullet \quad B_2$}{}}
To find the strong-coupling expansion of $B_2$, it is convenient to express (\ref{Q2}) as
\begin{align}
  B_2=   \iint_{-i \infty}^{i \infty}\frac{ds\,ds'}{(2\pi \ii)^2}\, \Big[B_{2,a}(s,s')+ B_{2,b}(s,s') \Big] \, , 
  \label{MBQ2}
\end{align}
where 
\begin{subequations}
\begin{align}
    \iint_{-i\infty}^{i \infty}\frac{ds\,ds'}{(2\pi \ii)^2 } \,B_{2,a}(s,s')&=\frac{64\sqrt{\lambda}}{3\pi}\sum_{k=1}^{\infty}k\; \textsf{Z}_{2k}^{(2)}\left(\widehat{\textsf{Z}}_{2k+1}^{(1)} -2\textsf{Z}_{2k+1}^{(1)}\right)~, \label{K1int} \\
    \iint_{-i \infty}^{i \infty}\frac{ds\,ds'}{(2\pi \ii)^2}\,B_{2,b}(s,s')&= -\frac{32 \sqrt{\lambda}}{3\pi}\sum_{k=1}^{\infty}k\, \textsf{Z}_{2k+1}^{(3)}\left(\widehat{\textsf{Z}}_{2k}^{(0)} -4\textsf{Z}_{2k}^{(0)}\right)~.
    \label{K2int}
\end{align}
\label{K12int}%
\end{subequations}
The explicit expressions of $B_{2,a}$ and $B_{2,b}$ can be found by first inserting the definitions \eqref{Zkp} and \eqref{Zhatkp} in the right-hand side of \eqref{K12int}, and then  using the Mellin-Barnes representation of the Bessel function and the integral representation of the Riemann $\zeta$-function. This leads to
\begin{subequations}
  \begin{align}
B_{2,a}(s,s')&= \frac{128 \sqrt{\lambda}}{3\pi} \sum_{k=1}^{\infty} k \;  \frac{\Gamma(k-s-1)\Gamma(k-s'-1)}{\Gamma(s+k+2)\Gamma(s'+k+3)} \times \nonumber \\
     & \quad\times\, \Gamma(2s+4)\Gamma(2s'+4) \zeta(2s+3) \zeta(2s'+3) \Big(\frac{\sqrt{\lambda}}{4\pi} \Big)^{2s+2s'+5} \big(4^{s'+1} -1  \big)~,\\[2mm]
B_{2,b}(s,s')&=-\frac{32 \sqrt{\lambda}}{3\pi} \sum_{k=1}^{\infty} k\;    \frac{\Gamma(k-s)\Gamma(k-s'-2)}{\Gamma(s+k+2)\Gamma(s'+k+3)} \,\times \nonumber \\
    &\quad \times \,\Gamma(2s+4)\Gamma(2s'+4) \zeta(2s+3) \zeta(2s'+3) \Big(\frac{\sqrt{\lambda}}{4\pi} \Big)^{2s+2s'+5} \big(4^{s'+2} -4 \big)\,.
\end{align}
\end{subequations}
After performing the sums over $k$ as shown in the ancillary {\tt Mathematica} file, a
drastic simplification occurs in the sum $B_{2,a}+B_{2,b}$, which becomes
\begin{align}
  &B_{2,a}(s,s')+B_{2,b}(s,s') = \label{Q2a+b} \\[2mm] &\quad=\frac{1}{3}\,
\frac{(4^{s'+1}-1) (s-s'-1) \zeta (2 s+3) \Gamma (-s) \Gamma(s+\frac{5}{2}) \zeta (2 s'+3) \Gamma (-s'-1) \Gamma (2 s'+4)}{2^{2 s+4 s'+1} \pi ^{2 s+2 s'+\frac{13}{2}}(s+s'+3) \Gamma (s'+3)} \lambda ^{s+s'+3}~.
\notag
\end{align}
Closing as always the integration contours counter-clockwise and summing the residues at the poles in the half-planes $\mathfrak{Re}(s)<0$ and $\mathfrak{Re}(s')<0$, we find
\begin{align}
   B_2 \underset{\lambda \to \infty}{\sim} \frac{8  \log (2)}{3 \pi^2 }\lambda+\frac{8  }{9 }-\frac{14   \zeta (3)}{3 }~.
   \label{SCQ2}
\end{align}
Again the details on the intermediate steps are given in the ancillary {\tt Mathematica} file. 

\subsubsection*{\texorpdfstring{$\bullet\quad B_3$}{}}
The strong-coupling behavior of $B_3$ can be readily obtained from the following strong-coupling expansions derived in \cite{Billo:2024ftq}, namely
\begin{align}
& \sum_{k=1}^\infty 2k\,\big(\mathsf{Z}_{2k}^{(2)}\big)^2 \underset{\lambda \rightarrow \infty}{\sim} ~\frac{1}{4}\log\lambda + \frac{1}{2}\gamma - \frac{1}{2}\log(4\pi)-\frac{1}{2}\zeta(3)+\frac{11}{12} ~,\\
&
-4\sum_{k=1}^{\infty}\,k\,\mathsf{Z}_{2k}^{(4)} \left(\widehat{\mathsf{Z}}_{2k}^{(0)}-4\,\mathsf{Z}_{2k}^{(0)}  \right) \,\underset{\lambda \rightarrow \infty}{\sim} \,\frac{3}{2}\,\zeta(3) ~,
\end{align}
together with the strong-coupling expansion of $\sum\limits_{k=1}^\infty 2k\,\mathsf{Z}_{2k}^{(2)}\;\widehat{\mathsf{Z}}_{2k}^{(2)}$ which we already computed for $A_2$ and given in \eqref{ZZstrongapp}. Combining these results we get 
\begin{align}
  B_3  \underset{\lambda \to \infty}{\sim} \frac{2\lambda}{\pi^2}\Big( \zeta(3)-\frac{4\log(2)}{3}\Big)+8 \log \Big(\frac{\lambda}{\pi^2}\Big)-14 \zeta (3)+\frac{88}{3}+16 \gamma -\frac{80\log(2)}{3}~.
  \label{SCQ3}
\end{align}
Adding up (\ref{SCQ1}), (\ref{SCQ2}) and (\ref{SCQ3}) we obtain the strong-coupling behavior of $\Delta_1^\D$ given in (\ref{Delta1strong}).

\section{Details on topological recursion method} \label{app:resolvent}

In this appendix, we provide more details on the topological recursion method that is used in the main text. The quantities of interest for us are the matrix-model correlators of $f(z_i)$ defined in \eqref{eq:fx}, which admit a topological expansion of the form
\begin{equation} \label{eq:W-funs1}
    \big\langle \prod_{i=1}^n f(z_i) \big\rangle_0^c = \sum_{g \geq 0} N^{2-2g-n} W_g^n(z_1,\cdots,z_n)~ .
\end{equation}
In the $\mathfrak{u}(N)$ Gaussian matrix model, the expansion coefficients $W_g^n$ can be explicitly determined using the topological recursion relation \cite{Eynard:2004mh,Eynard:2008we}, starting with the first case $W_0^1$. The first few terms can be found for example in the ancillary file of \cite{Chester:2020dja}. However, we are interested in the $\mathfrak{su}(N)$ Gaussian matrix model.\footnote{It is worth mentioning that for the leading and sub-leading orders in the large-$N$ expansion we consider here, the distinction between $\mathfrak{u}(N)$ and $\mathfrak{su}(N)$ turns out to be non relevant.} In the following we provide a systematic procedure to convert $\mathfrak{u}(N)$ correlators into $\mathfrak{su}(N)$ correlators.

The $\mathfrak{su}(N)$ vacuum expectation value of products of $f(z_i)$'s is defined as
\begin{equation}
\label{eq:suNcor}
    \big\langle\prod_{i=1}^nf(z_i)\big\rangle_0 = \frac{1}{\mathcal{Z}^{\mathfrak{su}(N)}}\int \!\prod_{u=1}^N da_u ~\Delta(a)\, \rme^{-\tr a^2} ~\delta\Big(\sum_{u=1}^N a_u \Big)
\prod_{i=1}^n \tr \exp\bigg(\sqrt{\frac{\lambda}{8\pi^2N}}\,a\,z_i\bigg)
\end{equation}
where $\mathcal{Z}^{\mathfrak{su}(N)}$ stands for $\mathfrak{su}(N)$ Gaussian matrix model partition function and $\Delta(a)$ is the Vandermonde determinant. We now use the Fourier representation of the $\delta$-function
\begin{equation}
    \delta \Big(\sum_{u=1}^N a_u \Big) =  \int_{-\infty}^\infty \frac{dp}{2\pi}\,  \exp \Big( \ii \, p \sum\limits_{u=1}^N a_u \Big) 
\end{equation}
and then shift $a_u\to a_u + \ii p/2$ to eliminate the linear term in $a_u$. In this way we produce an overall factor which is independent of $a_u$. After integrating over $p$, we obtain
\begin{equation}
    \big\langle \prod_{i=1}^n f(z_i)\big\rangle_0 =\rme^{-\frac{\lambda}{32\pi^2 N^2}\Big(\sum\limits_{i=1}^N z_i\Big)^2}\left[\frac{1}{\mathcal{Z}^{\mathfrak{u}(N)}}
    \int \!\prod_{u=1}^N da_u ~\Delta(a)\, \rme^{-\tr a^2}
    ~\prod_{i=1}^n f(z_i)\right]~.
\end{equation}
Notice that the quantity within square brackets in the right-hand side is precisely the correlator in the $\mathfrak{u}(N)$ Gaussian matrix model. By decomposing both sides into connected components and expanding at large $N$, we can recursively determine all $W_g^n$'s for the $\mathfrak{su}(N)$ Gaussian matrix model using the known $\mathfrak{u}(N)$ results. We now quote all the terms needed for this paper:  
\begin{equation}
    \begin{aligned} \label{eq:W-examples}
        W_0^1 (z_1)  &=- \frac{4\pi\, \ii }{z_1 \sqrt{\lambda}} \,J_1(x_1) ~, \\
         W_0^2 (z_1, z_2) &= \frac{\ii\,  z_1 z_2 \sqrt{\lambda} }{4\pi (z_1 + z_2)} \Big[ J_1\left(x_1 \right)
   J_2\left(x_2 \right)+J_2\left(x_1\right)
   J_1\left(x_2\right) \Big] ~, \\
W^3_0(z_1,z_2,z_3) &=  \frac{\ii\,z_1 \sqrt{\lambda}}{64\pi^2}J_1\left(x_1\right)  \bigg[ 8   \pi  
   J_1\left(x_2\right) J_1\left(x_3\right) - 4 \, \ii \, z_2 \sqrt{\lambda } \,   J_0\left(x_2\right)  J_1\left(x_3\right)\\
  & - \frac{\lambda}{2\pi}   \, z_2 z_3 J_0\left(x_2\right)
   J_0\left(x_3\right) + \frac{\lambda}{6\pi}   \, z_2 z_3 
   J_1\left(x_2\right) J_1\left(x_3\right)\bigg] + P(z_1,z_2,z_3)~, 
    \end{aligned}
\end{equation}
where $ x_i =    \frac{\ii \sqrt{\lambda}}{2\pi} z_i$ and $P(z_1,z_2,z_3)$ denotes the terms obtained by permuting $\{z_1,z_2,z_3\}$ in all possible ways. 

To apply the topological recursion, we need to express the derivatives of the partition function in terms of the connected correlators \eqref{eq:W-funs1}. Doing this for the $\mu/m$ mixed derivative, we find
\begin{align}
\partial_\mu^2\,\partial_m^2\log\mathcal{Z}^\DS\big|_{\D}
&= 4\Big[\big\langle\mathcal{M}_{2,\mathrm{F}}^\D\,\,\mathcal{M}_{2,\mathrm{A}}^\D\big\rangle-\big\langle\mathcal{M}_{2,\mathrm{F}}^\D\big\rangle\,\big\langle\mathcal{M}_{2,\mathrm{A}}^\D\big\rangle \Big] \\
&=\iint_0^\infty \!\frac{4 \,\omega\, \nu\, d \omega \,d \nu}{\sinh^2\omega \,\sinh^2 \nu}
\bigg\{ \Big[ 
\big\langle \rme^{-S^\D}f(2\ii \nu)^2\, f(2\ii\omega)\big\rangle_0^c - \big\langle \rme^{-S^\D} f(4\ii\nu)\, f(2\ii\omega)\big\rangle_0^c  \notag\\
&\quad +  2\,\big\langle \rme^{-S^\D} f(2\ii\nu)\big\rangle_0^c~ \big\langle \rme^{-S^\D} f(2\ii\nu) f(2\ii\omega)\big\rangle_0^c
   +(\omega \rightarrow - \omega) \Big] + (\nu \rightarrow -\nu )\bigg \}~.
   \label{eq:dmu2dm2int}
\end{align}
Expanding the exponential factors $\rme^{-S^\D}$ to the relevant order in the large-$N$ limit and using \eqref{eq:W-funs1}, we obtain the expression reported in \eqref{eq:F1intnew}. 

Similar considerations can be applied to the $\mu/b$ mixed derivative
\begin{align}
\partial_\mu^2\partial_b^2\log \mathcal{Z}^\DS\big|_{\D}=-4\Big[\big\langle\mathcal{C}_{\mathrm{F}}^\D\big\rangle-\big\langle\mathcal{M}_{2,\mathrm{F}}^\D\,\,\mathcal{B}_2^\D\big\rangle+\big\langle\mathcal{M}_{2,\mathrm{F}}^\D\big\rangle\,\big\langle\mathcal{B}_2^\D\big\rangle\Big]~.
\end{align}
Each term in this expression can again be written using connected correlators of $f$-functions. In particular, we have
\begin{equation} \label{eq:CFD}
-4\big\langle\mathcal{C}_{\mathrm{F}}^\D\big\rangle = 16N\Big(\zeta(3)-\frac{1}{3}\Big) + \!\int_0^\infty \!\frac{4\,\omega^2 \,d\omega}{\sinh^4 \omega} \big(\sinh (2 \omega)-2 \omega\big)\Big[\big\langle \rme^{-S^\D} f(2\ii\omega)\,\big\rangle^c_0 \,-\,N + (\omega \rightarrow -\omega)  \Big]~,
\end{equation}
and 
    \begin{align}
        \label{eq:MBD} &4\big\langle\mathcal{M}_{2,\mathrm{F}}^\D\,\,\mathcal{B}_2^\D\big\rangle-4\big\langle\mathcal{M}_{2,\mathrm{F}}^\D\big\rangle\,\big\langle\mathcal{B}_2^\D\big\rangle \notag \\
        &=\iint_0^\infty \frac{4\nu\,\big[\,\omega+\coth \omega \,(\omega \coth\omega-1)\big]d\omega\, d\nu }{\sinh^2 \omega \,\sinh^2 \nu} \bigg \{ \Big[   \big\langle \rme^{-S^\D} f(2\ii\nu)\, f(2\ii\omega)\big\rangle^c_0~ \big\langle \rme^{-S^\D} f(-2\ii\omega)\big\rangle^c_0 \notag\\&\quad 
        +\frac{1}{2} \big\langle \rme^{-S^\D} f(2\ii\nu) \,f(-2\ii\omega)\, f(2\ii \omega)\big\rangle^c_0 + (\omega \rightarrow -\omega)\Big] + (\nu \rightarrow -\nu) \bigg \}\\
        &\quad- \iint_0^\infty \frac{\nu\,\big(2\omega-\sinh2 \omega\big)d\omega \,d\nu }{\sinh^4 \omega \,\sinh^2 \nu} \bigg \{ \Big[ \big\langle \rme^{-S^\D} f(2\ii\nu) f(2\ii\omega)^2\big\rangle^c_0   -\big\langle \rme^{-S^\D} f(2\ii\nu) f(4\ii\omega)\big\rangle^c_0\notag\\ 
        &\quad+ 4 \big\langle \rme^{-S^\D} f(2\ii\nu) \,f(2\ii\omega)\big\rangle^c_0 +  2 \big\langle \rme^{-S^\D} f(2\ii\nu) \,f(2\ii\omega)\big\rangle^c_0 ~\big\langle \rme^{-S^\D} f(2\ii\omega)\big\rangle^c_0  + (\omega \rightarrow - \omega) \Big] \!+ (\nu \rightarrow - \nu) \bigg \}~. \notag
        \end{align}
After expanding the exponential factors $\rme^{-S^\D}$ and using \eqref{eq:W-funs1}, we obtain the expressions given in \eqref{eq:CDF} and \eqref{eq:M2F}. 

\section{\texorpdfstring{The Sp($N$) theory}{}}
\label{App:Sptheory}

In this appendix we analyze an $\cN=2$ SCFT with Sp$(N)$ gauge group and a matter content consisting of four fundamental hypermultiplets 
and one antisymmetric hypermultiplet with an $\mathrm{SU}(2)_L\times \mathrm{SO}(8)$ flavour symmetry. This theory, previously studied in \cite{Beccaria:2021ism,Beccaria:2022kxy,Behan:2023fqq,Alday:2024yax,Chester:2025ssu}, can be engineered in Type IIB string theory with $N$ D3-branes probing a $D_4$-singularity in F-theory.

We again consider mass and squashing deformations, adopting the same notation for the deformation parameters as in the \textbf{D}-theory. Specifically, we focus on the following quantities:
\begin{align}
\partial_\mu^2\,\partial_m^2\log\widetilde{\mathcal{Z}}^*\Big|_{\substack{m,\mu=0 \\ b=1}} \, , \qquad  \quad  \partial_\mu^2\partial_b^2\log \widetilde{\mathcal{Z}}^*\Big|_{\substack{m,\mu=0 \\ b=1}} \, , 
\end{align}
where $\widetilde{\mathcal{Z}}^{*}$ is the partition function of the deformed Sp($N$) theory. 
The first quantity was computed in \cite{Chester:2025ssu} using topological recursion; for completeness, we rederive their result using our full Lie-algebraic approach.

By exploiting supersymmetric localisation, the partition function $\widetilde{\mathcal{Z}}^{*} $ can be expressed as an integral over the eigenvalues $\tilde{a}_u$ of a matrix $\tilde{a}$ in the Lie Algebra $\mathfrak{sp}(N)$:
\begin{align}
\widetilde{\mathcal{Z}}^{*}=\int\!\prod_{u=1}^N d\tilde{a}_u ~ \rme^{-\frac{8\pi^2}{g_{_{\rm YM}}^2}\,\tr \tilde{a}^2} \, \big|Z_{\mathrm{1-loop}}\,Z_{\mathrm{inst}}\big|^2 \,\;,
\end{align}
with 
\begin{align}
    & |Z_{1-\text{loop}}|^2 = \Delta(\tilde{a})\;e^{-\widetilde{S}}\frac{\;\Upsilon'(0)^N\prod_{i<j=1}^N H_{\mathrm{v}}(\tilde{a}_{i}+\tilde{a}_j;b) H_{\mathrm{v}}(\tilde{a}_{ij};b)  }{ \prod_{i<j=1}^NH_{\mathrm{h}}(\tilde{a}_{i}+\tilde{a}_j,b,m)H_{\mathrm{h}}(\tilde{a}_{i}+\tilde{a}_j,b,-m)H_{\mathrm{h}}(\tilde{a}_{ij},b,m)H_{\mathrm{h}}(\tilde{a}_{ij},b,-m) } \nonumber \times \\
    & \quad \quad \quad \quad \quad \quad \quad \quad\times \frac{\prod_{i=1}^N H_{\mathrm{v}}(2\tilde{a}_i)}{ \prod_{F=1}^4\prod_{i=1}^N H_{\mathrm{h}}(\tilde{a}_i,b,\mu_{\mathrm{F}}) H_{\mathrm{h}}(\tilde{a}_i,b,-\mu_{\mathrm{F}})} \;.
    \label{SpZ1L}
\end{align}
Here $H_{\mathrm{v}}$ and $H_{\mathrm{h}}$ are the functions introduced in \eqref{Hfunctions}, $\Delta(\tilde{a})$ is the Vandermonde determinant and $\widetilde{S}$ is the matrix-model action of the massless Sp$(N)$ theory on the sphere \cite{Billo:2024ftq,Beccaria:2021ism}. 
In \eqref{SpZ1L}, $\mu_{\mathrm{F}}$ ($F=1, \ldots ,4$) are the masses of the four fundamental hypermultiplets, while $m$ is the mass of the anti-symmetric hypermultiplet. For simplicity, we take $\mu_{\mathrm{F}}= \mu$ for any $F$.

Following the same procedure as in Section\,\ref{subsecn:N2*}, we rewrite the partition function as \footnote{Decomposing $\tilde{a} = \tilde{a}^b T_b$ with $b=1,\dots,N(2N+1)$, where $T_b$ are the generators of $\mathrm{Sp}(N)$ in the fundamental representation normalized by $\tr(T_b T_c) = \tfrac{1}{2}\delta_{bc}$, the integration measure is
\begin{align*}
d\tilde{a} = \prod_{b=1}^{N(2N+1)} \frac{d\tilde{a}^b}{\sqrt{2\pi}}
\quad \text{so that} \quad \int \! d\tilde{a}~ \rme^{-\tr \tilde{a}^2} = 1~.
\end{align*}
}
\begin{align}
    \widetilde{\mathcal{Z}}^* = \int d\tilde{a} \; e^{-\text{tr}\tilde{a}^2 -\widetilde{S}^*}\;,
\end{align}
where 
\begin{align}
\widetilde{S}^*&= \widetilde{S} -N\log\Upsilon_b^\prime(0) -\sum_{i<j=1}^{N} \bigg[ \log H_{\mathrm{v}}\bigg(\sqrt{\frac{\lambda}{8\pi^2N}}\,\,(\tilde{a}_i+\tilde{a}_j);b\bigg)+\log H_{\mathrm{v}}\bigg(\sqrt{\frac{\lambda}{8\pi^2N}}\,\,\tilde{a}_{ij};b\bigg) \bigg]\notag  \nonumber\\
&-\sum_{i=1}^N\log H_{\mathrm{v}}\bigg(\sqrt{\frac{\lambda}{2\pi^2N}}\,\,\tilde{a}_{i};b\bigg) +8\sum_{i=1}^{N} \log H_{\mathrm{h}}\bigg(\sqrt{\frac{\lambda}{8\pi^2N}}\,\,\tilde{a}_{i};b,\mu\bigg) \nonumber \\ &+ 2\bigg[\sum_{i<j=1}^N  \log H_{\mathrm{h}}\bigg(\sqrt{\frac{\lambda}{8\pi^2N}}\,\,(\tilde{a}_{i}+\tilde{a}_j);b,m\bigg) + \sum_{i<j=1}^N  \log H_{\mathrm{h}}\bigg(\sqrt{\frac{\lambda}{8\pi^2N}}\,\,\tilde{a}_{ij};b,m\bigg)\;\bigg] \, , 
\label{SintSp}
\end{align}
with
\begin{align}
    \widetilde{S}= 4\sum_{k=1}^{\infty}(-1)^{k+1}\left(\frac{\lambda}{8\pi^2N}\right)^{k+1}(2^{2k}-1)\frac{\zeta(2k{+}1)}{k+1}\;\text{tr}\tilde{a}^{2k+2} \;.
\end{align}
This leads to the following expansion around $m, \mu=0$ and $b=1$:
\begin{align}
\widetilde{S}^{*} &= \widetilde{S}+m^2\,\widetilde{\mathcal{M}}_{2,\mathrm{A}}+\mu^2\,\widetilde{\mathcal{M}}_{2,\mathrm{F}}+\left[(b-1)^2-(b-1)^3\right]\widetilde{\mathcal{B}}_2+m^4\,\widetilde{\mathcal{M}}_{4,\mathrm{A}}+\mu^4\,\widetilde{\mathcal{M}}_{4,\mathrm{F}} \notag \\[1mm]
& \quad+m^2(b-1)^2\,\widetilde{\mathcal{C}}_{\mathrm{A}} +\mu^2(b-1)^2\,\widetilde{\mathcal{C}}_{\mathrm{F}} +(b-1)^4\,\widetilde{\mathcal{B}}_4+\ldots~.
\label{SSpSexp}
\end{align}
Here we report only the coefficients of the above expansion that are necessary for evaluating the integrated correlators of interest:
\begin{subequations}
\begin{align}
\widetilde{\mathcal{M}}_{2,\mathrm{A}}
    &= \frac{1}{2}\widetilde{\mathcal{M}}^\D_{2,\mathrm{A}} +N(N-1)(1+\gamma) \, ,
    \label{M2ASp}\\[1mm]
    \widetilde{\mathcal{M}}_{2,\mathrm{F}}&= \widetilde{\mathcal{M}}_{2,\mathrm{F}}^\D +4N(1+\gamma) \, , \label{M2FSp}\\[1mm]
\widetilde{\mathcal{B}}_{2}
    &=  N(2N+1)(1+ \gamma) -\frac{1}{6}\sum_{n=0}^{\infty}(-1)^n [4n\zeta(2n{+}1)+(2n+3)\zeta(2n{+}3)]\Big(\frac{\lambda}{2\pi^2N}  \Big)^{n+1} \! \tr \tilde{a}^{2n+2} \nonumber \\
    & \quad+ \frac{4}{3} \sum_{n=0}^{\infty}(-1)^n (2n\zeta(2n+1)-(2n+3)\zeta(2n+3)) \Big(\frac{\lambda}{8\pi^2N} \Big)^{n+1}\text{tr}\tilde{a}^{2n+2} \nonumber \\
    & \quad+ \frac{1}{2}\sum_{n=1}^{\infty}\sum_{k=0}^n (-1)^n \frac{(2n+1)!\zeta(2n+1)}{(2k)!(2n-2k)!} \Big(\frac{\lambda}{8\pi^2N} \Big)^{n}\text{tr}\tilde{a}^{2n-2k}\text{tr}\tilde{a}^{2k}  \, , 
    \label{B2Sp}\\[1mm]    
    \widetilde{\mathcal{C}}_{F}
    &= \frac{1}{4}\,\widetilde{\mathcal{C}}_{\mathrm{F}}^{\,\D} + N\bigg(\frac{1}{3}-\zeta(3) \bigg) \;,
    \label{CfSp}
\end{align}%
\end{subequations}
where $\widetilde{\mathcal{M}}^\D_{2,\mathrm{A}}$, $\widetilde{\mathcal{M}}^\D_{2,\mathrm{F}}$ and $\widetilde{\mathcal{C}}_{\mathrm{F}}^{\,\D}$ 
denote the same quantities as in (\ref{M2aDtr}), (\ref{M2fDtr}), and (\ref{CfDtr}), respectively, but expressed in terms of the Sp($N$) matrix $\tilde{a}$.

\subsection*{\texorpdfstring{Results for $\partial_\mu^2\,\partial_m^2\log\widetilde{\mathcal{Z}}^*$}{}}

At first we evaluate the mixed derivative 
\begin{align}
\partial_\mu^2\,\partial_m^2\log\widetilde{\mathcal{Z}}^*\Big|_{\substack{m,\mu=0 \\ b=1}}\,=\,4\Big[\big\langle\widetilde{\mathcal{M}}_{2,\mathrm{F}}\,\,\widetilde{\mathcal{M}}_{2,\mathrm{A}}\big\rangle_{\text{Sp}}-\big\langle\widetilde{\mathcal{M}}_{2,\mathrm{F}}\big\rangle_{\text{Sp}}\,\big\langle\widetilde{\mathcal{M}}_{2,\mathrm{A}}\big\rangle_{\text{Sp}} \Big] \;,
\label{intcorr1Sp}
\end{align}
where we have defined 
\begin{align}
    \langle f(\tilde{a})\rangle_{\text{Sp}}= \frac{\langle \;\rme^{-\widetilde{S}} ~f(\tilde{a})\rangle_{0,\text{Sp}}}{\langle  \;\rme^{-\widetilde{S}} \rangle_{0,\text{Sp}}},
\end{align}
with $\langle \, \cdot \, \rangle_{0,\text{Sp}}$ denoting the vacuum expectation value in the Sp($N$) $\mathcal{N}=4$ SYM theory.
In analogy with what done in Section\,\ref{secn:integrated}, we introduce the $\widetilde{\mathcal{P}}$ operators defined through 
\begin{align}
    \text{tr}\;\tilde{a}^k = \bigg( \frac{N}{2}  \bigg)^{\frac{k}{2}}\sum_{l=0}^{[\frac{k-1}{2}]} \sqrt{2(k-2l)} \binom{k}{l} \widetilde{\mathcal{P}}_{k-2l}  + \langle\text{tr}\;\tilde{a}^k   \rangle_{0,\text{Sp}} \;.
    \label{PSp}
\end{align}
This definition ensures that the $\widetilde{\mathcal{P}}$ operators are orthonormal in the Gaussian model at large $N$, namely
\begin{align}
    \langle \widetilde{\mathcal{P}}_{2k_1}\widetilde{\mathcal{P}}_{2k_2}\rangle_{0,\text{Sp}}= \delta_{k_1,k_2}+ O\big(1/N\big)\;.
\end{align}
Writing $\widetilde{\mathcal{M}}_{2,\mathrm{A}}$ and $\widetilde{\mathcal{M}}_{2,\mathrm{F}}$ in the $\widetilde{\mathcal{P}}$-basis, substituting the resulting expressions into \eqref{intcorr1Sp} and recalling that \cite{Billo:2024kri}
\begin{subequations}
    \begin{align}
\big\langle \widetilde{  \mathcal{P}}_{2k}\,\widetilde{\mathcal{P}}_{2\ell}\big\rangle^c_{\text{Sp}} &=\,\delta_{k,\ell}+
    \frac{\sqrt{k\;\ell}\,\,\,}{2N}\left(1+ 4 \mathsf{Y} \right)+O(1/N^2)~, \label{P2nP2mSp} \\
\big\langle  \widetilde{\mathcal{P}}_{2k}\,\widetilde{\mathcal{P}}_{2\ell}\,\widetilde{\mathcal{P}}_{2m}\big\rangle_{\text{Sp}}-\big\langle  \widetilde{\mathcal{P}}_{2k}\big\rangle_{\text{Sp}}\big\langle\widetilde{\mathcal{P}}_{2\ell}\,\widetilde{\mathcal{P}}_{2m}\big\rangle_{\text{Sp}}&=\, \sqrt{2}\left(\delta_{k,\ell}\,\mathsf{Y}_{2m}+\delta_{k,m}\,\mathsf{Y}_{2\ell} \right)+O(1/N)~,\label{PPPevenSp}
\end{align}
\label{vevsPSp}%
\end{subequations}
we obtain the following large-$N$ expansion
\begin{align}
\partial_\mu^2\,\partial_m^2\log\widetilde{\mathcal{Z}}^*\Big|_{\substack{m,\mu=0 \\ b=1}}\,=\sum_{g=0}^\infty N^{1-g}\,\mathcal{F}_g^{\text{Sp}} \;.
\label{dmudmlargeNSp}
\end{align}
After some algebra, one can show that
\begin{subequations}
       \begin{align}
          & \mathcal{F}_0^{\text{Sp}}= 2 \mathcal{F}_0^\D \label{F0SpF0D}\\
          & \mathcal{F}_1^{\text{Sp}}= 2 \mathcal{F}_1^{\D} + \Sigma_1\label{F1SpF1D}
       \end{align} 
\end{subequations}
where $\mathcal{F}_0^{\D}, \mathcal{F}_1^{\D}$ are the coefficients of large-$N$ expansion integrated correlator in the $\mathbf{D}$-theory, and 
\begin{align}
    \Sigma_1 =& \frac{64\,\pi}{\sqrt{\lambda}}\,\sum_{k=1}^\infty(-1)^k\,(2k)\,\,\mathsf{Z}_{2k}^{(2)}\,\sum_{\ell=1}^\infty(-1)^\ell\,(2\ell)\,\mathsf{M}_{1,2\ell}^{(1)} +16\sum_{k=1}^\infty\,(2k)\,\mathsf{Z}_{2k}^{(2)}\,\mathsf{M}_{0,2k}^{(2)} \nonumber \\ 
    & -16\,\sum_{k=1}^\infty(2k)\,\,\mathsf{Z}_{2k}^{(2)}\,\Big(\mathsf{Z}_{2k}^{(2)}-\widehat{\mathsf{Z}}_{2k}^{(2)}\Big) \; . 
\end{align}
The quantities  $\mathsf{M}_{m,n}^{(p)}$ are defined in 
\eqref{Mkp}, while $\mathsf{Z}_{m}^{(p)}$ and $\widehat{\mathsf{Z}}_{2k}^{(2)}$ are given in 
\eqref{Zkp} and \eqref{Zhatkp}, respectively. 
Combining (\ref{F0F1}) and (\ref{F0SpF0D}), we conclude that
$\mathcal{F}_0^{\text{Sp}}= F_1$, in agreement with what is reported in Eq.\,(B.11) of \cite{Chester:2025ssu}. Substituting the integral definitions in terms of Bessel functions, it is also straightforward to prove that $\mathcal{F}_1^{\text{Sp}}= F_2$, where $F_2$ is written in the ancillary {\tt Mathematica} file attached to \cite{Chester:2025ssu}.

 The strong-coupling expansion of $\mathcal{F}_0^{\text{Sp}}$ trivially follows from that of $\mathcal{F}_0^{\D}$ given in \eqref{F1strongfinal}. The one of $\mathcal{F}_1^{\text{Sp}}$ can be obtained from the strong-coupling expansion of 
 $\mathcal{F}_1^{\D}$ in \eqref{kF1D}, and that of $\Sigma_1$, 
 which is
\begin{align} \label{eq:sigma1}
 \Sigma_1& \underset{\lambda \to \infty}{\sim} 4+ 8 \log(2) -3\zeta(3) + \sum_{n=1}^{\infty} \frac{32\, n\, \Gamma \left(n-\frac{1}{2}\right) \Gamma \left(n+\frac{1}{2}\right)^2 \zeta (2 n+1) }{\pi ^{3/2} \;\Gamma (n)\;\lambda ^{n+\frac{1}{2}}} \; .
\end{align}
This can be derived following the same methods described in Appendix \ref{app:strong}. 
Putting everything together, we have
\begin{align}
\mathcal{F}_1^{\rm Sp}~\underset{\lambda \rightarrow \infty}{\sim}& -12\log\Big(\frac{\lambda}{\pi^2}\Big) +k_{\mathcal{F}_1^{\rm Sp}}-\log(2)\sum_{n=1}^\infty \frac{64\,n\,\Gamma(n-\frac{1}{2})^2\,\Gamma(n+\frac{3}{2})\,\zeta(2n+1)}{\pi^{7/2}\,\Gamma(n)\,\lambda^{n-1/2}}\notag\\
    &\quad+\sum_{n=1}^\infty \frac{64\,n\,\Gamma(n-\frac{1}{2})\,\Gamma(n+\frac{1}{2})^2\,\zeta(2n+1)}{\pi^{3/2}\,\Gamma(n)\,\lambda^{n+1/2}}\, , 
\end{align}
with
\begin{align}
    k_{\mathcal{F}_1^{\rm Sp}} = \frac{96 \zeta (3)}{5}-\frac{182}{5}-24 \gamma +40 \log (2) ~. 
\end{align}
This result is not only in complete agreement with \cite{Chester:2025ssu}, but also analytically fixes the constant $k_{\mathcal{F}_1^{\rm Sp}}$ which in that reference was only estimated numerically.

\subsection*{\texorpdfstring{Results for $\partial_\mu^2\partial_b^2\log \widetilde{\mathcal{Z}}^*$}{}}

Following the same procedure, we can also obtain the large-$N$ expansion of 
\begin{align}
\partial_\mu^2\partial_b^2\log \widetilde{\mathcal{Z}}^*\Big|_{\substack{m,\mu=0 \\ b=1}}=-4\Big[\big\langle\widetilde{\mathcal{C}}_{\mathrm{F}}\big\rangle_{\text{Sp}}-\big\langle\widetilde{\mathcal{M}}_{2,\mathrm{F}}\,\,\widetilde{\mathcal{B}}_2\big\rangle_{\text{Sp}}+\big\langle\widetilde{\mathcal{M}}_{2,\mathrm{F}}^\D\big\rangle_{\text{Sp}}\,\big\langle\widetilde{\mathcal{B}}_2^\D\big\rangle_{\text{Sp}}\Big]~,
\label{intcorr2Sp}
\end{align}
which takes the form
\begin{align} \partial_\mu^2\,\partial_b^2\log\widetilde{\mathcal{Z}}^*\Big|_{\substack{m,\mu=0 \\ b=1}}\,=\sum_{g=0}^\infty N^{1-g}\,\widetilde{\mathcal{F}}_g^{\text{Sp}} \;.
\label{dmudblargeNSp}
\end{align}
The first two coefficients are 
\begin{subequations}
\begin{align}
    & \widetilde{\mathcal{F}}_0^{\text{Sp}}= 2 \widetilde{\mathcal{F}}_0^{\D} \, , \nonumber \\
    & \widetilde{\mathcal{F}}_1^{\text{Sp}}= 2 \widetilde{\mathcal{F}}_1^{\D} + \Sigma_1+ \widetilde{\Sigma}_1\;,
\end{align}
\end{subequations}
where the new quantity $\widetilde{\Sigma}_1$ is
\begin{align}
\widetilde{\Sigma}_1 =  \frac{1}{3}\bigg(\frac{\lambda}{\pi^2}+4\,\bigg)\textsf{Z}_0^{(4)} + \frac{4\,\lambda^{1/2}}{3\pi}\,\textsf{Z}_1^{(3)} -16\,\zeta(3) \;.
\end{align}

The strong-coupling expansion of $\widetilde{\mathcal{F}}_0^{\text{Sp}}$ readily follows from that of $\widetilde{\mathcal{F}}_0^{\D}$ given in \eqref{eq:Ftitle}, 
while the strong-coupling expansion of $\widetilde{\mathcal{F}}_1^{\text{Sp}}$ is obtained by adding those of $\widetilde{\mathcal{F}}_1^{\D}$, given in \eqref{eq:Ftitle}, and of $\Sigma_1$, given in \eqref{eq:sigma1}, and the strong-coupling expansion of $\widetilde{\Sigma}_1$ which is
\begin{align}
    \widetilde{\Sigma}_1 & \underset{\lambda \to \infty}{\sim} \frac{8}{3}-16 \zeta(3)\; .
\end{align}
Remarkably, at strong-coupling, $\widetilde{\Sigma}_1$ simply behaves only as a constant.
The final result is
\begin{align}
\widetilde{\mathcal{F}}_{1}^{\text{Sp}}~\underset{\lambda \rightarrow \infty}{\sim}&
    4\log\Big(\frac{\lambda}{\pi^2}\Big) +k_{\widetilde{\mathcal{F}}_{1}^{\text{Sp}}}-\log(2)\sum_{n=1}^\infty \frac{64\,n\,\Gamma(n-\frac{1}{2})^2\,\Gamma(n+\frac{3}{2})\,\zeta(2n+1)}{\pi^{7/2}\,\Gamma(n)\,\lambda^{n-1/2}}\notag\\
    &\quad+\sum_{n=1}^\infty \frac{64\,n\,\Gamma(n-\frac{1}{2})\,\Gamma(n+\frac{1}{2})^2\,\zeta(2n+1)}{\pi^{3/2}\,\Gamma(n)\,\lambda^{n+1/2}} \, , 
\end{align}
where
\begin{align}
    k_{\widetilde{\mathcal{F}}_{1}^{\text{Sp}}} = -\frac{136 \zeta (3)}{5}+\frac{406}{15}+8 \gamma -\frac{40 \log (2)}{3} ~. 
\end{align}

Finally, we can also consider the large-$N$ expansion with fixed YM coupling. Using the strong-coupling expressions we derived in this appendix and exploiting the SL$(2, \mathbb{Z})$-completion in terms of non-holomorphic Eisenstein series as proposed in Section\,\ref{secn:verystrong}, we obtain the modular invariant expression given in \eqref{d2bd2muexpandedESp} of the main text.

\bibliography{main.bib}

\end{document}